\documentclass[12pt]{spieman}
\usepackage[titletoc,title]{appendix}
\usepackage{graphicx}
\graphicspath{{Figures/}{../Figures/}} 
\usepackage{rotating} 
\usepackage{subfiles} 
\usepackage{amsmath} 
\usepackage{xr} 
\usepackage{siunitx} 
\usepackage{placeins} 

\title{Design and Performance of the Prototype Schwarzschild-Couder Telescope Camera}

\author[0]{C.~B.~Adams}
\author[1]{G.~Ambrosi}
\author[2]{M.~Ambrosio}
\author[2]{C.~Aramo}
\author[3]{T. ~Arlen  }
\author[4]{W.~Benbow}
\author[1,5]{B.~Bertucci}
\author[6,7]{E.~Bissaldi}
\author[8,9]{J.~Biteau}
\author[10]{M.~Bitossi}
\author[2]{A.~Boiano}
\author[2]{C.~Bonavolont\`a}
\author[11]{R.~Bose}
\author[8,12]{A.~Bouvier}
\author[13]{M.~Buscemi}
\author[0]{A.~Brill}
\author[14]{A.~M.~Brown}
\author[11]{J.~H.~Buckley}
\author[15]{R.~Canestrari}
\author[16]{M.~Capasso}
\author[1]{M.~Caprai}
\author[17]{P.~Coppi}
\author[18]{C.~E.~Covault}
\author[19,20]{D.~Depaoli}
\author[6,7]{L.~Di~Venere}
\author[11]{M.~Errando}
\author[21]{S.~Fegan}
\author[16]{Q.~Feng}
\author[1,5]{E.~Fiandrini}
\author[22]{A.~Furniss}
\author[23]{M.~Garczarczyk}
\author[24]{A.~Gent}
\author[6,7]{N.~Giglietto}
\author[6,7]{F.~Giordano}
\author[25]{E.~Giro}
\author[26]{R.~Halliday}
\author[8]{O.~Hervet}
\author[4,27]{G.~Hughes}
\author[28,13]{S.~Incardona}
\author[29]{T.~B.~Humensky}
\author[1]{M.~Ionica}
\author[30]{W.~Jin}
\author[8,31]{C.~A.~Johnson}
\author[32]{D.~Kieda}
\author[33]{F.~Krennrich}
\author[8,34]{A.~Kuznetsov}
\author[35]{J.~Lapington}
\author[7]{F.~Licciulli}
\author[6,7]{S.~Loporchio}
\author[28,13]{G.~Marsella}
\author[2]{V.~Masone}
\author[24,36]{K.~Meagher}
\author[36]{T.~Meures}
\author[36]{B.~A.~W.~Mode}
\author[37]{S.~A.~I.~Mognet}
\author[16]{R.~Mukherjee}
\author[38]{A.~Okumura}
\author[6,7]{F.~R.~Pantaleo}
\author[39,10]{R.~Paoletti}
\author[20]{F.~Di~Pierro}
\author[0]{D.~Ribeiro}
\author[36]{L.~Riitano}
\author[4]{E.~Roache}
\author[35]{D.~Ross}
\author[40]{J.~Rousselle}
\author[10]{A.~Rugliancich}
\author[30]{M.~Santander}
\author[8,41]{M.~Schneider}
\author[42]{H.~Schoorlemmer}
\author[43]{R.~Shang}
\author[43]{B.~Stevenson}
\author[39,10]{L.~Stiaccini}
\author[38]{H.~Tajima}
\author[36]{L.~P.~Taylor}
\author[35]{J.~Thornhill}
\author[1,5]{L.~Tosti}
\author[28,13]{G.~Tripodo}
\author[1,44]{V.~Vagelli}
\author[45,2]{M.~Valentino}
\author[36]{J.~Vandenbroucke}
\author[43]{V.~V.~Vassiliev}
\author[46]{S.~P.~Wakely}
\author[47]{J.~J.~Watson}
\author[42]{R.~White}
\author[48,49]{P.~Wilcox}
\author[8]{D.~A.~Williams}
\author[43,50]{M. ~Wood}
\author[43]{P.~Yu}
\author[51]{A.~Zink}

\affil[0]{Physics Department, Columbia University, New York, NY 10027, USA}
\affil[1]{INFN Sezione di Perugia, 06123 Perugia, Italy}
\affil[2]{INFN Sezione di Napoli, 80126 Napoli, Italy}
\affil[3]{Department of Physics and Astronomy, University of California, Los Angeles, CA 90095, USA  }
\affil[4]{Center for Astrophysics | Harvard \& Smithsonian, Cambridge, MA 02138, USA}
\affil[5]{Dipartimento di Fisica e Geologia dell’Universit\`a degli Studi di Perugia, 06123 Perugia, Italy}
\affil[6]{Dipartimento Interateneo di Fisica dell’Universit\`a e del Politecnico di Bari, 70126 Bari, Italy}
\affil[7]{INFN Sezione di Bari, 70125 Bari, Italy}
\affil[8]{Santa Cruz Institute for Particle Physics and Department of Physics, University of California, Santa Cruz, CA 95064, USA}
\affil[9]{Now at: Universit\'e Paris-Saclay, CNRS/IN2P3, IJCLab, 91405 Orsay, France}
\affil[10]{INFN Sezione di Pisa, 56127 Pisa, Italy}
\affil[11]{Department of Physics, Washington University, St. Louis, MO 63130, USA}
\affil[12]{Now at: Verily Life Sciences, South San Francisco, CA 94080, USA}
\affil[13]{INFN Sezione di Catania, 95123 Catania, Italy}
\affil[14]{Dept. of Physics and Centre for Advanced Instrumentation, Durham University, Durham DH1 3LE, United Kingdom}
\affil[15]{INAF IASF Palermo, 90146 Palermo, Italy}
\affil[16]{Department of Physics and Astronomy, Barnard College, Columbia University, NY 10027, USA}
\affil[17]{Department of Astronomy, Yale University, New Haven, CT 06520, USA}
\affil[18]{Department of Physics, Case Western Reserve University, Cleveland, Ohio 44106, USA}
\affil[19]{Dipartimento di Fisica dell’Universit\`a degli Studi di Torino, 10125 Torino, Italy}
\affil[20]{INFN Sezione di Torino, 10125 Torino, Italy}
\affil[21]{LLR/Ecole Polytechnique, Route de Saclay, 91128 Palaiseau Cedex, France}
\affil[22]{Department of Physics, California State University - East Bay, Hayward, CA 94542, USA}
\affil[23]{Deutsches Elektronen Synchrotron (DESY), Platanenallee 6, D-15738 Zeuthen, Germany}
\affil[24]{School of Physics \& Center for Relativistic Astrophysics, Georgia Institute of Technology, Atlanta, GA 30332-0430, USA}
\affil[25]{INAF Osservatorio Astronomico di Padova, 35122 Padova, Italy}
\affil[26]{Dept. of Physics and Astronomy, Michigan State University, East Lansing, MI 48824, USA}
\affil[27]{CTAO, Saupfercheckweg 1, 69117 Heidelberg, Germany}
\affil[28]{Dipartimento di Fisica e Chimica "E. Segr\`e", Universit\`a degli Studi di Palermo, via delle Scienze, 90128 Palermo, Italy}
\affil[29]{Science Department, SUNY Maritime College, Throggs Neck, NY 10465}
\affil[30]{Department of Physics and Astronomy, University of Alabama, Tuscaloosa, AL 35487, USA}
\affil[31]{Now at: NextEra Analytics, St. Paul, MN 55107, USA}
\affil[32]{Department of Physics and Astronomy, University of Utah, Salt Lake City, UT 84112, USA}
\affil[33]{Department of Physics and Astronomy, Iowa State University, Ames, IA 50011, USA}
\affil[34]{Now at: Apple Inc., Cupertino, CA 95014, USA}
\affil[35]{Space Research Centre, University of Leicester, University Road, Leicester, LE1 7RH, United Kingdom}
\affil[36]{Department of Physics and Wisconsin IceCube Particle Astrophysics Center, University of Wisconsin, Madison, WI 53706, USA}
\affil[37]{Pennsylvania State University, University Park, PA 16802, USA}
\affil[38]{Institute for Space--Earth Environmental Research and Kobayashi--Maskawa Institute for the Origin of Particles and the Universe, Nagoya University, Nagoya 464-8601, Japan}
\affil[39]{Dipartimento di Scienze Fisiche, della Terra e dell'Ambiente, Universit\`a degli Studi di Siena, 53100 Siena, Italy}
\affil[40]{Subaru Telescope NAOJ, Hilo HI 96720, USA}
\affil[41]{Univerity of Maryland}
\affil[42]{Max-Planck-Institut für Kernphysik, P.O. Box 103980, 69029 Heidelberg, Germany}
\affil[43]{Department of Physics and Astronomy, University of California, Los Angeles, CA 90095, USA}
\affil[44]{Agenzia Spaziale Italiana, 00133 Roma, Italy}
\affil[45]{CNR-ISASI, 80078 Pozzuoli, Italy}
\affil[46]{Enrico Fermi Institute, University of Chicago, Chicago, IL 60637, USA}
\affil[47]{Deutsches Elektronen-Synchrotron, Platanenallee 6, D-15738 Zeuthen, Germany}
\affil[48]{School of Physics and Astronomy, University of Minnesota, Minneapolis, MN 55455, USA}
\affil[49]{Department of Physics and Astronomy, St. Cloud State University, St. Cloud, MN, 56301}
\affil[50]{Now at: Facebook Inc. Menlo Park, CA 94025}
\affil[51]{Friedrich-Alexander-Universit\"at Erlangen-N\"urnberg, Erlangen Centre for Astroparticle Physics, D 91058 Erlangen, Germany}

\begin{document} 
\maketitle

\begin{abstract} \label{Abstract}
The prototype Schwarzschild-Couder Telescope (pSCT) is a candidate for a medium-sized telescope in the Cherenkov Telescope Array. The pSCT is based on a novel dual mirror optics design which reduces the plate scale and allows for the use of silicon photomultipliers as photodetectors. 

The prototype pSCT camera currently has only the central sector instrumented with 25 camera modules (1600 pixels), providing a 2.68$^{\circ}$ field of view (FoV). The camera electronics are based on custom TARGET (TeV array readout with GSa/s sampling and event trigger) application specific integrated circuits. Field programmable gate arrays sample incoming signals at a gigasample per second. A single backplane provides camera-wide triggers. An upgrade of the  pSCT camera is in progress, which will fully populate the focal plane. This will increase the number of pixels to 11,328, the number of backplanes to 9, and the FoV to 8.04$^{\circ}$. Here we give a detailed description of the pSCT camera, including the basic concept, mechanical design, detectors, electronics, current status and first light.

\end{abstract}

\keywords{Cherenkov Telescope Array, Instrumentation, Imaging Atmospheric Cherenkov Telescopes, prototype Schwarzschild-Couder Telescope, very-high-energy gamma-ray astronomy, silicon photomultipliers}

{\noindent \footnotesize\textbf{*}L.~P.~Taylor,  \linkable{ltaylor23@wisc.edu} }

\begin{spacing}{2}

\section{Introduction}\label{Introduction}
The current energy frontier for high-energy gamma-ray astronomy is around 50~TeV and increased detection area is vital to future high-energy gamma-ray instruments. Along with ground-based extensive air shower arrays (e.g., HAWC or LHAASO), large arrays of imaging atmospheric Cherenkov telescopes (IACTs) are the most promising instruments to push the energy frontier to higher energies \cite{actis2011design}.

Earth's atmosphere is opaque to very-high-energy (VHE; $\ge$100~GeV) photons and the fluxes are too small for satellite detection; thus, they must be detected from the ground through indirect means.  VHE gamma rays initiate extensive air showers. The shower constituent particles move faster than the speed of light in the atmosphere and thus emit Cherenkov radiation.  Showers reach their maximum development at around 10~km above sea level. Cherenkov angles range from 0.8$^{\circ}$ at 10~km above sea level to 1.4$^{\circ}$ at sea level. For vertical showers these opening angles produce a Cherenkov light pool with a radius of about 120~m at 1500~m above sea level \cite{2018NIMPA}. IACTs record the Cherenkov light produced by these showers. At energies greater than 100~GeV, cosmic rays are much more abundant than gamma-ray events and hence are the primary backgrounds at these energies \cite{2018NIMPA}.

The Cherenkov Telescope Array (CTA) is a ground-based observatory for VHE gamma rays. The CTA ``alpha configuration''  will have two sites, one in the northern hemisphere in La Palma, Spain, and one in the southern hemisphere in Paranal, Chile. The array will cover 3~\si{km^2} in the south and 0.25~\si{km^2} in the north and will have telescope spacings of 100--300~m. 

In order to cover a wide energy range, CTA will use three different sizes of telescopes: small-sized (SST), medium-sized (MST), and large-sized (LST). These three sizes are designed to access different energies between 20~GeV and 300~TeV.  The northern site will have 4 LSTs and 9 MSTs. The southern site is planned to have 14 MSTs, and 37 SSTs \cite{consortium2017science}. 

Since the highest energy photons from extragalactic sources are absorbed by collisions with the extragalactic background light, and the preponderance of Galactic sources are in the southern hemisphere, only the southern site includes the SSTs which are needed for good sensitivity at 100~TeV and above. Together, these sites will cover the whole sky and will have approximately an order of magnitude greater sensitivity than current instruments \cite{consortium2017science}.  

\begin{figure}[ht]
    \centering
	\includegraphics[width=\textwidth]{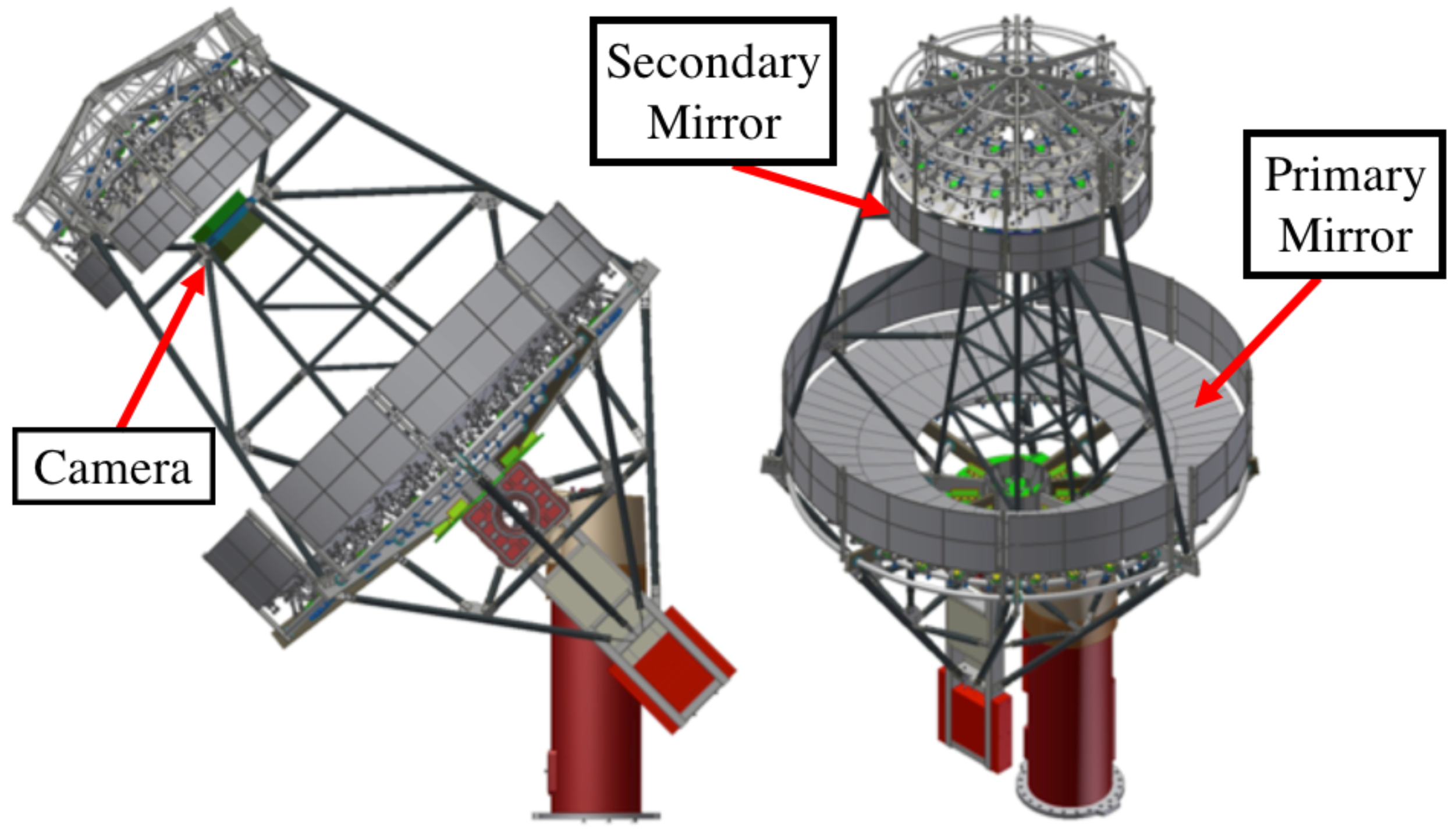}
	\caption{Artists rendition of the prototype Schwarzschild-Couder Telescope. Schwarzschild-Couder optics utilize two mirrors and a curved focal plane to achieve excellent resolution and sensitivity. The Primary (9.66~m) and Secondary (5.4~m) Mirrors are labeled above along with the location of the camera. Figure taken from \cite{Wachtendonk2018}.}
	\label{fig:TelescopeImage}
\end{figure}

\begin{figure}[ht]
    \centering
	\includegraphics[width=\textwidth]{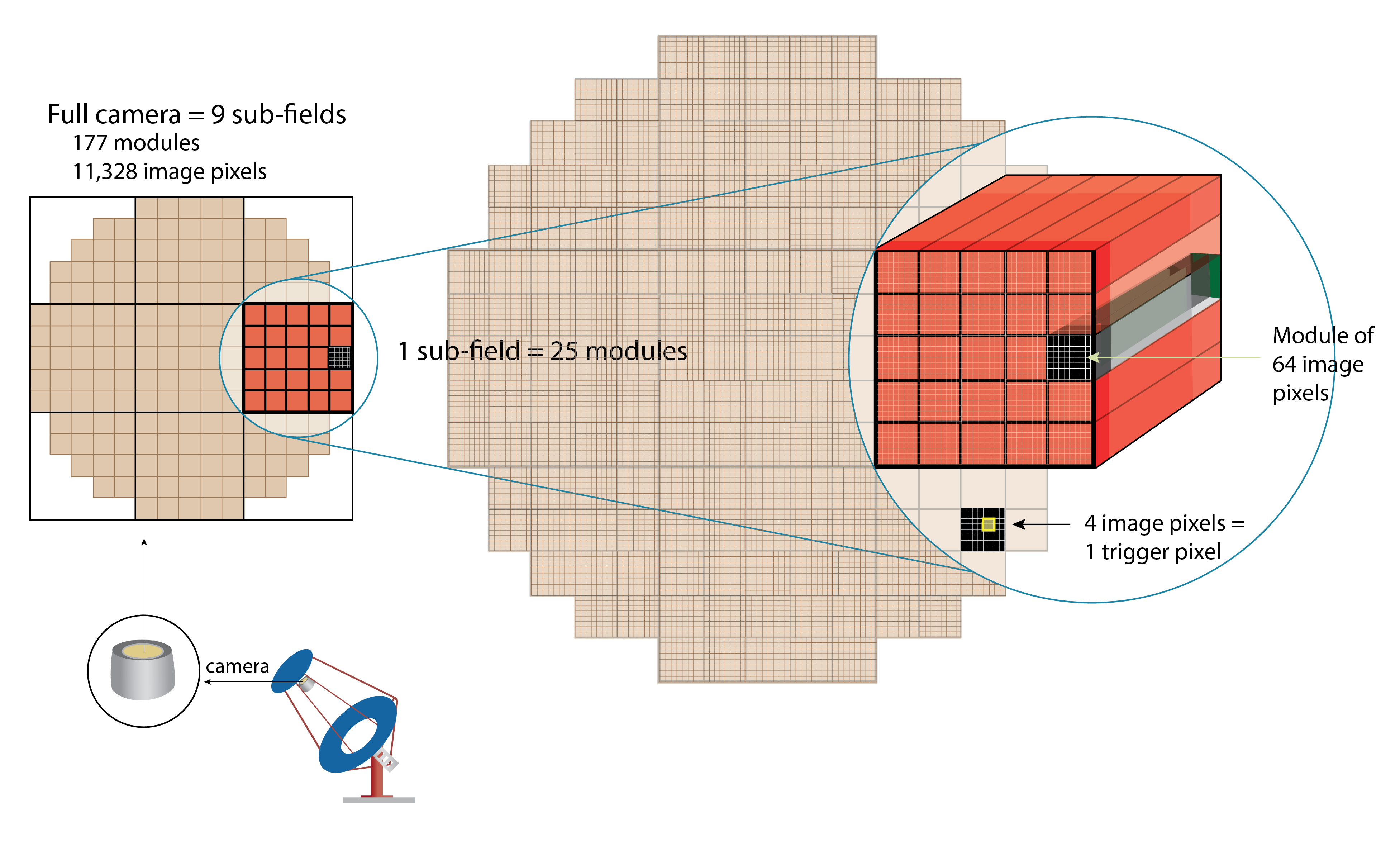}
	\caption{The pSCT camera has a hierarchical design. The full camera is comprised of 9 sectors, each of which can hold up to 25 camera modules (see Section \ref{Modules}). The sectors in each corner of the camera are equipped with fewer modules so that the entire camera has 177 modules. Each module has 64 image pixels so that the full camera will have 11,328 pixels. The current pSCT camera has only the center sector populated with modules.}
	\label{fig:cta_cameraDesign}
\end{figure}

\begin{sidewaysfigure}[p]
    \centering
	\includegraphics[width=\textwidth]{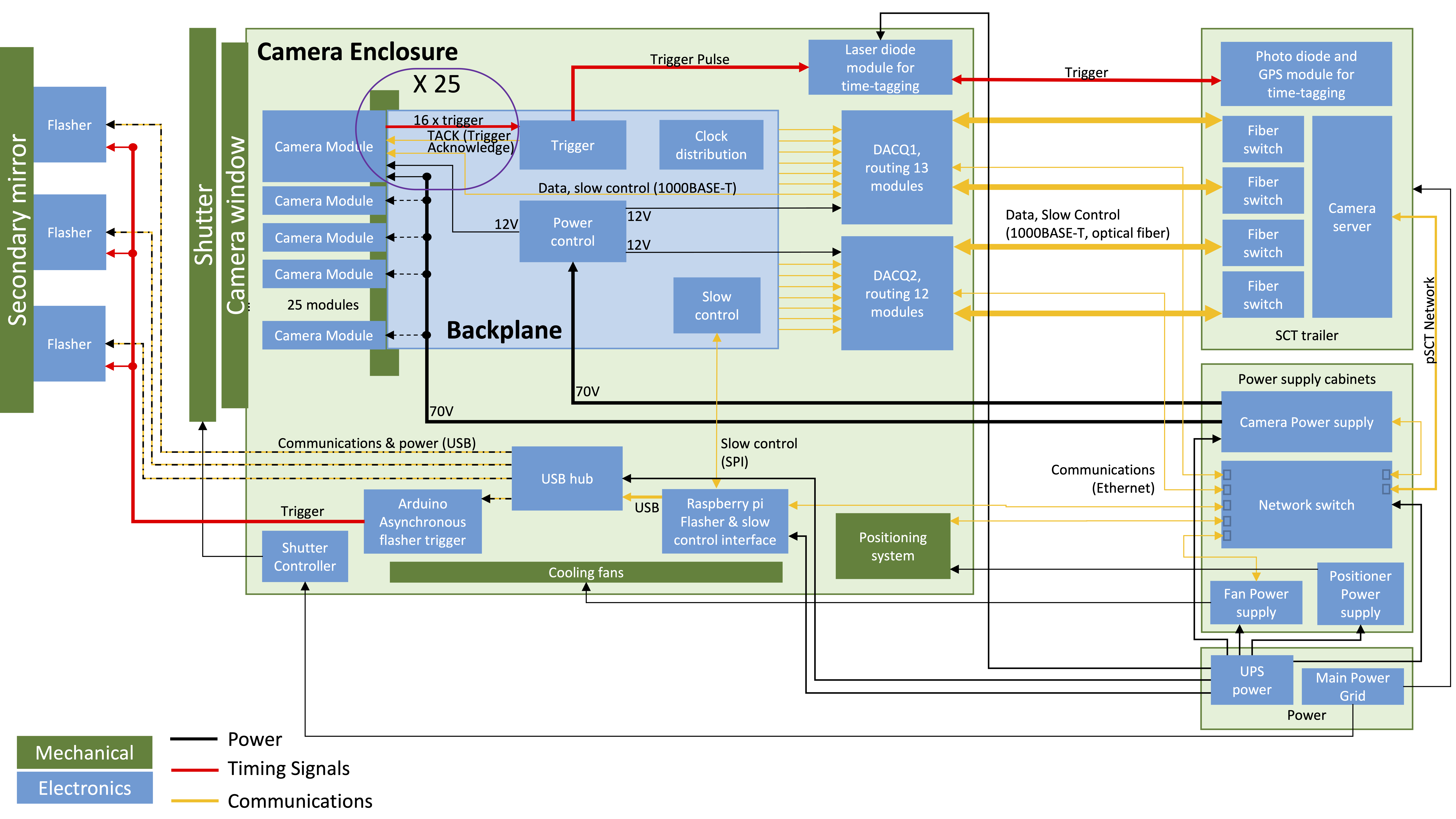}
	\caption{Block diagram of all main camera systems. Mechanical elements are shown in green and electrical elements in blue. Power connections are black, timing signals red, and communication signals in gold. The network switch and camera server is connected directly to the pSCT network.}
	\label{fig:block_diagram}
\end{sidewaysfigure}

The prototype Schwarzschild-Couder telescope is a candidate for an MST for CTA and is located at the Fred Lawrence Whipple Observatory in southern Arizona, USA. An artists rendition of the pSCT is shown in Figure \ref{fig:TelescopeImage}.

The pSCT will be sensitive to gamma-rays with energies between 100~GeV and 10~TeV. Most IACTs use single-mirror Davies-Cotton or parabolic optics; however, the pSCT uses a novel dual-mirror design with a 9.66~m primary mirror and a 5.4~m secondary mirror. Schwarzschild-Couder optics produce an excellent optical point spread function, a wide field of view, and a much smaller plate scale than traditional Davies-Cotton optics \cite{vassiliev2007schwarzschild}. The small plate scale (1.625~mm per minute of arc \cite{vassiliev2017prototype}) means that silicon photomultipliers (SiPMs) can be used in lieu of traditional photomultiplier tubes (PMTs) \cite{otte2015development}.

The pSCT camera uses Hamamatsu and FBK SiPMs and custom TARGET (TeV Array Readout with GSa/s sampling and Event Trigger) ASICs. SiPM pixels are much smaller than traditional PMTs and together these many small pixels are expected to provide a much higher resolution air shower image. This improved image resolution means reduced uncertainty in gamma-ray direction and energy resolution and better background rejection.

The pSCT camera uses a modular design with space for 9 backplanes and 177 modules (see Figure \ref{fig:cta_cameraDesign}). The current pSCT camera has a 2.68$^{\circ}$ field of view and 1600 pixels. A pSCT camera upgrade will increase the field of view to 8.04$^{\circ}$ and the total number of pixels to 11,328. Figure \ref{fig:block_diagram} shows a full block diagram of the current pSCT camera. The camera shares common components (front-end electronics and backplanes) with the Compact High Energy Camera (CHEC) which is being developed for the dual-mirror CTA SSTs \cite{Zorn:2019hgk}.

\section{Mechanical Design} \label{Mechanical Design}
The pSCT camera is located between the Primary and Secondary Mirrors (see Figure \ref{fig:TelescopeImage}). The camera structure is designed to hold up to 177 camera modules. Each module contains front-end electronics as well as a focal-plane module. Once installed, the focal plane modules together form a curved focal plane which faces the Secondary Mirror. The camera can be moved via an alignment system (see Section \ref{Alignment System and Motors}) in order to achieve the optimum distance and orientation between the focal plane and the secondary mirror (determined by the Schwarzschild-Couder optics). The focal plane and camera electronics are protected from the elements by a retractable shutter (see Section \ref{Shutter}) and the electronics are cooled by a chiller system (see Section \ref{Cooling System}).

\subsection{Structure} \label{Camera Structure}

\begin{figure}[p]
    \centering
    \includegraphics[width=0.75\textwidth]{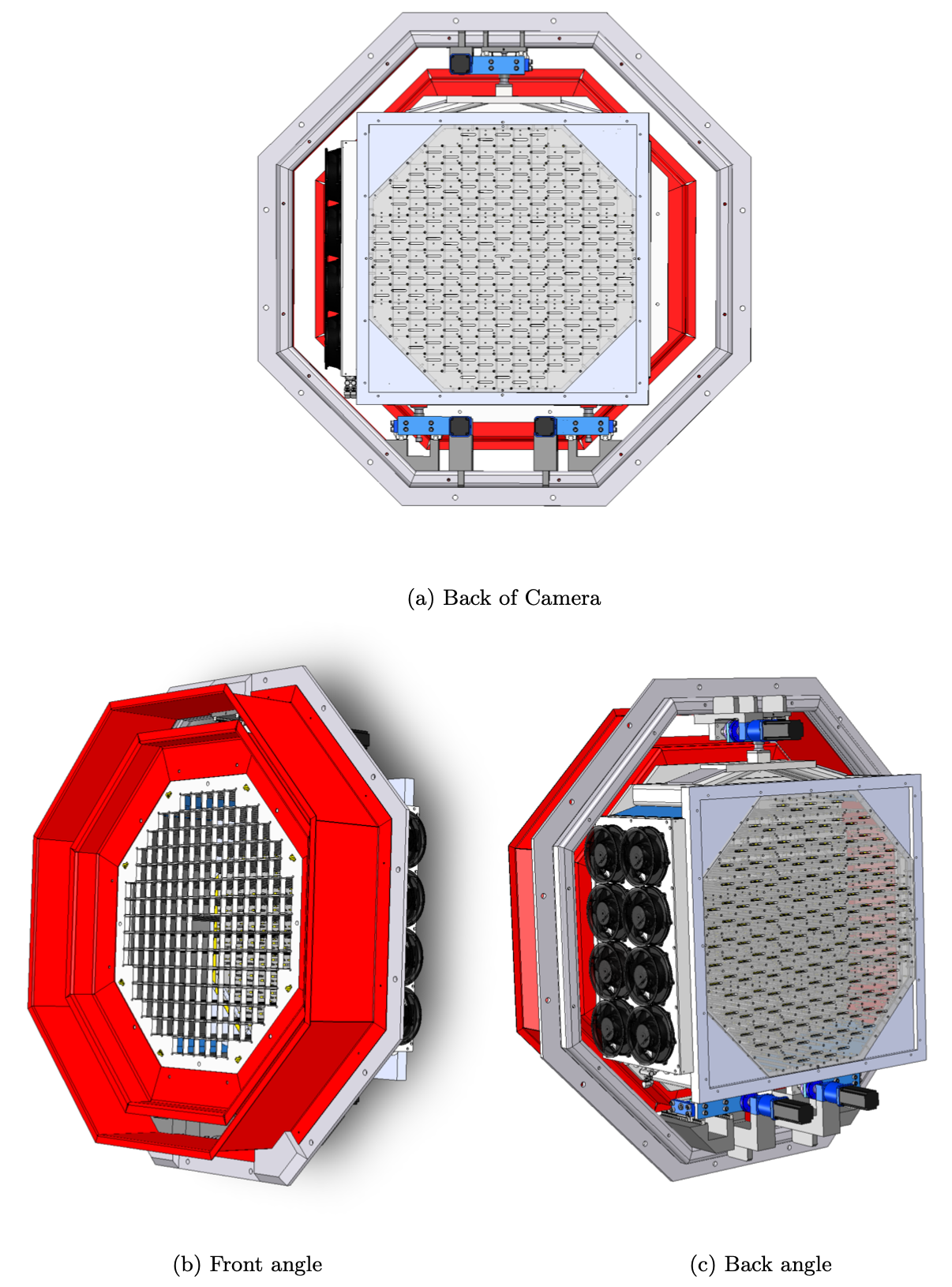}
    \caption{Mechanical structure of the camera. The outer structure (shown in grey) connects to the shroud and the telescope itself. The inner structure (containing the lattice) can be moved relative to the outer structure via the alignment system described in Section \ref{Alignment System and Motors}. The shroud and window frame are in red, the motors in blue, and fans in black (visible on the left side of view (c)). Modules are inserted into the inner camera structure from the front and are secured through screws in the back bulkhead. Back end electronics (not shown) are also mounted to the back bulkhead and connect with each module through a backplane connector.}
    \label{fig:CameraDrawing}
\end{figure}

The pSCT camera is comprised of an outer structure and an inner structure (see Figures \ref{fig:CameraDrawing} and \ref{fig:MotorDiagram}). The inner structure is designed to hold the camera modules (see Section \ref{Modules}) and can be moved relative to the outer structure for the purposes of optical alignment (see Section \ref{Alignment System and Motors}). The inner structure contains a front aluminum lattice and a back bulkhead which are connected with internal carbon fiber rods to reduce thermal variations in the separation between the lattice and the bulkhead. The lattice has been machined precisely to provide a flat front surface and to hold the focal plane component of each camera module (see Section \ref{FPM Design}) in the correct position.

The fully populated camera (weight approximately 300 kg) was lifted into the telescope in the summer of 2018 with a crane. Once in position, the outer structure of the camera was bolted to the mounting ring of the telescope. The outer structure is connected rigidly to the telescope and to the outer shroud (which protects the camera from the elements).

Modules are installed through the front lattice into the gaps between the carbon fiber rods and connect to backend electronics through two holes in the back bulkhead. One hole is for the module connector (which connects the module to its backplane) and the other is for the securing screw (which holds the module in place). This securing screw can be used to pull each camera module into its backplane connector. When tightened the securing screw keeps the modules tight against the front face of the lattice, ensuring that the module is at the correct location. Modules can be extracted by loosening the securing screw, and pushing to eject the module. This system allows there to be effectively zero clearance between the SiPMs on adjacent modules, while still maintaining the ability to remove individual camera modules for servicing.

\subsection{Alignment System and Motors} \label{Alignment System and Motors}

The interior of the pSCT camera (including the focal plane) can be moved relative to the outer structure. This alignment system allows the focal plane to be precisely positioned in situ in the correct orientation and at the correct distance to the secondary mirror, as required by SC optics. Initial optical alignment of the pSCT was reported in \cite{adams2020verification}.

\begin{figure}[ht]
    \centering
    \includegraphics[width=\textwidth]{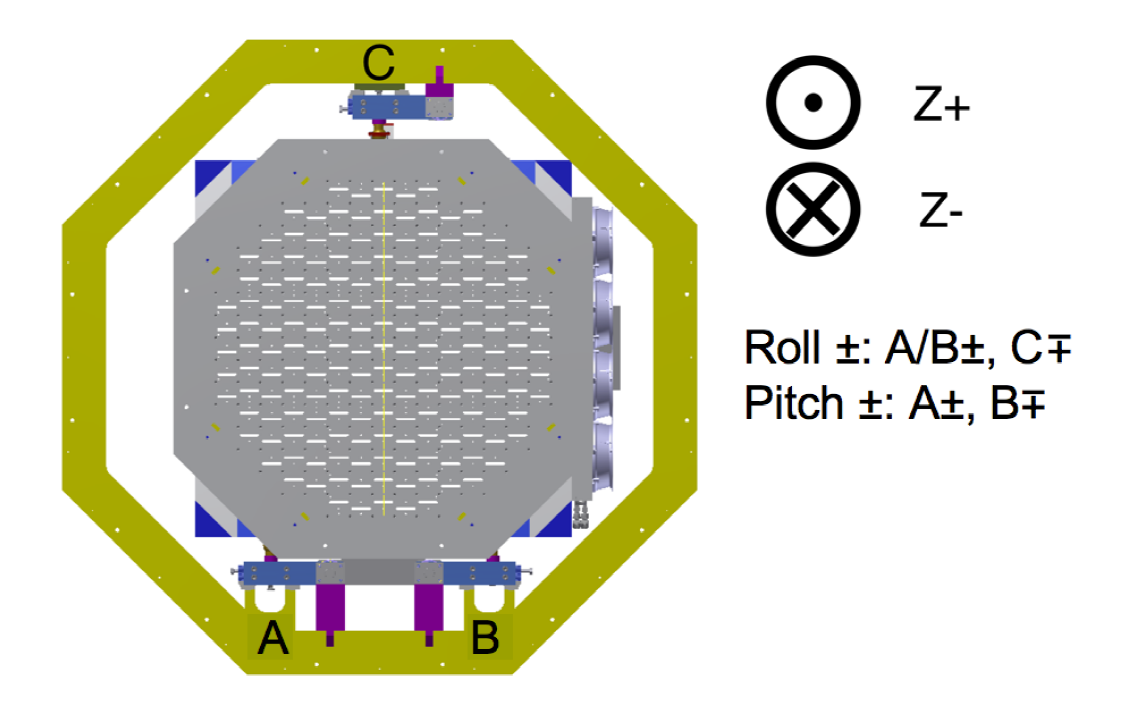}
	\caption{Motor locations in the pSCT camera as shown from the front. The yellow octagon is the outer structure of the camera and is connected with bolts to the telescope. The inner structure is movable via the three motors labeled A, B and C. Motion of the camera is described with a right-hand coordinate system with the z-direction labeled above along the optical axis of the telescope. z-motion of the camera requires that all motors move in unison. Pitch (rotation about the y axis) requires motors A and B moving in opposite directions. Roll (rotation about the x axis) requires motors A and B moving together and motor C moving in the opposite direction.}
	\label{fig:MotorDiagram}
\end{figure}

The inner and outer structures of the pSCT camera are connected with three ball pin joints. The inner structure is positioned entirely with these joints and can be moved relative to the outer structure using a camera alignment system. Figure \ref{fig:MotorDiagram} shows both parts of the camera structure as well as the location of the motors used in the alignment system.

The camera can be moved in the z-direction or rotated using the appropriate pattern of three motor assemblies. Each motor assembly contains a step motor which turns a motor driver screw. This screw pushes or pulls a flange which is rigidly connected to the ball pin assembly. The motion of the ball pin assembly is restricted by a rail system to a single direction, in this case in the z-direction (along the optical axis). One step of the motor system corresponds to 12.7~microns of motion along the specified axis. The total range of motion along the z-axis is 5.08~cm.

The camera can also be moved along the x- and y-axes by physically rotating the alignment screws. When the horizontal alignment screws are loose the camera can be moved along a track perpendicular to the rails. Similarly, vertical alignment screws can control the vertical motion of the camera. The motion in these directions is continuous rather than occurring in steps, and the total range of motion is 2.5~cm in both of these directions. 

The motor assemblies are controlled by an Arduino located in the motor electronics box. Each motor can be moved independently or in conjunction with other motors. By moving specific motors in unison the camera can either be moved in the z-direction (along the optical axis) or rotated. Figure \ref{fig:MotorDiagram} shows which motors are used to achieve each type of movement.

Once the camera has been fully aligned, the vertical and horizontal alignment screws can be tightened, keeping the camera in place indefinitely.

\subsection{Shutter} \label{Shutter}

\begin{figure}[ht]
    \centering
    \includegraphics[width=\textwidth]{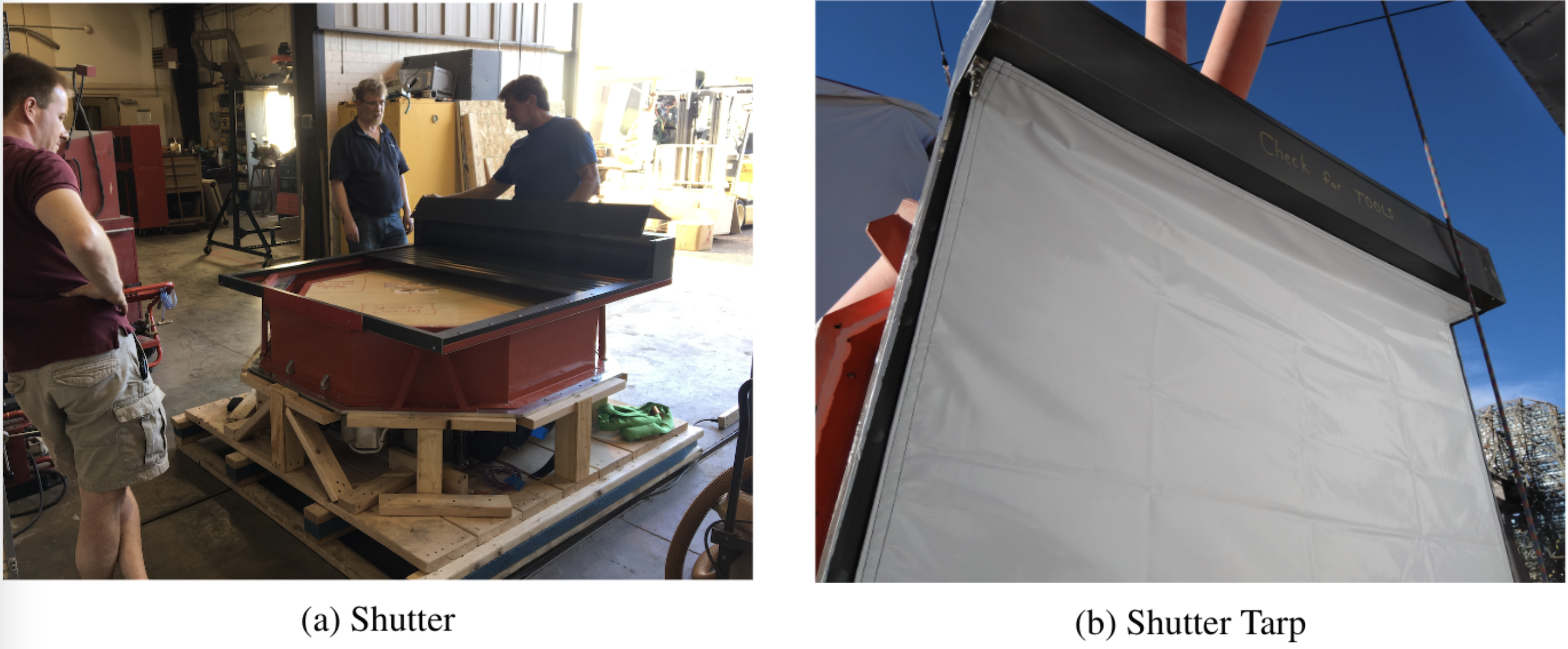}
    \caption{The focal plane of the camera is shaded and protected from the elements by a commercial AluTech rollup shutter. (a) Shows the shutter installed on the camera prior to its installation. The shutter (black) is mounted onto the camera (red) with four cantilever triangles (red). This photo shows the shutter slats partially rolled up so that the shutter is halfway closed. (b) Shows the shutter tarp (white) which has been fixed to the face of the shutter (black) with Velcro to provide additional water protection.}
    \label{fig:Shutter}
\end{figure}

A shutter, mounted to the front of the camera enclosure, is used to protect and shade the camera in daylight. After the camera and front shroud were installed into the telescope, the shutter was lifted into place and mated to the front shroud of the camera using four triangle brackets to cantilever the shutter weight.

The shutter is a 142 x 142 cm  motorized commercial rollup shutter (AluTech, Inc.) with manual override and remote control. A photograph of the shutter installed onto the camera is shown in Figure \ref{fig:Shutter}.

The shutter is composed of interlocking slats which form a continuous curtain that rolls up onto a mechanical spool when in the open position. In order to close the shutter, the motor drives the curtain sheet to slide along a continuous, brush-hair-sealed slot in the 5x2.5~cm extruded aluminum support frame. The motor uses 120~V, 60~Hz AC power.

The shutter support frame is bolted to an interface frame which is mounted on the camera using four cantilever triangles. The interface frame provides stiffness to the commercial shutter as well as the transition between the camera octagonal front shroud and the square camera shutter. The interface frame also provides a weatherproof seal to the back of the shutter. The shutter can be controlled in four different ways: web page, handheld remote control, a local mechanical switch, or using a manual crank.

The camera shutter uses a Simu RTS Radio Micro Receiver 2008191 shutter motor controller, which   can be controlled using a handheld (RF) remote as well as with an external waterproof  SPDT switch mounted on the side of the camera. The Simu controller is also  interfaced to a ControlByWeb X-301 Remote control Dual Relay Controller.  Two commercially available proximity sensors (DC 3 Wire 6-36V PNP IR Photoelectric Sensor Switch; E18-B03P1)  are interfaced to the sense inputs of the X-301 and are used to sense both open and closed positions of the shutter curtain.  The X-301 provides a simple web interface for operation of the Simu shutter motor controller over the web, as well as a method of reading the status of the shutter position (open or closed). 

A waterproof tarp cover for the front of the camera shutter was constructed for an extra measure of water protection for the camera. The waterproof camera shutter tarp mounts on Velcro strips directly onto the front shutter frame.  The  camera shutter tarp can be quickly installed or removed (less than one minute once the operator is in front of the camera).

\subsection{Cooling System} \label{Cooling System}

\begin{figure}[ht]
    \centering
	\includegraphics[width=\textwidth]{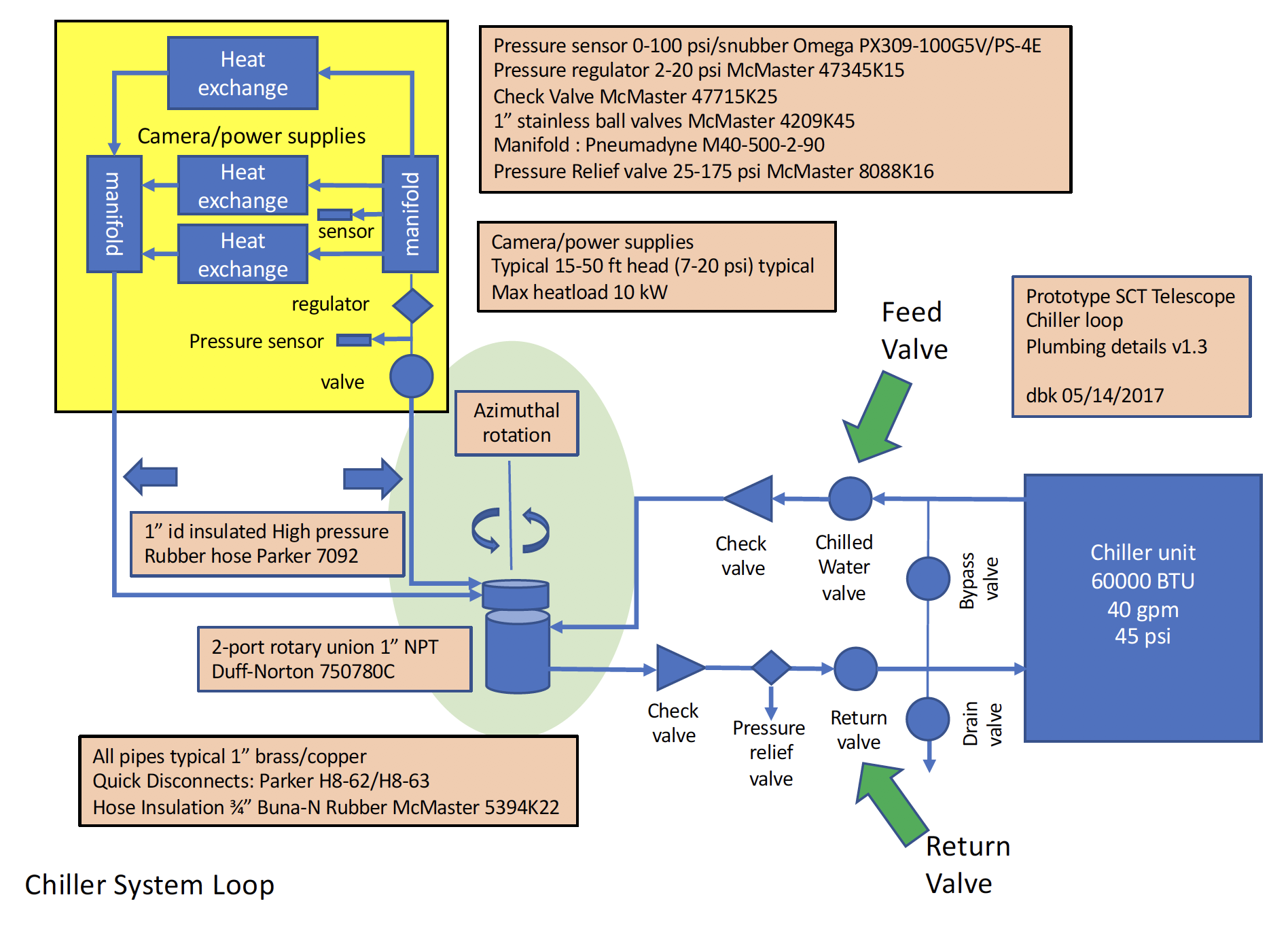}
	\caption{The pSCT chiller block diagram. The chiller unit cools a mixture of water and propylene glycol which is then pumped into the camera through a feed valve. The chilled coolant mixture is circulated through the right side of the camera. Fans blow across the chilled pipes, blowing this cold air through the camera electronics. The coolant is then returned to the chiller unit to be chilled again.}
	\label{fig:Chiller}
\end{figure}

The camera cooling system is comprised of a chiller (detailed in Figure \ref{fig:Chiller}) which chills and circulates coolant and a set of fans coupled to a heat exchanger which blow cooled air through the camera. The primary purpose of this cooling system is to remove the excess heat which is generated by the camera and module electronics during data taking. The cooling system cannot hold the camera at a pre-determined temperature. Individual modules contain a separate cooling system for their SiPMs. Modules are equipped with a Peltier controller which can be used together to further stabilize the temperature of the focal plane.

The camera cooling system was designed to provide sufficient cooling capability for a fully populated focal plane, but during the prototype phase, the heat exchangers and cooling fans within the camera were sized to accommodate only the partially filled focal plane. The remainder of the cooling system was constructed to accommodate the cooling requirements of the full focal plane installation. 

The cooling system uses a Dimplex SVO-5001M 60000~BTU/hour (17.6 kW) modified for delivery of 40~U.S. gallons per minute (151 lpm) at 45~psi (310 kPa), filled with coolant that is a 50-50 mixture of propylene glycol and water. The Dimplex chiller is permanently mounted on a separate concrete pad from the SCT telescope. Chilled coolant and return lines run underground between the chiller unit and the telescope through insulated rubber hoses (1" (2.54 cm) inner diameter, Parker 7092 hose wrapped in 1/2"  (1.27 cm) thick Buna-N/PVC closed cell insulation). A series of manual valves and check valves, located near the chiller unit, allow the chiller unit to be emptied or run in a local test loop to bypass the telescope.

The insulated chiller lines run vertically through the telescope pedestal axis, and connect to a two-port rotary union (Duff-Norton 750780C) which is located on the azimuthal axis and is directly driven by the azimuthal telescope head through a connecting rod with a pair of U-joints. The rotary union provides a rotating joint for the chilled coolant and return feed, thereby eliminating twisting of the chiller lines under telescope azimuthal rotation. The chiller lines continue up through a flexible hose/cable manager (IGUS R18840 energy chain) which bridges the variable gap between telescope positioner and the Primary mirror. The chiller coolant feed and return lines run along the  telescope tower arms. The chilled coolant feed line passes through a manifold mounted midway between the primary mirror and the secondary mirror, bolted to a tower strut.  The chilled coolant manifold  provides shutoff valve, check valve, and a pressure regulator to reduce the chilled coolant from the unregulated pressure of 40--45~psi (275-310 kPa) to a regulated pressure of approximately 3~psi (20 kPa). The pressure regulator is needed to compensate for the varying coolant pressure in the chiller lines caused by different elevation pointings of the telescope. The pressure regulator ensures the pressure to the heat exchangers in the camera remain constant, regardless of camera elevation. 

The chiller manifold also provides a parallel (shunt) coolant flow to supply chilled coolant to a pair of heat exchangers located in the power supply cabinets, located behind the SCT camera. Separate pressure regulators are used to adjust the relative chilled coolant flow between the various chilled coolant loops; pressure and temperature sensors are also located in the chilled coolant manifold, allowing measurement of the coolant pressure before and after the pressure regulator, as well as coolant temperature and temperature of the power supply cabinets. The pressure and temperature sensors are read out by a ControlByWeb X-320 temperature/analog logging controller, which provides a web interface for remote monitoring. The return coolant manifold is also located on the tower mid-way between the primary and secondary struts, but on a diagonally opposite strut. 

The prototype SCT camera uses eight Pabst EBM 6314 fans coupled to a Super Radiator Coils Model 38 ST WC heat exchanger to cool the camera. The heat exchanger is rated at 7000~BTU/hour (2~kW) which is more than sufficient to cool the 25 camera modules, as each module dissipates less than 20~W each when observing.

The camera power supplies, including the front-end electronics (FEE) power, the SiPM power, the fan power supply, and the power supply for the camera alignment system (See section \ref{Alignment System and Motors}), are physically located in two independent, weather-tight power supply cabinets located approximately 1~m behind the back of the SCT camera. The two sealed power supply cabinets each contain one 19-inch rack.  The racks are used to hold the rack-mounted power supplies, as well as other rack mounted equipment, such as a managed ethernet switch for communications and control of various camera and telescope subsystems. Each power supply cabinet is cooled by the chilled coolant feed using a  Lytron 6220G1SB heat exchanger and fanpack. Each Lytron 6220G1SB  heat exchanger is capable of handling 6832~BTU/hr (2~kW) with 2~ U.S. gallons per minute (7.5 lpm) coolant flow at 1~psi (6.9 kPa). 

\begin{figure}[ht]
    \centering
	\includegraphics[width=0.5\textwidth]{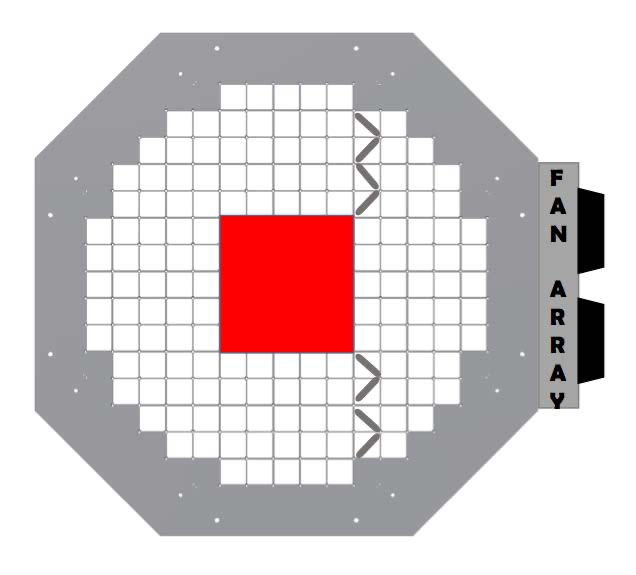}
	\caption{Air baffle orientation diagram. Baffles are long, rectangular aluminum pieces which are inserted into unused module slots at a 45$^{\circ}$ angle. The baffles are arranged in such a way that air from the fans is directed through the modules in the camera. These baffles will be removed for the camera upgrade to make room for the modules of the fully populated focal plane. The camera upgrade will also include additional fans in order to cool the increased number of modules.}
	\label{fig:Baffles}
\end{figure}

Because the camera is not fully populated at this time, a series of baffles is used to force air through just the sector currently containing modules (see Figure \ref{fig:Baffles}). The first baffle is a single, large sheet and fits onto the front of the camera. It blocks the unused module holes in the front of the camera while leaving the central sector open so that modules can be inserted. This keeps cooled air from escaping out of the front of the camera. The other baffles are small rectangular pieces which are inserted at a 45$^{\circ}$ angle into unused module slots. They are secured in a way similar to camera modules, with a screw through the back plate of the camera. The baffles are designed to direct air from the fans through the inserted modules. Figure \ref{fig:Baffles} shows the orientation of the baffles currently being used in the camera. 

\begin{figure}[ht]
    \centering
	\includegraphics[width=\textwidth]{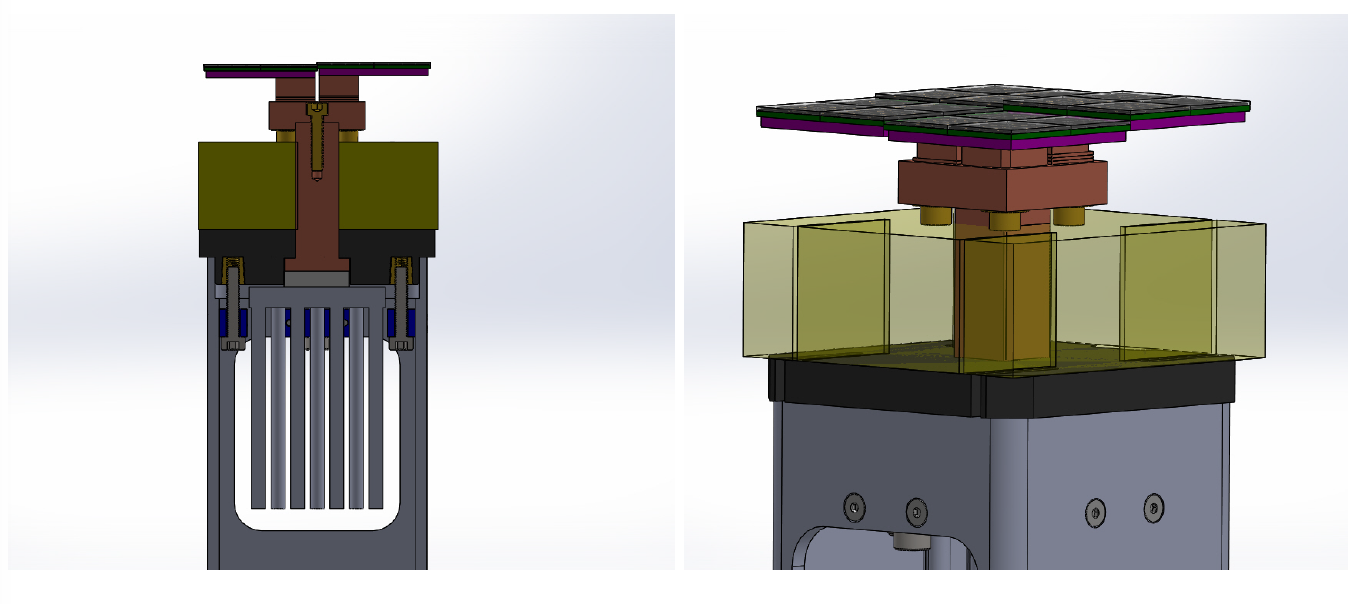}
	\caption{Two perspectives of an FPM. Each FPM is divided into four quadrants. Each quadrant is composed of 16 image pixels and has a z position achieved by placing shims between the copper post and the copper block on the bottom of the PCB for each quadrant. Yellow insulating foam surrounds the copper post which attaches the light sensors to a Peltier element. On the opposite side of the Peltier element is a heat sink.}
	\label{fig:FPM}
\end{figure}

\begin{figure}[ht]
    \centering
	\includegraphics[width=0.7\textwidth]{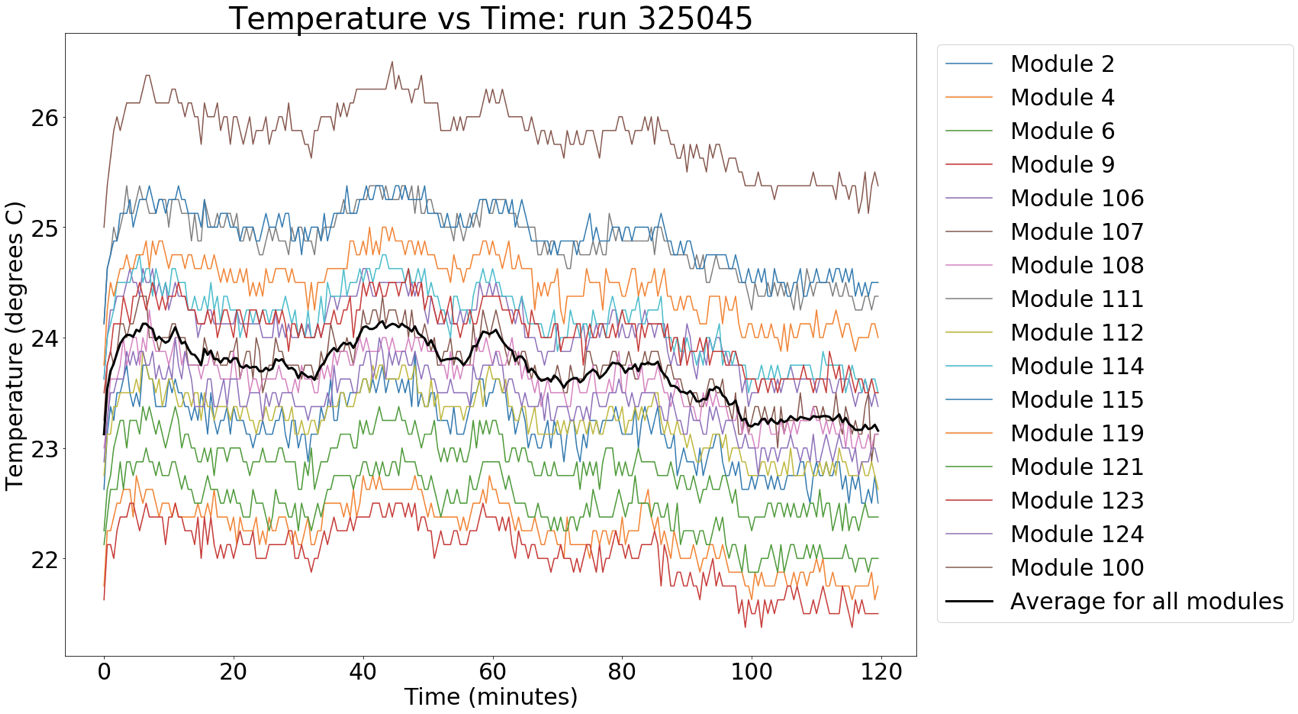}
	\caption{Temperature vs time for a two hour run taken onsite during the summer of 2018 (prior to installation of the Peltier coolers and control). The ambient temperature over the course of the run was 22.8 C - 23.9 C.  Module temperature is recorded on the module's front end electronics. The temperature of each individual FEE is constant to within 1$^{\circ}$ over the full two hour period. Typical data runs last one hour meaning that individual modules do not experience significant changes in temperature within a single run. The spread of temperatures across modules is due to their physical position in the camera and is consistent over time. Additionally, Each module has an independent lookup table for temperature-dependent variables to account for the 4$^{\circ}$ spread between modules.}
	\label{fig:TempVSTime}
\end{figure}

The focal plane photon sensors and the module electronics are thermally isolated from one another. This is achieved by inserting insulating foam between each focal-plane module (FPM) and its FEE. A thermoelectric (Peltier) element transfers heat from the FPM connected to its cold side to a heat sink connected to its hot side. This is shown in Figure \ref{fig:FPM}. The heat sink is cooled by the air flow of the chilled coolant cooling system, alongside the module electronics. A Peltier controller will be used to stabilize the temperature of the focal plane. Each module's electronics are stable to within 1$^{\circ}$\,C when using the chiller system (see Figure \ref{fig:TempVSTime}). Using the Peltier controller, the focal plane temperature stability has been measured in the lab to be within 0.1$^{\circ}$\,C but the Peltier controller has not yet been tested onsite. As a result the in situ focal plane temperature stability is unknown.

\section{Modules} \label{Modules}
The camera is capable of housing 177 modules (see Figure \ref{fig:cta_cameraDesign}). Each module consists of the front-end electronics which are housed inside an aluminum module cage and a focal-plane module which mates to the FEE. The modules are inserted into the front aluminum lattice until the focal plane module fiducial surface is flush with the lattice. In order to fit the modules into the camera's back aluminum plate, modules in the same row have alternating orientations (they are rotated 180$^{\circ}$ from their neighbor) while those in the same column have the same orientation.

\subsection{FPM Design} \label{FPM Design}

\begin{figure}[ht]
    \centering
	\includegraphics[width=0.7\textwidth]{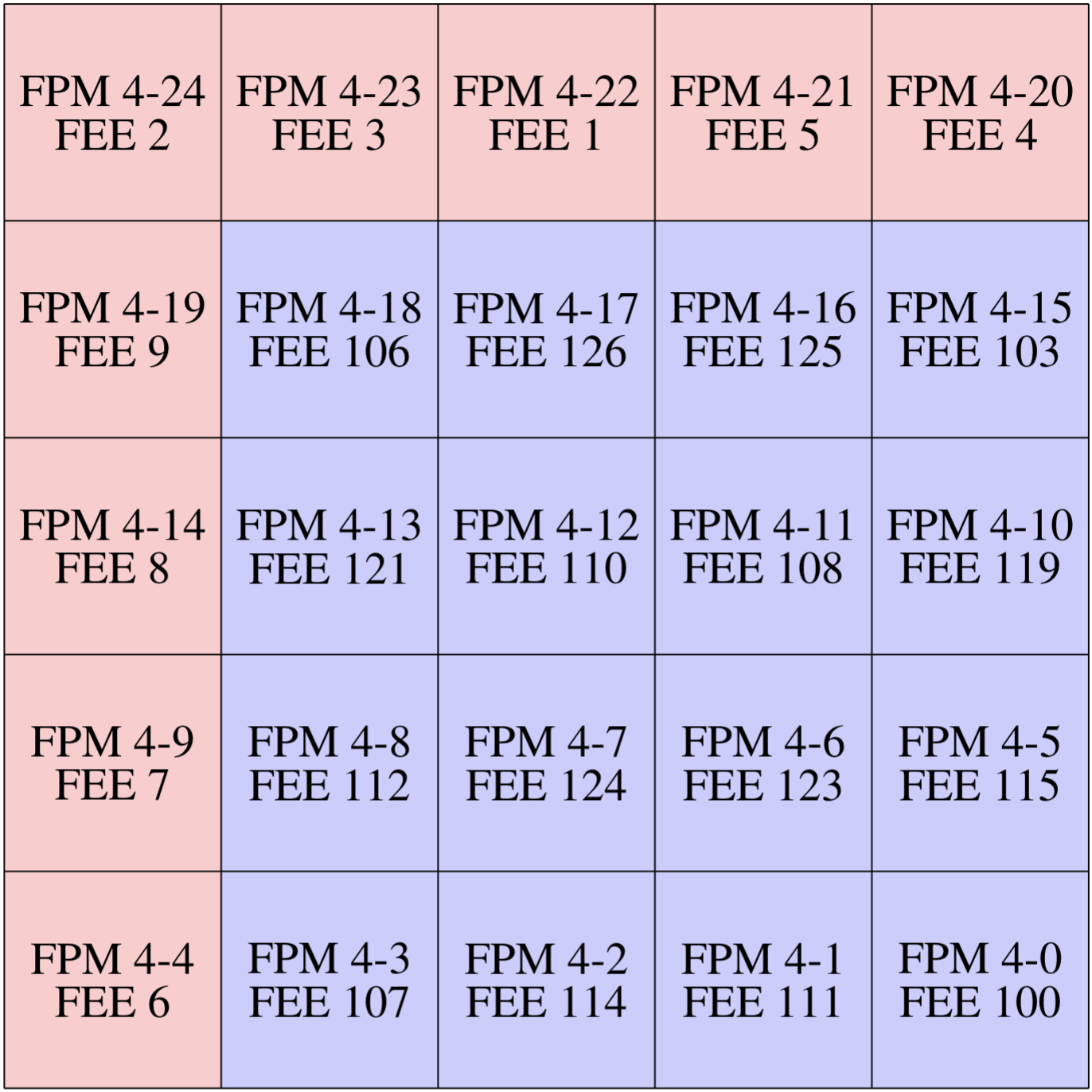}
    \caption{The pSCT focal plane is currently equipped with 25 modules in a square grid configuration. Sixteen of the modules are equipped with Hamamatsu photon detectors S12642-0404PA-50(X) (see Figure \ref{fig:HamamatsuSensor}). The remaining nine modules are equipped with the NUV-HD3 sensors produced by FBK. The locations of these modules are shown as viewed facing toward the sky, with blue indicating Hamamatsu sensors and red FBK sensors. For each slot the FPM number is listed along with the FEE number for the module in that slot.}
	\label{fig:CurrentCameraMapping}
\end{figure}

\begin{figure}[ht]
    \centering
	\includegraphics[width=\textwidth]{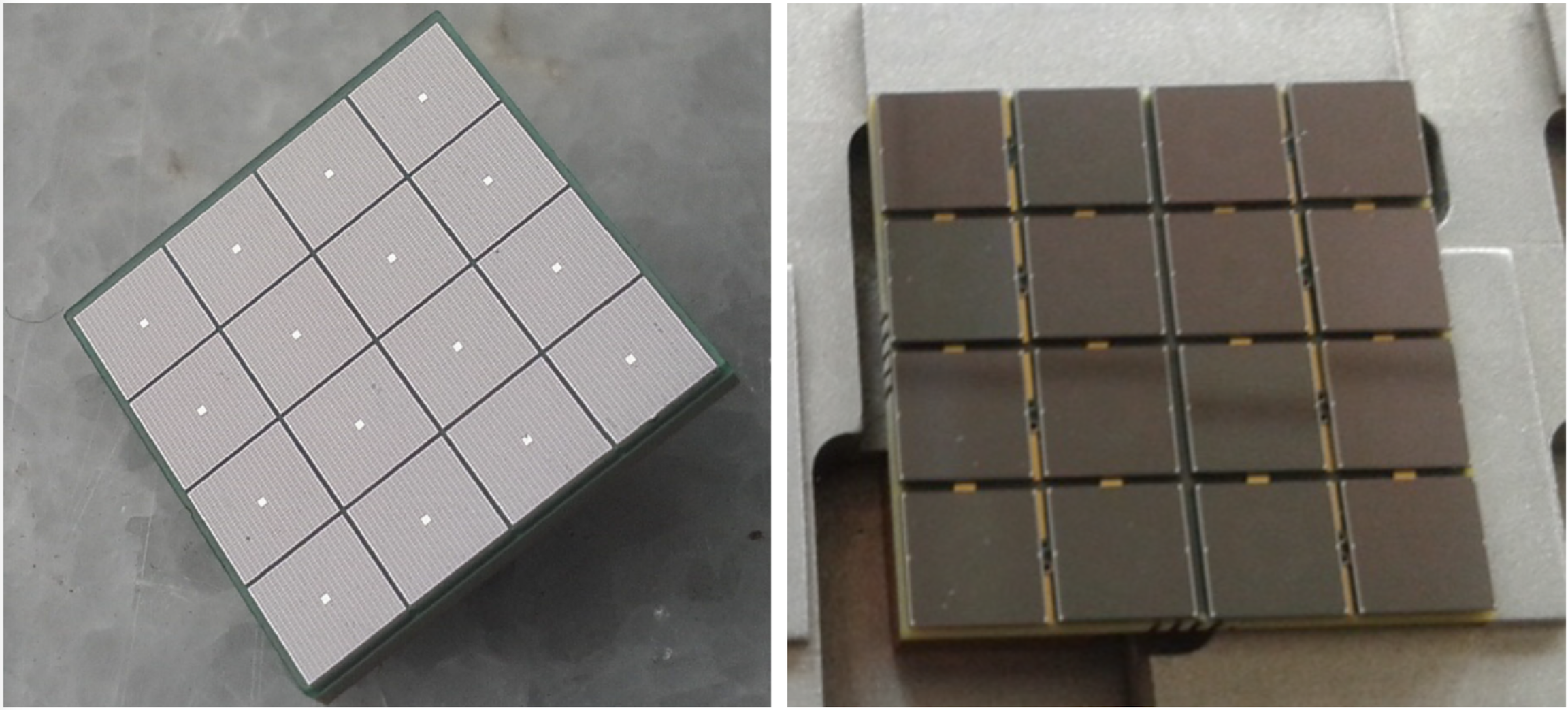}
	\caption{On the left is a photo of a Hamamatsu MPPC S12642-0404PA-50(X) sensor used in the pSCT camera. Shown is a single tile as delivered by the manufacturer. Each tile has 16 3x3~mm SiPMs which are grouped into four 6x6~mm image pixels. Each tile corresponds to one Trigger Pixel and A single FPM uses 16 of these tiles \cite{Otte:2015}. A single Hamamatsu tile measures 13x13~mm. On the right is a photo of an FBK sensor used in the pSCT camera. Shown is a single tile. Each tile has 16 6x6~mm SIPMs each of which correspond to a single image pixel. Each tile corresponds to 4 Trigger Pixels and a single FPM uses 4 of these tiles. A single FPM tile measures 26x26~mm}
	\label{fig:HamamatsuSensor}
\end{figure}

The pSCT focal plane is made up of two types of photosensors. The locations of these photosensors in the focal plane is shown in Figure \ref{fig:CurrentCameraMapping}. Sixteen modules are equipped with Hamamatsu photon detectors S12642-0404PA-50(X). One Hamamatsu tile is shown in Figure \ref{fig:HamamatsuSensor}. One tile consists of 16 SiPMs in a square grid configuration each with dimensions 3x3~mm$^{2}$. One FPM is equipped with 16 Hamamatsu tiles and each tile corresponds to one trigger pixel. 

Nine modules are equipped with third generation near ultra violet high density SiPMs (NUV-HD3) which were produced by Fondazione Bruno Kessler (FBK) in collaboration with Istituto Nazionale di Fisica Nucleare (INFN). One FBK unit consists of 16 SiPMs in a square grid configuration each with dimensions 6x6~mm$^{2}$ \cite{ambrosi2017silicon}. One camera module is equipped with four FBK units with each unit containing four trigger pixels.

\begin{figure}[ht]
    \centering
	\includegraphics[width=0.7\textwidth]{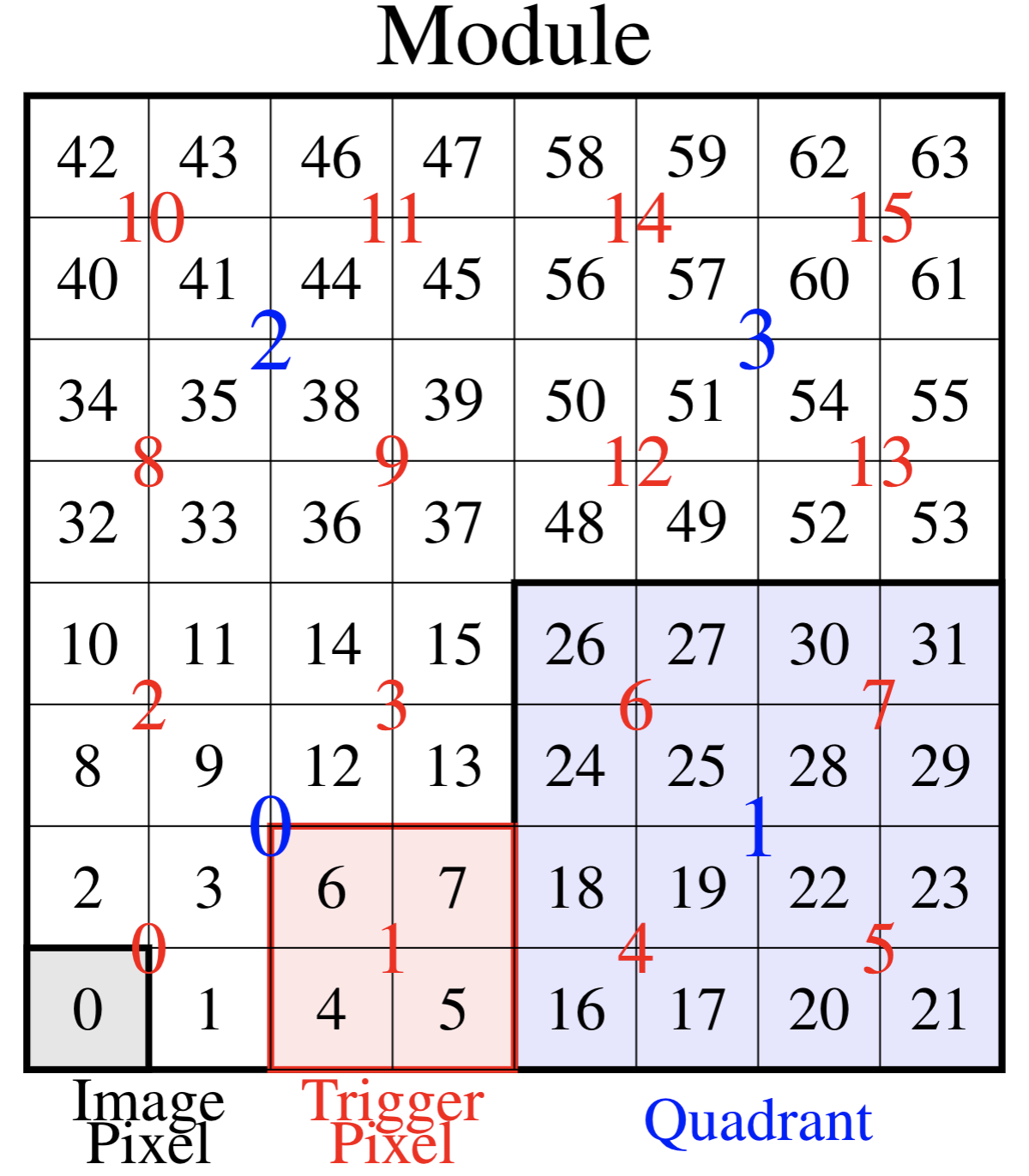}
	\caption{One module contains 64 image pixels (black) numbered 0--63. A square group of four image pixels makes up one trigger pixel (red). Thus one module contains 16 trigger pixels. A square group of 4 trigger pixels makes up one quadrant (blue). Thus one quadrant contains 16 image pixels and there are four quadrants in a module. Image pixels are associated with Channels, trigger pixels with trigger groups, and quadrants with ASICs in the software.}
	\label{fig:PixelLayout}
\end{figure}

For either type of photosensor, a single FPM contains 64 image pixels. One square group of 4 image pixels makes up one trigger pixel. Thus one FPM contains 16 trigger pixels. One square group of 4 trigger pixels makes up one quadrant of the FPM. Figure \ref{fig:PixelLayout} shows how quadrants, trigger pixels, and image pixels are laid out in a single FPM. Quadrants are mounted onto a printed circuit board (PCB) that has a tapped copper block on the other side. These quadrants are bolted onto the top of a copper post. This is shown in Figure \ref{fig:FPM}.  The copper post passes through a layer of insulating foam and is secured at its base by a plastic base plate.

When installed, the base plate mates securely into a slot in the front camera lattice and is pulled flush to the front surface of the lattice by springs in the module housing, positioning the photosensors precisely in the desired position in the focal-plane surface. 

Trigger pixels are placed at unique z-positions in relation to the quadrant they are in, in order to produce a smoothly varying convex focal plane required by Schwarzschild-Couder (SC) optics \cite{vassiliev2007schwarzschild}. The final z-positions of the quadrants are determined by the height of the copper post and the thickness of shims between the post and each of the four PCBs. The bottom of the copper post is connected to the cool side of a Peltier thermoelectric element (TE) and acts as a thermal conductor between the sensors and the TE. On the warm side of the TE a heat sink is connected. A thermistor is mounted to the back of each quadrant and is capable of providing temperature feedback.

The sides of the quadrants are aligned to the edge of the plastic base plate. This base plate is then pushed onto the FEE aluminum module cage and secured with screws. The pitch between modules is 54~mm. The optimal sag of the focal plane, $z_f$, has been derived through simulations and is given by equation \ref{eq1},

\begin{equation} \label{eq1}
    \frac{1}{F} z_f(v) = k_1 v + k_2 v^2
\end{equation}

where $v$ (given by equation \ref{eq2}) is related to the cylindrical coordinate $r_f$ of the surface of the focal plane.

\begin{equation} \label{eq2}
v = (\frac{r_f}{F})^2
\end{equation}

The constants $k_1$ and $k_2$ in equations \ref{eq1} and \ref{eq2} have values: $k_1 = ?0.8327$ and $k_2 = 4.9950$.

This results in a focal plane which is nearly parabolic. If the curved focal plane is assembled from photon detectors of size $\Delta$ in tiled arrangement with the center of the detector
placed exactly at the optimal coordinates ($r_f$, $z_f$), then the edges of the detector will be misplaced from their optimal location by $\Delta Z_f$ (given by equation \ref{eq3}).

\begin{equation} \label{eq3}
\Delta Z_f = \pm k_1 (\frac{r_f}{F}) \Delta = \pm 3.023 \text{mm} (\frac{\Theta}{4 \text{deg}}) (\frac{\Delta}{52 \text{mm}})
\end{equation}

\begin{figure}[ht]
    \centering
	\includegraphics[width=0.7\textwidth]{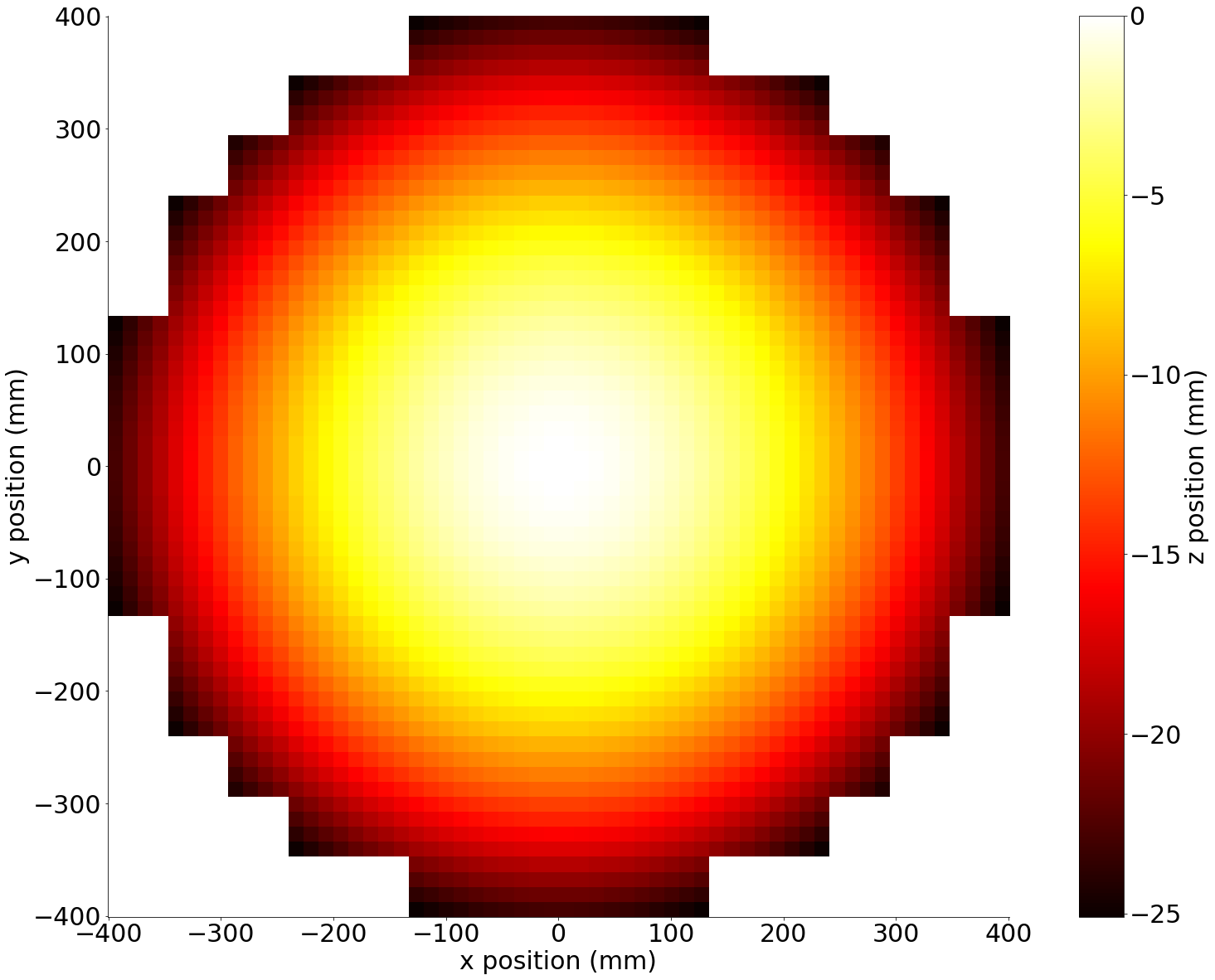}
	\caption{Camera Focal Plane. Colormap of the position of each trigger pixel in the camera. This plot is from the front perspective of the camera with the positive z axis toward the secondary mirror. The z-position of each quadrant is determined by shims placed between the SIPMs and the copper post. Trigger pixels within a quadrant are fixed at unique z-positions in relation to the quadrant. The final product is an approximation of a smoothly varying convex surface.}
	\label{fig:FocalPlanePosition}
\end{figure}

SC optics require positional uncertainties along the optical axis of less than 1.2~mm and uncertainties perpendicular to the optical axis of less than 0.27~mm \cite{nieto2017prototype}. The precision of the placement of the center of each image pixel is better than 100~$\mu$m in the z-direction (parallel to the optical axis). In the X-Y plane (perpendicular to the optical axis) the precision is better than 200~$\mu$m. The positions of each pixel are shown in Figure \ref{fig:FocalPlanePosition}. The on-axis point-spread-function (PSF) of the current pSCT camera has been measured to be 2.8 arcminutes. The off-axis PSF has not yet been measured \cite{adams2020verification}.

Each quadrant is connected to the FEE with a micro-coaxial ribbon cable with 50~$\Omega$ impedance. These cables provide power to the SiPMs and send the SiPM signals to ASICs located on the FEE. Each image pixel is connected to an ASIC channel of the same number. Additionally, the ribbon cables provide connectivity between each quadrant?s thermistor and a microcontroller, located on the FEE, which is responsible for temperature stabilization (see Section \ref{FEE Design}). Temperature stabilization is required because the SiPM gain is temperature dependent. Currently, temperature stabilization is achieved only through the chiller system. Stabilization including the thermistor is currently underway and is expected to be achieved during the camera upgrade. 

The FEE also provides power to the TE. Further temperature stabilization of the focal plane is achieved through insulation of the sides of the camera walls and airflow cooling the module heatsinks (see Section \ref{Cooling System}).

The breakdown voltage of the Hamamatsu sensors has a median of 64~V \cite{ieeenss7431131}. The breakdown voltage of the FBK sensors is approximately 27~V. Every SiPM will be operated at approximately 3~V above breakdown (this corresponds to a gain of approximately $1.6 \times 10^6$ for Hamamatsu sensors and approximately $2 \times 10^6$ for FBK sensors at a temperature of $20^{\circ}$C). To achieve this bias voltage for each SIPM, a uniform 70~V (which is applied to all pixels in a sector) is combined with a trim voltage (adjustable for individual pixels). The 70~V and trim voltage both come from the module FEE (See Section \ref{FEE Design}).

\subsection{FEE Design} \label{FEE Design}

The FEE part of the camera module generally handles the readout and control of the FPM. The main functionalities it combines are:
\begin{itemize}
    \item amplification and digitization of FPM signals,
    \item control of SiPM bias voltage,
    \item temperature monitoring and control of FPM,
    \item low-level trigger generation,
    \item waveform data packaging and transfer to storage.
\end{itemize}

\begin{figure}[ht]
    \centering
	\includegraphics[width=\textwidth]{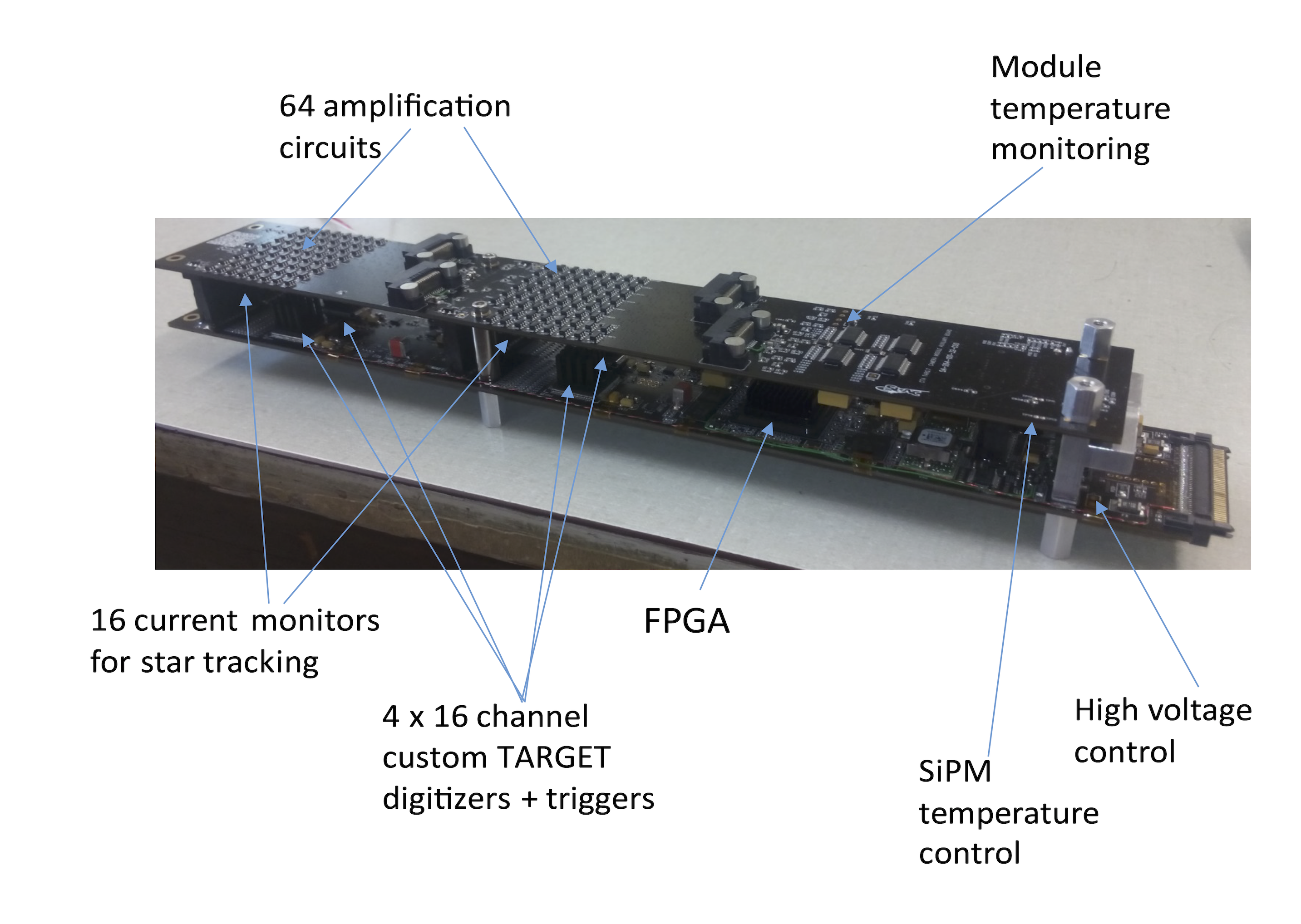}
	\caption{The pSCT front-end electronics. The auxiliary board is stacked on top of the primary board and they are connected with high-speed connectors. The primary board connector on the right-hand side connects the module to a backplane when it is mounted in the camera.}
	\label{fig:FEE_Picture}
\end{figure}

The electronics of the FEEs are distributed over two circuit boards, the primary and the auxiliary board. Their dimensions are approximately 5~cm x 28~cm  and they are stacked, the auxiliary board on top of the primary board, to fit a 5~cm x 5~cm x 45~cm aluminum housing. A photograph of the stacked boards is shown in Figure \ref{fig:FEE_Picture}. The boards are interconnected through Samtec QTH series high-speed board-to-board connectors. The primary board connects directly to the backplane, when mounted in the camera. The FPM connects to the auxiliary board through 4 ribbon cables, 1 per quadrant. Those cables carry ground, a 70~V supply to the cathode of the SiPMs, the SiPM anode output signals, and the connection to thermistors on the FPM quadrant. 

The auxiliary board contains most of the analog processing for the incoming 64 FPM signals. This includes a pulse-shaping circuit for each channel which shortens the SiPM pulses through a high pass filter, and 16 current sensors which read the combined current through groups of four SiPM pixels. 

\begin{figure}[ht]
    \centering
	\includegraphics[width=\textwidth]{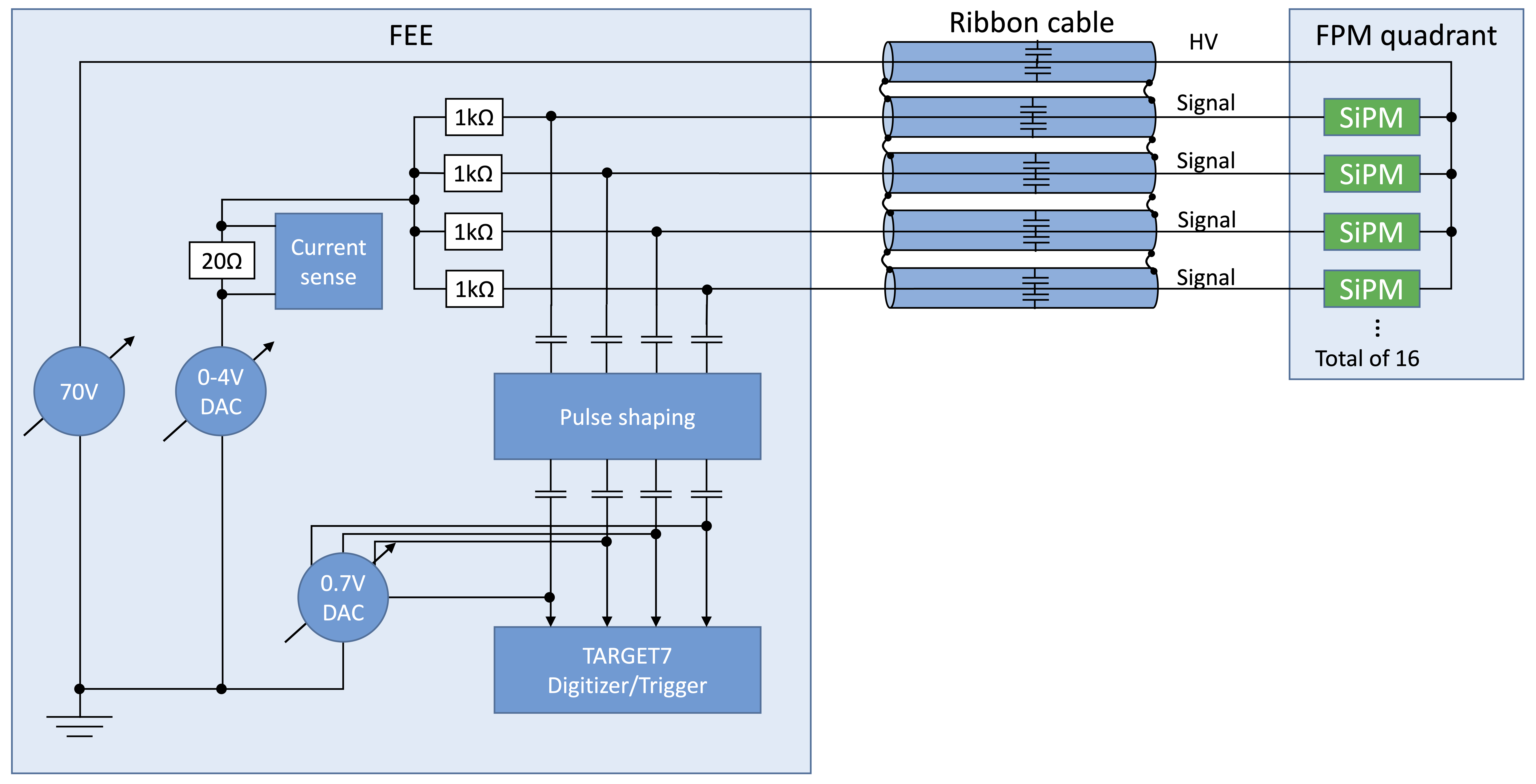}
	\caption{A block diagram of the signal chain between SiPM sensors of one FPM quadrant and a single TARGET7 readout chip (of four in one module). The $0-4\,$V DAC adjusts the bias voltage across groups of $4$ SiPMs for differences in their breakdown voltage. The $0.7\,$V DAC provides a pedestal offset to the signals entering the digitizer chip.}
	\label{fig:FEE_signal_chain}
\end{figure}

The current pSCT has only one sector populated with modules with a bias voltage of 70~V. Modules which include FBK sensors have an extra voltage regulator located on the FEE which converts this bias voltage from 70~V to 35~V. The trim voltage across the SiPMs can be regulated for groups of four image pixels by use of digital to analog converters (DACs) which can set the SiPM anode voltage between 0 and 4~V in steps of 1~mV. The cathode voltage is fixed to the bias voltage; this feature is used to compensate for production variances of the SiPMs which affect the breakdown voltages of the sensors. The groups have been chosen such that SiPMs with similar breakdown voltages are part of the same bias regulation group. The DAC can sink a maximum current of 50~mA from the four SiPM pixels. The maximum SiPM current is limited to 128~mA for each module. If this limit is reached the module will be shut off and all trim voltages set to zero. A block diagram which illustrates the signal processing, current readout and biasing of a group of four SiPMs is shown in Figure \ref{fig:FEE_signal_chain}.

The auxiliary board also manages the FPM temperature readout and control through a micro-controller to achieve stable SiPM temperatures. For that purpose a full PID regulation scheme is implemented in the micro-controller. The micro-controller will monitor the temperature of the four quadrants by measuring the resistance of the thermistors inside the FPM and will regulate the current through the Peltier element, to supply the necessary cooling.

The primary board contains most of the digital control of the FEE as well as readout and triggering circuits for the SiPM signals. The communication with and control of all devices within the FEE, as well as connection to the backplane and data acquisition (DACQ) boards, is established by use of a field programmable gate array (FPGA). It uses I2C, SPI and custom protocols to configure and read FEE components. Digitized data are managed within the FPGA and transferred to the DACQ boards via gigabit Ethernet links, if requested by an incoming trigger signal. 

\subsubsection{Module Sampling and Trigger System} \label{Module Sampling and Trigger System}

The primary board contains four TARGET7 chips (the seventh generation the ``TeV Array Readout with GSa/s sampling and Event Trigger''). This chip has been designed to sample, digitize and trigger on 16 incoming channels \cite{bechtol2012target} \cite{albert2017target} \cite{tibaldo2015target}. It has a dynamic range of approximately 2~V and samples the input waveform's charges every nanosecond, so at a speed of 1~GSa/s (see Appendix \ref{Appendix:ASIC} for a full description of this sampling system). The charge samples are held in an analog storage buffer within the chip which can hold up to 16~$\mu$s of samples before having to overwrite. Inside the buffer, samples are grouped into 512 blocks of 32. These blocks can be randomly accessed and transferred to the Wilkinson digitizer of the chip by use of an individual block address.

At the same time, a trigger system inside the chip monitors groups of 4 channels for a threshold crossing of their analog sum. If the threshold is crossed, a \textit{module trigger} (roughly 10~ns long trigger pulse) is sent directly to the backplane. \textit{module triggers} are evaluated by the backplane to produce \textit{backplane triggers} (see Section \ref{Backplane} for a description of the backplane trigger system). In case of a trigger request from the backplane, the FEE gets a timestamp of the data to be read out. The time stamp can be translated into the appropriate address within the analog buffer and data can be digitized and sent to the FPGA for transfer to storage. Currently, the typical length of a readout is 4 blocks of 32 samples. This highly integrated circuit allows us to implement a trigger system, sampling and readout of 64 high-speed channels within the given tight space at a reasonable power consumption of approximately 13~W per module.

The input signals to the TARGET7 chip are AC coupled to the SIPM output and an offset of 0.7~V is applied to them by DAC to optimize the usage of the dynamic range of the digitizer. For calibration, this pedestal offset can be varied through the entire voltage range of the digitizer.

\section{Backend Electronics} \label{Backend Electronics}
The focal plane of the pSCT is divided into 9 sectors, each of which will have its own backplane PCB and two DACQ boards. Each sector can hold a maximum of 25 modules. Corner sectors only hold 13 modules. All 9 backplanes will connect with a single telescope-wide trigger system (see Section \ref{DIAT Board}) and a single housekeeping/command computer (currently a Raspberry Pi with Ethernet interface).

Currently, only one sector of the pSCT is populated with modules, requiring only one backplane and two DACQ boards for full operation. An upgrade of the camera is currently in progress which will fully populate the camera.

\subsection{Backplane} \label{Backplane}

The current pSCT camera contains one backplane mounted to the back bulkhead. When modules are inserted into the camera (see Section \ref{Camera Structure}) they are connected to the backplane via a backplane connector. The modules are held in place via a screw which holds the backplane, module, and back bulkhead together. 

The high-speed serial data from the camera modules are routed through impedance matched low voltage differential signal (LVDS) lines to the backplane via a single connector.   The same connector includes the trigger signals from each module, and the Trigger Acknowledge (TACK) messages needed to localize the triggering data in the analog pipeline. This connector also carries the 12~V DC power for the module, distributed by the backplane.

The backplane also provides housekeeping and power supply management functionality.  Each backplane has a single 70~V input DC converter board that produces the 12~V DC input to the backplane. A second mezzanine power supply module is used to derive all of the other voltage levels needed by the backplane. Putting these power supply components on mezzanine cards allows relatively low-lifetime components (e.g., electrolytic capacitors) to be replaced without pulling the main backplane motherboard. A housekeeping FPGA (HKFPGA) is used to control critical power-up sequencing of the various voltage levels required by the FPGAs and FEEs, and to reduce the current inrush. Current shunts and ADCs are used to monitor voltages and currents. Power field effect transistors (FETs) are used to control power to individual FEE modules and the DACQ boards. The HKFPGA also communicates with the trigger FPGA (TFPGA), allowing slow-control commands for configuring and monitoring the trigger logic. A single SPI link to each BP HKFPGA allows communication through a small housekeeping/command computer (currently a Raspberry Pi with Ethernet interface, see Section \ref{Raspberry Pi}). 

An upgrade to the pSCT is currently underway and will increase the number of backplanes from one to nine, one for each sector of the focal plane. Each backplane will connect to two DACQ boards (see Section \ref{DACQ Boards}). Backplanes in the same row will have alternating orientations (they are flipped 180$^{\circ}$ with respect to their neighbor) while backplanes in the same column will have the same orientation.

\subsubsection{Backplane Triggers}

One of the principal responsibilities of the backplane is to provide the camera trigger logic and time synchronization functions (working with the telescope's time-tagging system; see Section \ref{Time-Tagging}). Each camera module covers 64 SiPM image pixels and provides 16 trigger signals to the backplane each coming from one trigger pixel (called \textit{module triggers}; see Section \ref{Module Sampling and Trigger System}). The programmable pulse width set by the FEE electronics directly determines the coincidence resolving time of the pattern trigger. The pulse width of the \textit{module trigger} is currently set to $\sim$10~ns and no additional discriminator or one-shot components intervene to modify the pulse width. A minimal overlap of approximately 1.5~ns of these FEE trigger pulses in the TFPGA is sufficient to generate a coincidence. 

The backplane forms \textit{backplane triggers} by using a single TFPGA to form coincidences from the 400 \textit{module trigger} inputs for the 25 modules in each backplane. A \textit{backplane trigger} is only produced when three adjacent trigger pixels produce a \textit{module trigger} at the same time. A pixel is considered adjacent if it is orthogonally-adjacent or diagonally-adjacent (i.e. a pixel can have up to eight adjacent pixels).

The TFPGA runs on an external 125~MHz clock, from which one derives an internal 250~MHz clock for the synchronous logic. However, the TFPGA logic actually allows one to further interpolate this 4~ns time resolution, giving coincidence resolving times (minimum trigger pulse overlap) down to 1.5~ns, and time stamping of triggers to 1~ns. This time precision is achieved by dividing each trigger input into 4 pipelines (A, B, C, D) with  phase offsets of 0~ns, 1~ns, 2~ns, and 3~ns (respectively) with respect to the 250~MHz clock.  The 400 inputs are latched on each of the 4 phase offsets, and then delivered to the coincidence logic.  The coincidence logic (e.g., to derive 3 adjacent hits) is exactly duplicated for each pipeline, and implemented as synchronous logic running at the 4 different clock phases.   

If any of the coincidence logic pipelines give a high value, a \textit{backplane trigger} is generated. We note that the DIAT communication protocol (not used in the prototype) includes synchronization words to ensure that the phase of the 8~ns clock is common between the FEE modules and TFPGA.  In the pSCT camera, the TFPGA provides a similar mechanism to send sync messages to the FEEs. This 1~ns time is latched by the first pipeline to trigger, time-tagging the coincidence to a similar precision. The hit pattern is also latched on a trigger, and the time and hit pattern are sent to the DIAT interface board on the backplane which serializes the data to forward through a high-speed serial link to the DIAT. 

\subsection{DACQ Boards} \label{DACQ Boards}

Data acquisition is accomplished with two custom Data ACQuisition (DACQ) boards. Each board hosts a 16 channel 1~Gbit Ethernet switch implemented in an FPGA. The boards are based on the commercial White Rabbit design by Seven Solutions, but were modified to provide electrical and mechanical interfaces needed for compatibility with the backplane motherboard. For the pSCT application, a number of features of these boards are not used and they function as simple network switches to route the data from the FEE modules to a single 1~GigE network connection.  These boards were originally designed for the CHEC/GCT camera which used an identical backplane design, but was populated with 32 camera modules (see \cite{2018NIMPA} for a more detailed description).  Commands to and data from the FEEs are sent in UDP packets managed by the switch. The LVDS data lines between the FEEs and DACQ boards are aggregated into two connectors on the backplane and delivered via a ribbon cable to the DACQ boards. The power to the DACQ boards can be controlled via the HKFPGA to allow for remote rebooting.  The 125~MHz backplane clock is also distributed to the DACQ boards.

\subsection{Raspberry Pi} \label{Raspberry Pi}

The purpose of the Raspberry Pi is to establish communications to the backplane HKFPGA, to the TFPGA, and to the flasher modules (see Section \ref{Flashers}), none of which can be accessed via the available Ethernet connections to the camera (see Figure \ref{fig:block_diagram}). A backplane control program establishes an SPI link to the HKFPGA of the backplane and, among other things, allows the user to:
\begin{itemize}
    \item switch and monitor supply voltages for camera modules and measure their individual power consumption,
    \item send synchronization messages to the camera modules, 
    \item enable and monitor trigger channels from the camera modules,
    \item control the DACQ board power,
    \item monitor backplane power voltages, currents, and temperatures.
\end{itemize}
The Raspberry Pi also contains software to configure the 3 flashers as well as their trigger system via a serial interface over USB. The trigger system allows the user to choose a rate and time frame for triggers to be sent to the flashers.

\subsection{Camera Trigger Time-Tagging} \label{Time-Tagging}

\begin{figure}[ht]
    \centering
    \includegraphics[width=0.95\textwidth,angle=0]{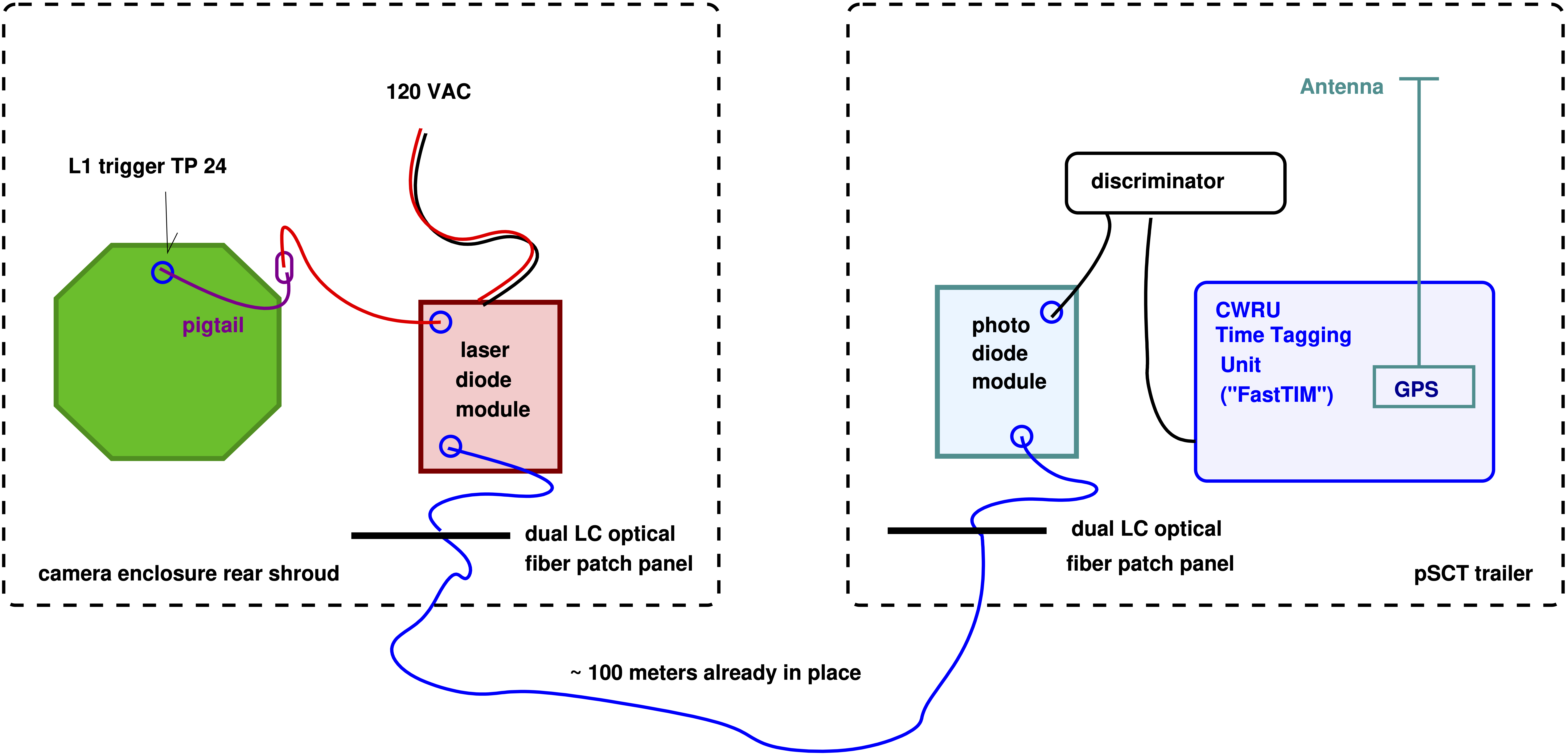}
    \caption{Conceptual schematic of the stand-alone camera event trigger time-tagging system for the pSCT. The Level-1  \textit{backplane trigger} output is used to trigger a pulsed laser diode located at the back of the telescope.  The laser pulse  propagates via fiber optic cable to a photodiode detector which generates a GPS time-stamp accurate to better than 5~ns.}
    \label{fig:ttag_concept}
\end{figure}

In the final SCT the DIAT board will provide essential GPS time-tagging and a synchronized 62.5~MHz clock to each backplane and FEE module. The DIAT will be installed during the Camera Upgrade (see Section \ref{Camera Upgrade}. In lieu of a DIAT board, we have installed a stand-alone backplane event time-tagging system that will operate on the pSCT independent of the backplane. Figure \ref{fig:ttag_concept} shows a conceptual schematic of how the time-tagging system operates. The system is divided into two modules, the laser-diode module which is located in the camera rear shroud and a photo-detector/GPS module which is located in the telescope trailer. The two modules are connected by about a hundred meters of fiber optic communications cable. We use a laser-diode (850~nm, q-switch) to generate fast pulses at up to 5~kHz from camera \textit{backplane triggers}. The output is fed to a discriminator, and then the logic pulse is generated. We latch a GPS time-stamp to each pulse using custom-developed software/firmware 2-in-1 programmable logic device (PLD) implemented on the Xilinx 7200 based Zedboard. The system provides stable timestamps for every camera event trigger to a GPS accuracy of better than 5~ns as measured on the laboratory test-bench. 

\section{Additional Systems} \label{Auxiliary Systems}
\subsection{Flashers} \label{Flashers}

\begin{figure}[ht]
\centering
	\includegraphics[clip, trim={15cm 5cm 0 0}, width=0.5\textwidth]{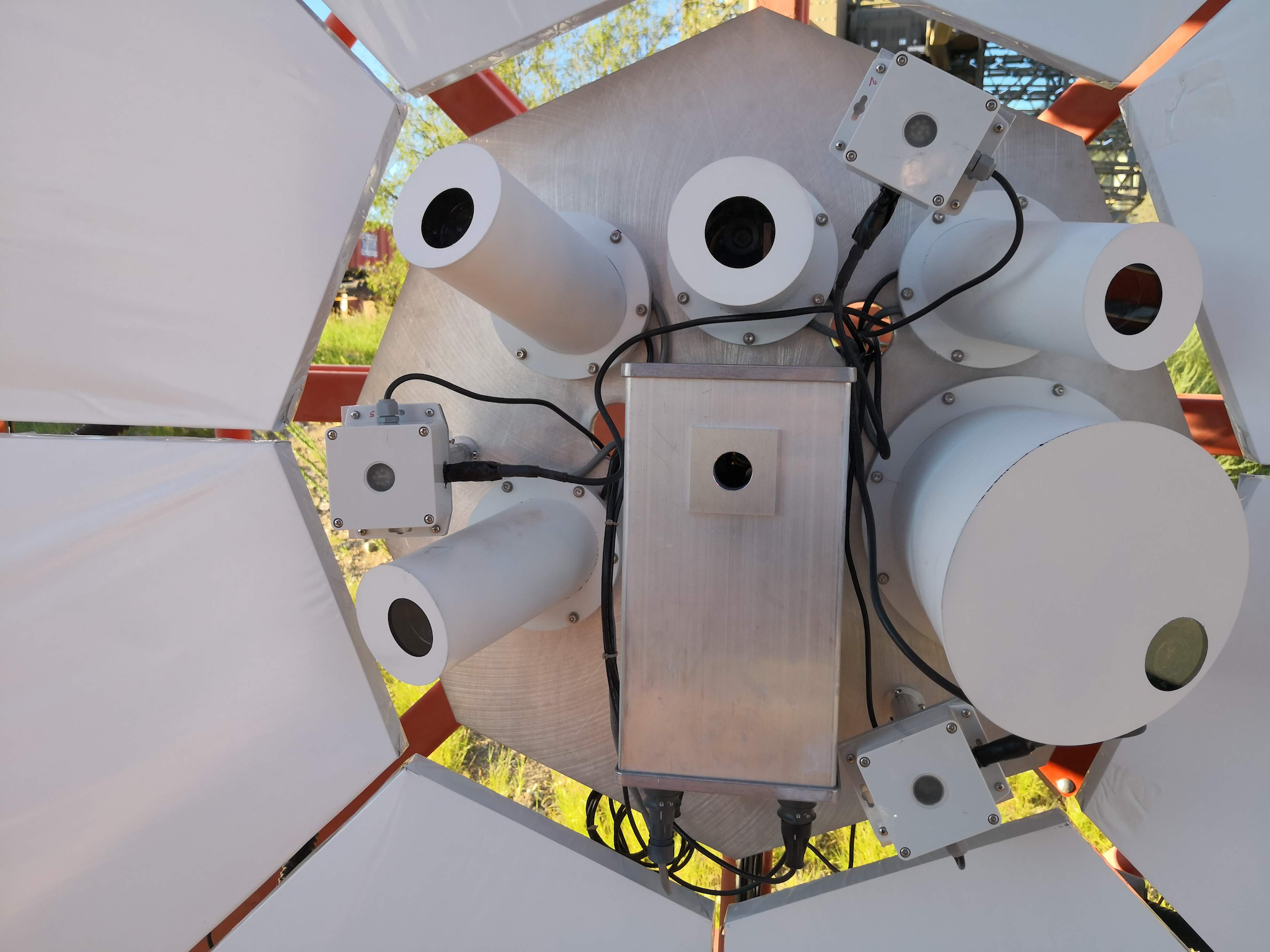}
	\caption{Optical Table located at the center of the secondary reflector. The flashers are the square object mounted along the exterior. From top to bottom the flashers have potentiometer settings 2.5~k$\Omega$, 1.8~k$\Omega$, 4~k$\Omega$. The flasher set to 4~k$\Omega$ is also equipped with a 10\% neutral density filter.}
	\label{fig:OpticalTable}
\end{figure}

A photograph of the LED flashers as they are currently installed on the optical table of the pSCT is shown in Figure \ref{fig:OpticalTable}. This LED flasher system is based on the calibration system of the CHEC-M camera \cite{Brown2015}, and based on the design for the VERITAS LED flasher system \cite{Hanna2010}. The pSCT primarily uses these flashers to provide consistent triggers at a known rate to the camera. A variety of waveform amplitudes can be achieved by adjusting which LEDs in the flasher are used during data taking.

The LED flasher consists of two PCB boards stacked together, linked with a 14-way Samtec SFMC connector. The two boards are the programmable system-on-chip ``PSoC board'' and the ``LED board.''

The LED board is the upper board, and hosts ten 3~mm LEDs providing light in the UV range (400$\pm$2.5~nm). Each LED has a viewing angle of 15$^\circ$. The light from the LEDs in each flasher unit passes through a 20$^\circ$ circle pattern diffuser.  

Three entire assemblies of the flasher are mounted on the optical table at the center of the secondary reflector, as shown in Figure \ref{fig:OpticalTable}. Each flasher is powered through a USB serial connector connection to the Raspberry Pi  (via a USB hub) located in the camera enclosure. Using the Raspberry Pi, these USB connections can be used to turn individual LEDs on and off in order to produce a variety of LED patterns. 

The flasher receives a TTL trigger pulse through a LEMO connector. After the rising edge of the TTL trigger is received at the PSoC board, an active-high LED pulse is generated by a NOT gate, two AND gates, a potentiometer and a 1~pF capacitor, with the pulse width determined by the setting of a potentiometer. 

Currently, triggers to the flashers are provided by an Arduino seated in the back of the camera. The rate and duration of the trigger signal are communicated to the Arduino from the Raspberry Pi. 

The current through an individual LED within a flasher is controlled by a resistance in series with the LED. Each flasher on the pSCT has 5 ``bright'' LEDs (L1, L3, L4, L6, and L8), each of which is connected in series with a corresponding resistor of 80~$\Omega$. The other 5 LEDs are dimmer than these ``bright'' LEDs but each has a unique resistor value. The LEDs L2, L5, L7, L9, and L10 are connected in series with resistors of 100~$\Omega$, 110~$\Omega$, 120~$\Omega$, 140~$\Omega$, and 130~$\Omega$, respectively. These higher resistor values lead to a lower current. 

\begin{figure}[ht]
    \centering
	\includegraphics[width=0.7\textwidth]{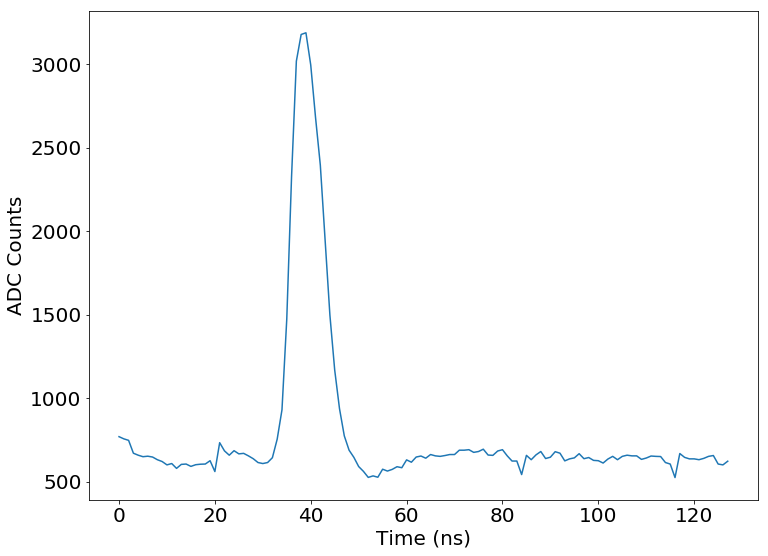}
	\caption{Flasher pulse taken onsite using a flasher on the optical table with the potentiometer set at 6~k$\Omega$. No pedestal or baseline subtraction has been applied.}
	\label{fig:FlasherPulse}
\end{figure}

The output light level of an LED depends on the total charge from an active-high LED pulse, which is controlled by both the current (set by the resistance in series) and the pulse width (set by the potentiometer resistance). After an LED flasher unit is assembled, only the pulse width is adjustable through the potentiometer. A higher potentiometer resistance leads to a wider pulse, and therefore a higher output light level. 
In the current design, the potentiometer can be set only when the flasher is disassembled and uninstalled from the optical table shown in Figure \ref{fig:OpticalTable}. The current potentiometer values for each flasher on the pSCT are noted in the caption. An example raw waveform, from a channel of pSCT from a flasher pulse generated with a 6~k$\Omega$ potentiometer setting, is shown in Figure \ref{fig:FlasherPulse}. 

\begin{figure}[ht]
    \centering
	\includegraphics[width=0.7\textwidth]{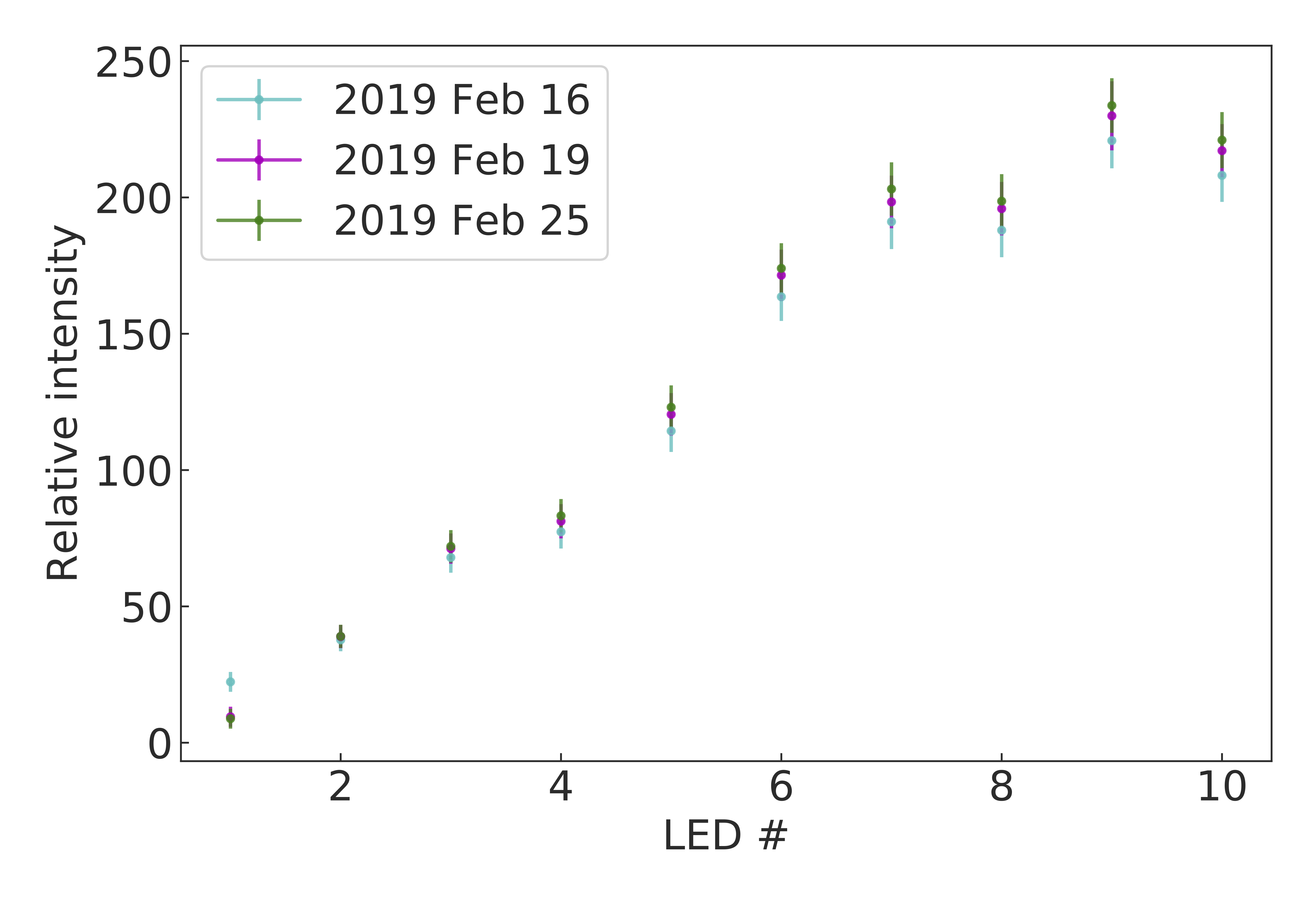}
	\caption{Relative intensity levels for individual flasher LEDs measured in the laboratory. The absolute scale of the relative intensity corresponds to the approximate number of photoelectrons per pixel expected in the camera running on the telescope, accounting for pixel geometry and estimated photon detection efficiency. Each individual LED was flashed independently and the intensity was measured on three different dates. The LEDs in a single flasher each have a different relative output. Here the LEDs are numbered according to this output with 1 being the lowest and 10 the highest.}
	\label{fig:FlasherLevelsIndividualLEDs}
\end{figure}

The relative output light levels for each LED are fixed as the resistance in series is constant. We performed three measurements of the light output of each individual LED on different dates in a similar, controlled environment in the lab, as shown in Figure \ref{fig:FlasherLevelsIndividualLEDs}. The results show good agreement between different measurements for the brighter LEDs. The fluctuation in the LED with the lowest intensity may be attributed to the PMT used in the test. 

When a combination of different LEDs is turned on, the total output light level is not equal to the simple sum of those from individual LEDs when they are turned on by themselves. Small changes in the circuit performance when multiple LEDs are used impact the behaviour of the flasher. 

\begin{figure}[ht]
    \centering
	\includegraphics[width=0.7\textwidth]{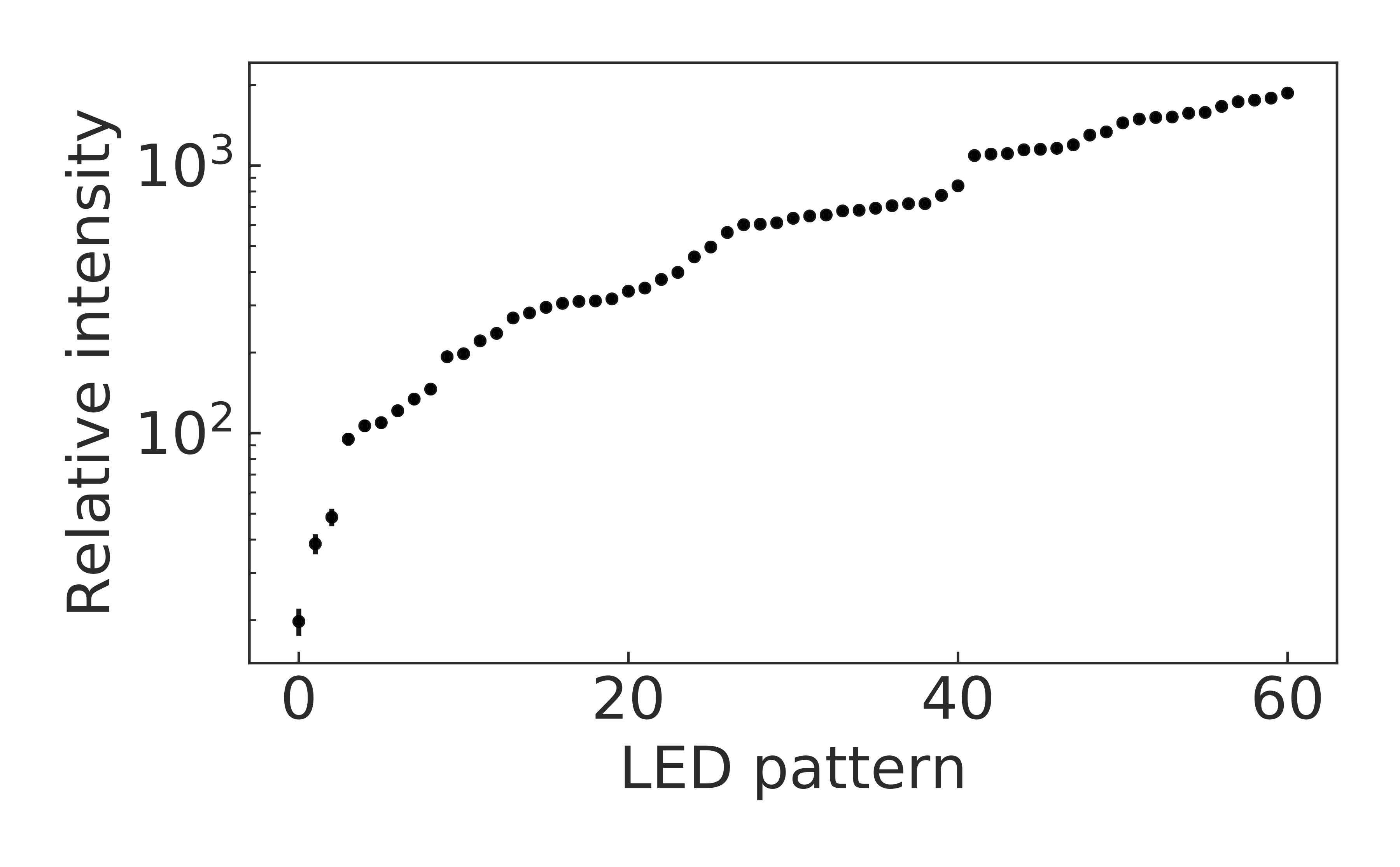}
	\caption{Relative intensity levels for several combinations of flasher LEDs measured in the laboratory. The absolute scale of the relative intensity corresponds to the approximate number of photoelectrons per pixel expected in the camera running on the telescope, accounting for pixel geometry and estimated photon detection efficiency. A selection of patterns were chosen which combine one or more Flasher LEDs. The intensity of these patterns were measured. Here the patterns have been ordered according to their intensity with 1 being the lowest and 60 being the highest.}
	\label{fig:FlasherLevels}
\end{figure}

We show an example of the relative light levels from one flasher unit with the potentiometer set to 2.5~k$\Omega$ in Figure \ref{fig:FlasherLevels}. More combinations of LED on/off patterns can be used to achieve finer increments in light levels. The LED patterns shown are sorted by the light output. The light level is scaled so that it approximately corresponds to the number of photoelectrons (PE value) that a pSCT image pixel detects if the flasher is mounted on the pSCT (at a distance of $\sim$1.86~m). Note the lowest order of magnitude of the output light levels (in this case, roughly between 8 and 80~PEs) is only covered by a few of the weakest patterns, while higher output levels are densely sampled. 

The relative dynamic range covered by one flasher is slightly over two orders of magnitude. The full width at half maximum of the trace measured with the test PMT from all flasher pulses (with a 2.5~k$\Omega$ potentiometer setting) is well under 10~ns. The pulses have a rise time under 4~ns and a somewhat slower decay (under 6~ns).   

\subsection{Camera Power Supply} \label{Camera Power Supply}

Two NEMA weatherproof enclosures mounted in the telescope behind the camera contain rack-mounted power supplies and switches for the camera.  The devices in the enclosures receive 120~V, 60~Hz power.  One Acopian power supply provides power to the camera fans.  Another power supply (Wiener PL506 series) provides power for all other camera components, including silicon photomultipliers and electronics.  This power supply (PBN506 3U RASO) has a height of 3U, a total capacity of 3~kW, and interior space for six power supply modules, each providing an independent power channel.  It currently contains three modules (model PBX506-EX) each operated at 70~V with 8~A capacity and $\sim$90\% efficiency.  One of the three modules powers all 25 FEE modules. Each module consumes approximately 13~W under normal operation. This channel is isolated from the others for noise reduction.

The other two modules are tied together in parallel and power all other camera components (front-end electronics, backplane, and Peltier system).  This module is also operated at 70~V. DC-DC converters on the backplane convert the 70~V to 12~V for use within the camera (see Section \ref{Backplane}).

\subsection{Camera Servers and Software} \label{Server and Software}

The event data stream generated by the camera is sent through the DACQ boards to a data server (DS) located in the pSCT server room over two 9/125 single-mode fiber lines (one per board). The connection is carried through a D-Link DGS-1510-28X 10 Gbps network switch. 

The DS, a rack-mounted 2U custom server from Aberdeen LLC, runs on a 10-core Intel Xeon E5 processor at 2.6~GHz equipped with 64~GB of DDR4 2133~Mhz RAM on eight channels. The server, which is responsible for the data taking and general operation of the camera, also acts as a local data repository with 48~TB of storage space in a RAID6 configuration. The DS is equipped with a dual-port, 10~Gbps Intel X520 network interface card.

The main software components running in the DS are the slow-control (SC) and run-control (RC) systems. The slow control system is responsible for powering and monitoring hardware components of the camera such as the backplane, FEE modules, chiller, fans, and the SiPM temperature controllers.

The SC software suite is written in Python following a server-client model. The server runs as a daemon with continuous connection to the hardware components (each with a specific class providing access to its low-level functionality), while a Python client serves as a user interface that can send commands to the server or receive updates from it. High-level commands are defined by combining several low-level functions on multiple components into a single step, allowing for simplicity of control from the client side. The server-client communication uses the Protocol Buffers serialization library,\cite{Varda2008} while commands and configuration settings are defined and stored in YAML files.

\begin{figure}[ht]
    \centering
    \includegraphics[width=\textwidth,angle=0]{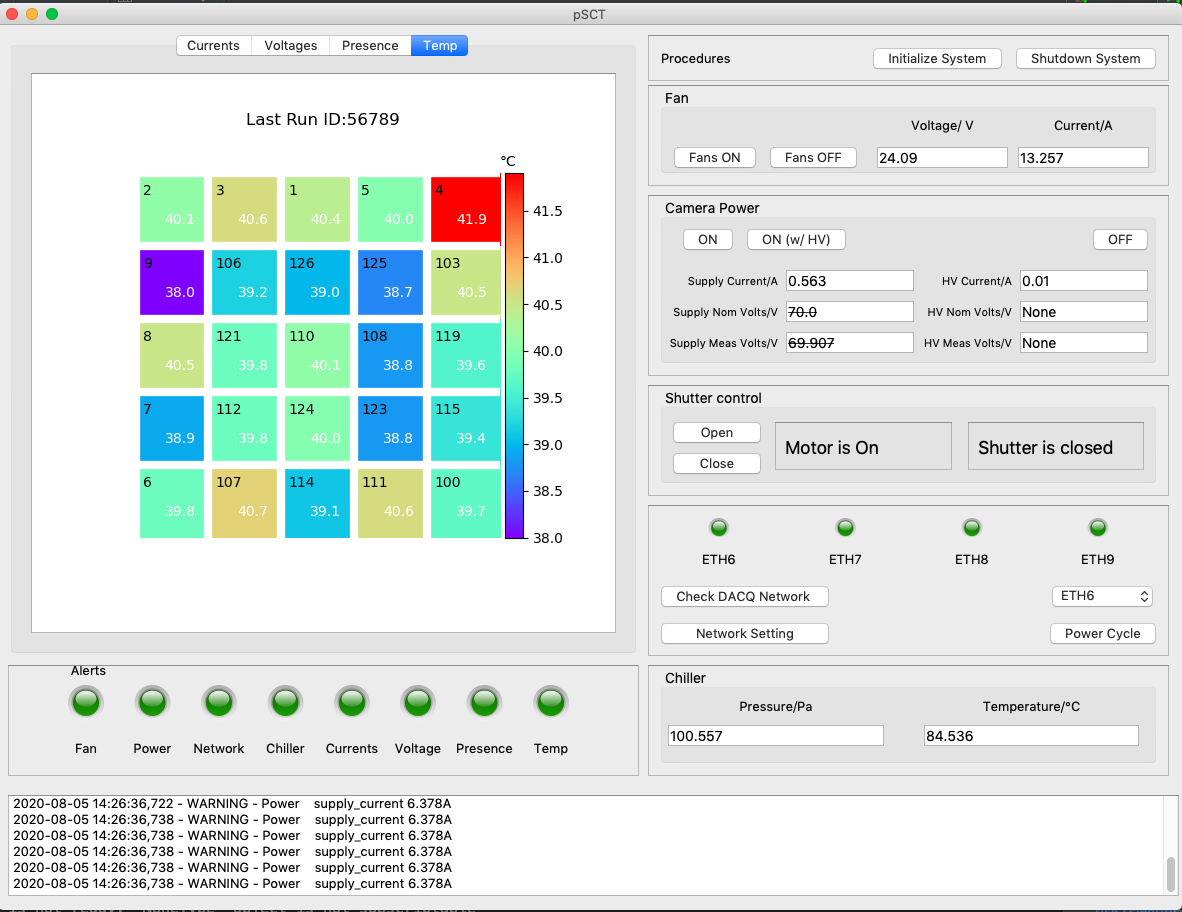}
    \caption{First version of the slow control GUI. Values are displayed at the FEE module level in this view.}
    \label{f:SCGUI}
\end{figure}

The SC system allows the creation of alerts, where nominal and alert values for monitored variables (such as currents, voltages, etc.) are defined so that an alert can be issued if a threshold value is crossed. Two implementations of the SC user interface (UI) exist: a command-line UI that provides fast access to the SC functions and is meant mainly for troubleshooting purposes, and a basic graphical UI (GUI) meant for standard data-taking operations (see Figure \ref{f:SCGUI}). The GUI is written in python and designed using PyQt5. The SC control server logs its output to a local log file and will also save monitoring variables to a local MySQL database.

The RC system of the pSCT follows a similar design to the SC system. Its main functions are to load the configuration settings of the FEE modules ahead of a data-acquisition run; to start, monitor and stop runs; and to manage the recording of the physics-level and calibration data to disk. The RC system runs on a daemon server written in Python which connects to the FEE modules using the \texttt{TargetDriver} and \texttt{TargetIO} libraries developed by the CHEC camera group \cite{2018NIMPA}. The \texttt{TargetDriver} library is responsible for control and readout, while \texttt{TargetIO} reads/writes event data from the modules. The RC system is implemented as a state machine, with definitions that follow current CTA requirements on the possible camera states and allowed transitions.  

Event data is sent from the camera to the DS as asynchronous UDP packets (2 packets per event per module) which are buffered and assembled into full events based on their event trigger ID. Events are subsequently written to disk in FITS \cite{2010AA} files using the \texttt{TargetIO} library.

The data throughput that the DS must handle depends on the length of the readout waveforms, which can be modified, and the trigger rate, which depends on the trigger thresholds. While first tests (e.g., Figure \ref{fig:FlasherPulse}) have been performed using a 128-sample waveform, it is expected that a 64-sample waveform will be used in standard data-taking mode. The waveform digitizer dynamic range requires two bytes per sample for storage, which added to a five-byte header per waveform results in a waveform size of 133~bytes, or 200~kB for a full-camera readout of 1,600 imaging pixels. A trigger rate of 1~kHz would result in a throughput of 1.6~Gbps, well suited for the prototype network and computing speeds.

\section{Performance} \label{Performance}
pSCT performance testing focused on elements which were relevant to the camera upgrade. This includes software testing, waveform calibration, and trigger performance. Further performance testing is planned after the camera upgrade has been completed.

\subsection{Software Tests} \label{Software Tests}

\begin{figure}[ht]
    \centering
    \includegraphics[width=\textwidth]{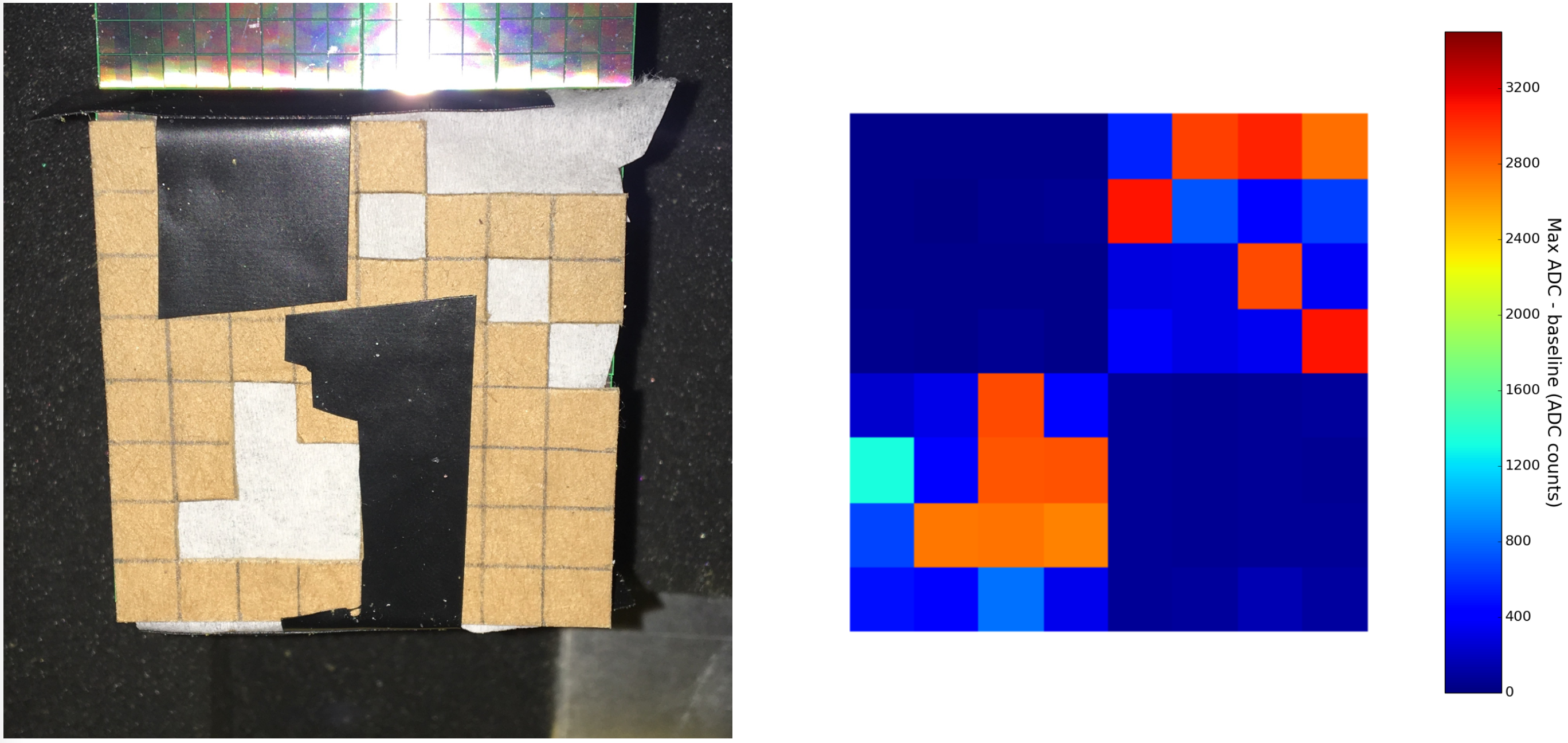}
    \caption{The face of a single module FEE was first covered with a tissue (which allows light to pass through) and then with a cardboard mask. The mask was affixed to the FEE with black electrical tape. An LED was flashed on the prepared FEE. The left shows a picture of the physical setup of the module. The right shows the heatmap of average waveform height in different channels. Some cross-talk exists between adjacent pixels in the same quadrant. However, it is clear that the ASIC channels are connected to the correct Pixel IDs, as those pixels exposed to LED pulses have significantly larger average waveform heights.
    \label{fig:MaskTest}}
\end{figure}

An initial software test was checking that the pixels in the software were mapped correctly to their physical location in the camera. The camera can be easily run with a subset of modules. Because of this, we were easily able to identify that the module location in the software matched its physical location. To test the channel locations within a module, a mask test was done. This test consisted of masking off a subset of channels within a module and then flashing the module with an LED. The masked pixels were exposed to significantly less light allowing us to differentiate between masked and unmasked pixels. A heatmap (shown in Figure \ref{fig:MaskTest}) of all the pixels in the module showed that the pixels' location in software correctly matched their physical location.

A second software test was to check that the flashers could consistently trigger the camera when used. To test this a rate scan was run while flashers were on. The flashers were run at 10~Hz and the rate scan shows a clear plateau at 10~Hz, indicating that there is a wide range of thresholds over which the flasher will trigger the camera. Results of the flasher test are shown in Figure \ref{fig:RateScanShutter}

\subsection{Waveform Calibration} \label{Waveform Calibration}

Inside the TARGET7 chip, the data path (the path of the recorded charge from the sampling array through the analog buffer to digitization) is subject to variations in physical routing of the connections and variations in the capacitance of the storage buffer elements (see Appendix \ref{Appendix:ASIC} for a more detailed discussion). Therefore, the buffer elements that have been used need to be recorded with every waveform and appropriate voltage to ADC count calibration needs to be applied on a sample by sample basis. The current strategy for calibration includes two steps: pedestal subtraction and ADC to voltage conversion via a lookup table.

For pedestal subtraction, a database needs to be created from pure electronic noise data, to prevent bias from dark SiPM pulses. This can be achieved by using a software trigger to produce waveforms with the SiPM bias voltage disabled. The first digitized block of a waveform shows a systematic elevation in ADC counts which needs to be accounted for when constructing the database and in calibration.

Once the pedestal subtraction has been applied, the final step is to convert waveforms from ADC counts to voltage via a lookup table which needs to be created from waveforms, recorded at different input pedestal voltages. Generally, for Hamamatsu sensors, the conversion for all cells is close to 2 counts for every 1~mV or 8~ADC counts at the peak of a 1~PE pulse. FBK sensors have double the electronics gain of Hamamatsu sensors and therefore the conversion is close to 16~ADC counts per 1~PE pulse. Once the ADC conversion is calibrated, the charge can be found by integrating over the waveform.

\subsection{Noise and Trigger Performance} \label{Noise and Trigger Performance}

The trigger path of the camera is responsible for processing incoming signals quickly using higher-level logic to determine if the charges should be digitized and read-out. On the module level this is done with two inverting amplifiers per pixel and an inverting summing amplifier per trigger pixel. These are used to amplify and take the analog sum of a group of four pixels (called a trigger pixel). If this sum surpasses a selected threshold then the module produces a trigger which is sent to the backplane.

Rate scans are one way in which to measure the effect of changing this threshold. A trigger is generated when the input signal crosses the specified threshold. A rate scan consists of systematically changing this threshold (starting at a high value and decreasing in regular steps) and measuring the trigger rate at each threshold. The trigger rate is measured by recording the TACK rate. When the threshold reaches the amplitude of an incoming signal the rate rapidly increases. As the threshold is decreased further the rate will increase again as the system begins to trigger on noise. Currently, the rate saturates at approximately 5~kHz due to an enforced $200 \mu s$ dead time. This dead time was implemented in order to ensure that modules would not trigger on noise produced by the module's own TACK signal.

\begin{figure}[ht]
    \centering
    \includegraphics[width=0.7\textwidth] {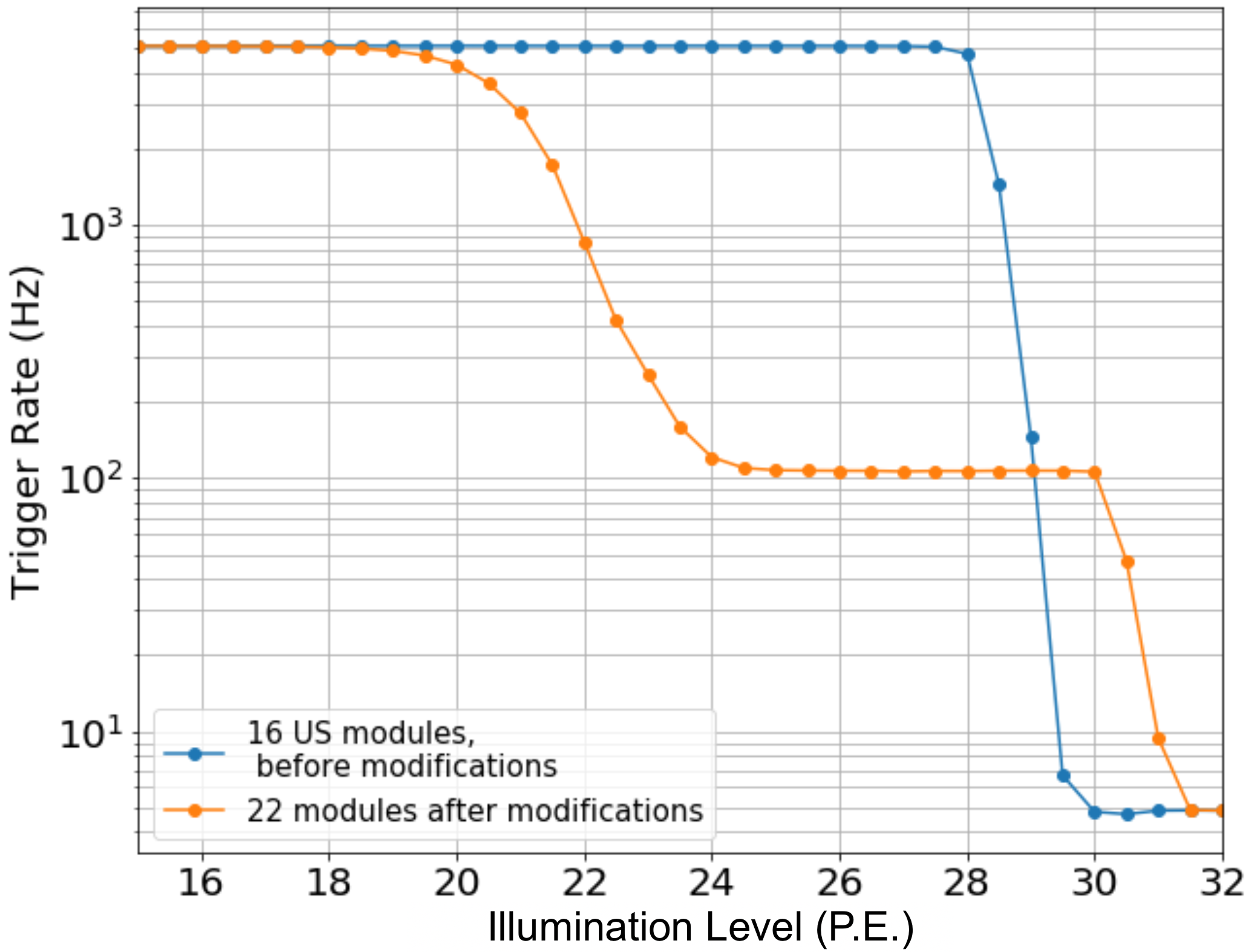}
	\caption{Trigger noise reduction. Prior to noise reduction steps the trigger noise was high enough that it saturated the trigger system. After these noise-reducing modifications the trigger rate was reduced to 22~PE for a 1~kHz trigger rate \cite{Wachtendonk2018}.}
	\label{fig:TriggerNoise}
\end{figure}

\begin{figure}[ht]
    \centering
	\includegraphics[width=0.7\textwidth]{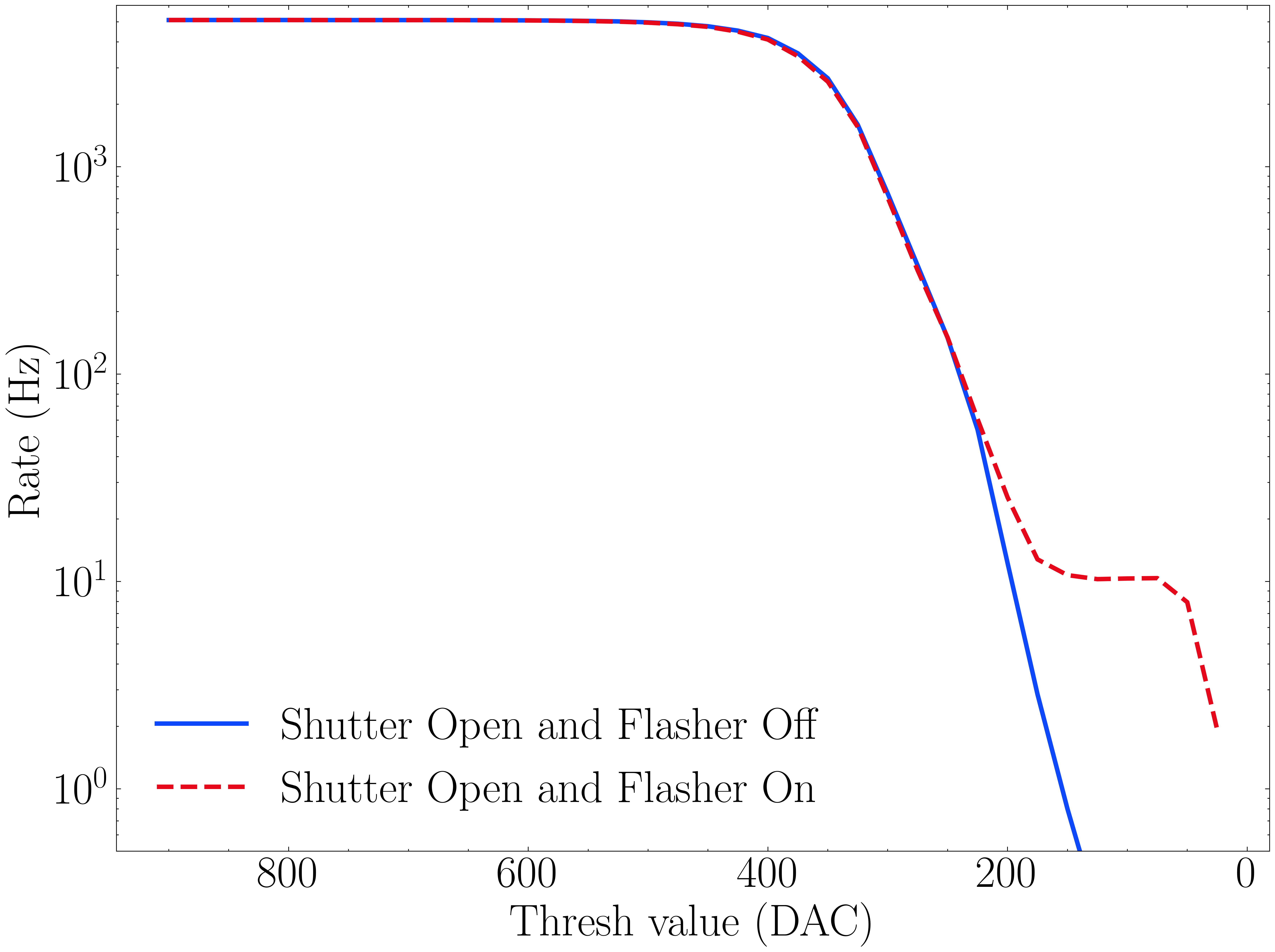}
	\caption{Rate scans with flasher on and off. Rate scans included all modules and were taken under typical operating conditions in the same night. The rate scan taken with the flasher on shows a clear plateau at 10 Hz (the same rate at which the flasher was run).}
	\label{fig:RateScanShutter}
\end{figure}

\begin{figure}[ht]
    \centering
	\includegraphics[width=0.7\textwidth]{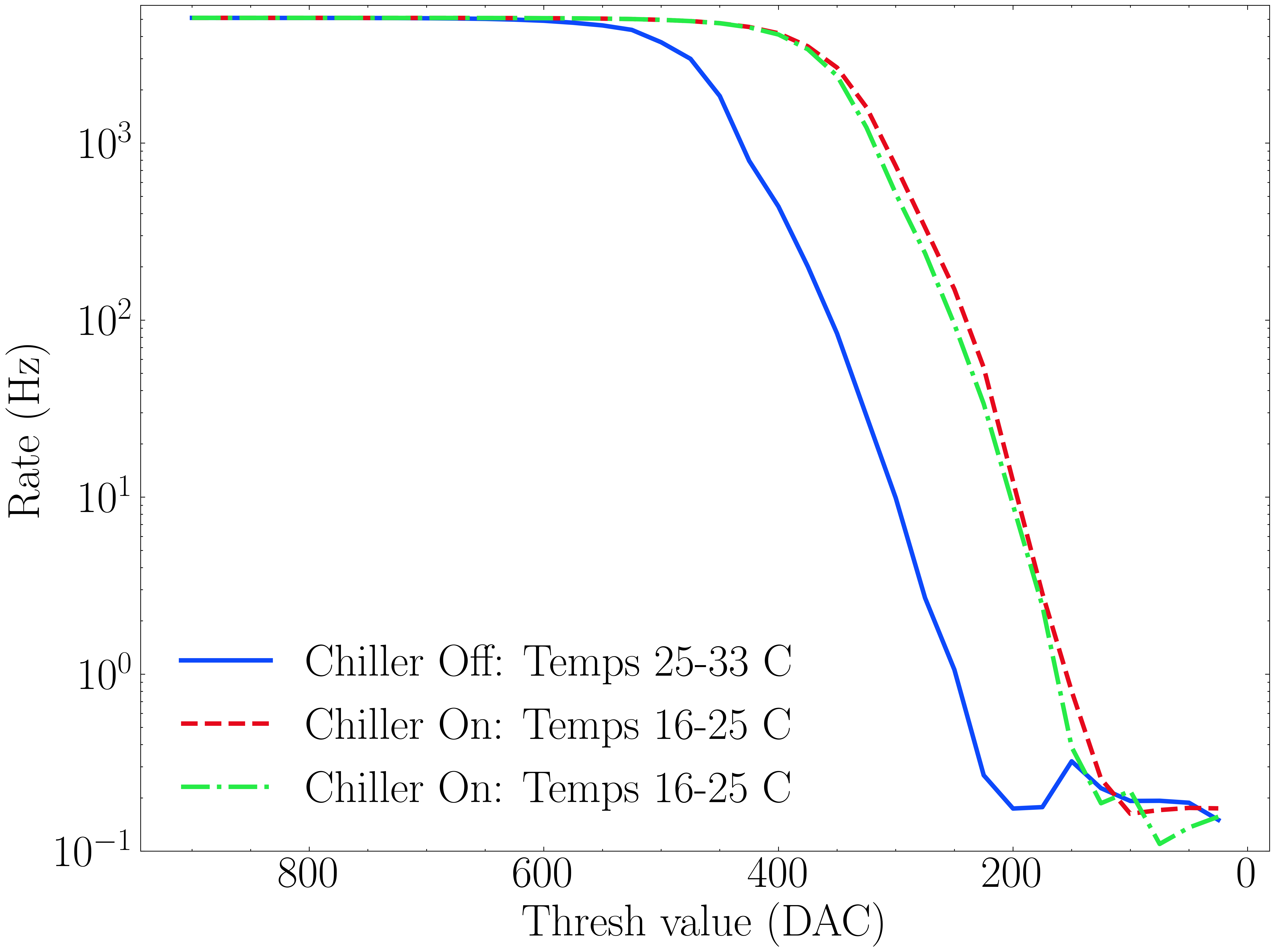}
	\caption{Rate scans included all modules and were taken with the shutter open on the same night so that ambient conditions were the same. One rate scan was taken with the chiller off, resulting in warmer average module temperatures. Two rate scans were taken with the Chiller on resulting in colder temperatures. Rate scans with a lower temperature are shifted to lower Thresh values (corresponding to a higher physical threshold).}
	\label{fig:RateScanChiller}
\end{figure}

Because the incoming signal travels through three inverting amplifiers before being compared to this threshold, the way in which the threshold is specified is unintuitive. The threshold level is set by the Thresh parameter. A low Thresh value corresponds to a physically high threshold while a high Thresh value corresponds to a physically low threshold \cite{Wachtendonk2018}. US and INFN SiPMS have different gains (190~ADC~ns and 300~ADC~ns respectively) which means that the conversion between Thresh value and photoelectrons for both modules is different. For rate scans which include only one type of module (Figure \ref{fig:TriggerNoise}) it is straightforward to make this conversion, however for rate scans which include a mix of modules (Figures \ref{fig:RateScanShutter} and \ref{fig:RateScanChiller}) it is impractical to make this conversion.

Noise in the trigger path can cause the module to trigger even when there is no signal present. Such noise was found to be caused by radiative coupling with voltage regulators present on the module FEEs and with the aluminum housing within which the FEEs are fixed. The voltage regulators were redesigned and the aluminum housing was replaced with a fiberglass alternative in all modules, significantly reducing the noise on the trigger path \cite{Wachtendonk2018}. Figure \ref{fig:TriggerNoise} shows this reduction.

The nominal maximum rate at which the camera can be operated is 1000~Hz. The nominal maximum throughput that can be sustained is 4x1~Gbps (This corresponds to four 1 Gbps lines, two in each DACQ board). The real maximum rate and maximum throughput have not yet been measured.

Once installed in the telescope, further rate scans were taken in order to determine how the night sky background rate and inherent trigger noise rate compare. This was achieved by taking rate scans with the shutter closed (no night sky photons) and with the shutter open under similar circumstances. Figure \ref{fig:RateScanShutter} shows these rate scans. 

It was also found that the temperature of the camera has a strong effect on the rate at a given Thresh value. The colder the temperature the higher the trigger rate for a given Thresh. This is either due to an increase in noise or a change in the threshold with temperature. Figure \ref{fig:RateScanChiller} shows two rate scans taken at different temperature ranges to illustrate this effect.

\subsection{Camera Images} \label{Camera Images}

\begin{figure}[ht]
    \centering
	\includegraphics[width=0.7\textwidth]{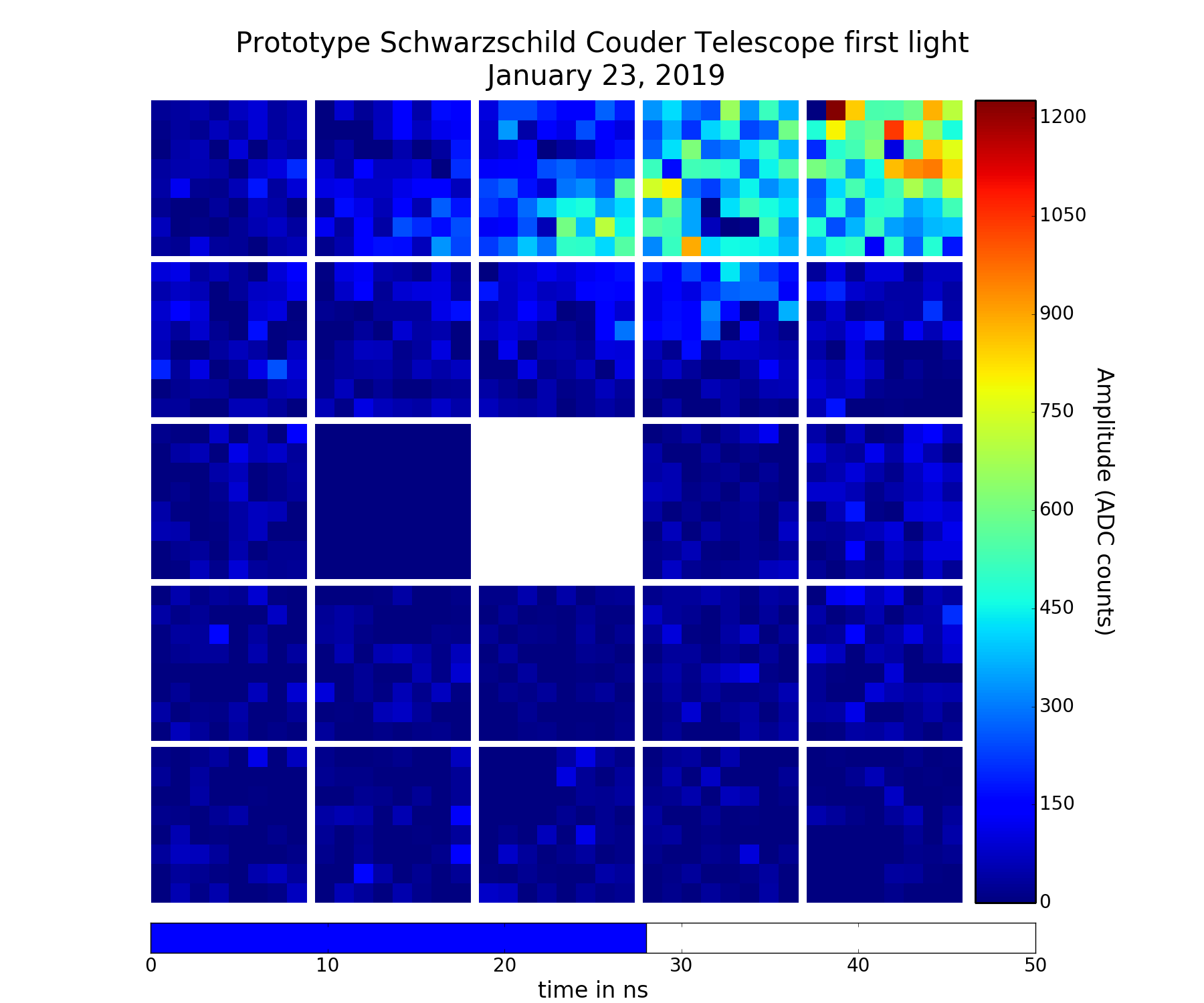}
	\caption{First light from the pSCT. First light and first shower candidates were achieved on January 23, 2019. This is a snapshot of a single event. The center module is blank because that position is temporarily occupied by an alignment calibration module.  The module in the center left is dark because it failed to connect on this run.}
	\label{fig:FirstLight}
\end{figure}

On January 23, 2019 the pSCT camera and optics were simultaneously uncovered for the first time. First light and first shower candidates were achieved on this date. Two runs were taken, one with and one without simultaneous flashers. One event is shown in Figure \ref{fig:FirstLight}.

\begin{figure}[ht]
    \centering
	\includegraphics[width=0.7\textwidth]{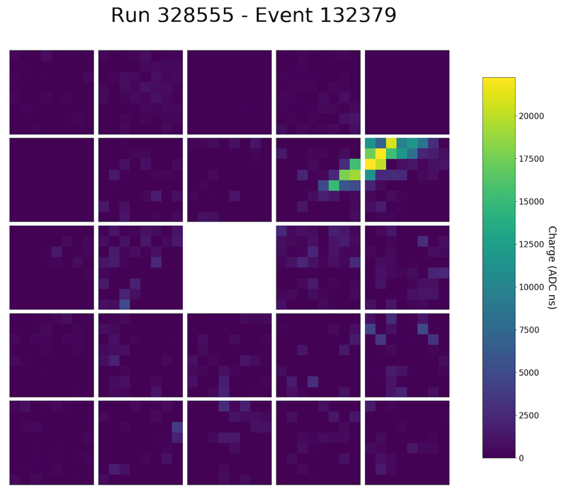}
	\caption{First confirmed gamma-ray-like event recorded by the pSCT. This event was taken on January 17, 2020 with simultaneous observation with VERITAS. This event was confirmed as a gamma-ray via timing coincidence with simultaneous VERITAS observation.}
	\label{fig:FirstGammaRay}
\end{figure}

At this time the optics were not fully aligned, leaving some trace optical effects in the first light data. The optical alignment was completed in December of 2019 and observations of the Crab Nebula began at that time. In January of 2020 the pSCT recorded its first gamma-ray-like shower. This event is show in Figure \ref{fig:FirstGammaRay}. Observations in January of 2020 led to a detection of the Crab Nebula at a significance of 8.6 sigma \cite{adams2021detection}.

\section{Camera Upgrade} \label{Camera Upgrade}
\subsection{Upgrade Outlook} \label{Upgrade Outlook}

The camera is scheduled to be upgraded by December of 2022. The upgraded camera will have a fully populated focal plane, increasing the number of pixels from 1600 to 11,328 and will increase the field of view from 2.7$^{\circ}$ to 8.0$^{\circ}$. This will require a total of 9 backplanes (each with two DACQ boards) which will require synchronized timing between them. A DIAT system (See section \ref{DIAT Board}) will be used to achieve this and will ensure camera-wide triggers between backplanes (see Section \ref{DIAT Board}). The backplane PCB will also be redesigned as its current form factor was intended to accommodate 32 modules as opposed to the 25 modules per sector required for the pSCT. Each backplane will control up to 25 FEE modules (corner backplanes control only 13 modules. See Fig \ref{fig:cta_cameraDesign}).The 9 backplanes will be mounted to the back bulkhead of the camera with their associated DACQ boards mounted on top of them.

The camera upgrade will represent a substantial reconfiguration in terms of the electronics components utilized. Instead of a mixture of FBK NUV-HD and Hamamatsu SiPMs, the entire camera will use the third generation FBK NUV-HD SiPMs. These sensors demonstrate substantial improvements over those currently installed, including improved photon detection, lower temperature-dependence, and lower breakdown voltage. The FEE modules will use a new design that incorporates next-generation TARGETC ASICs for digitization and sampling, and a separate TARGET 5 Trigger Extension ASIC (T5TEA) for triggering. This new design is expected to significantly reduce trigger noise \cite{tibaldo2015target}. The PCBs for the new FEE modules are also improved over older designs. 

Another new custom ASIC called the SMART chip is being developed to connect the SiPMs to the FEE modules. This chip will allow for improved control over SiPM bias voltages as well as pulse shaping and amplification directly at the SiPM sensors. 

Auxiliary systems such as the heat management system and the camera shutter will also be improved during the upgrade. The heat management system which currently accommodates 25 modules will be improved to accommodate the 177 modules of the full pSCT camera. The LED flashers will also be upgraded to improve performance and trigger synchronously with the backplanes \cite{meuresICRC2019}.

\subsection{DIAT Board} \label{DIAT Board}

The Distributed Intelligent Array Trigger (DIAT) is a custom-designed board module for GPS time-tagging, timing synchronization, hit pattern pre-processing, and distributed trigger formation \cite{2018NIMPA.891....6D}.  The DIAT has been specifically designed to optimize the performance of the SCT and will replace the current Time-Tagging system (see Section \ref{Time-Tagging}). 

By design, the DIAT provides five key features to the operation of the SCT camera and telescope:

\begin{enumerate}
    \item It provides a GPS-disciplined 62.5~MHz clock signal to each of nine backplanes, actively synchronizing the internal 125~MHz clock signals distributed to every FEE detector module within the camera.
    \item It collects low-level trigger data from each of up to nine backplane modules within the camera to form a telescope trigger.
    \item It provides nanosecond accuracy GPS time-tagging for all telescope trigger events.
    \item It also provides rapid, intelligent pre-processing of all low-level trigger information, resulting in a digital hit-pattern to indicate trigger pixels above the threshold.
    The DIAT uses these patterns to calculate lower order moments which can then be shared with neighboring telescopes to form regional triggers, thus reducing the rate of spurious telescope camera triggers. These moments are also sufficient to veto cosmic-ray induced air showers which can further reduce triggers \cite{2018NIMPA.891....6D}. 
    \item Finally, it can also be programmed to enable synchronized timing and cross-triggers with any neighboring telescopes.
\end{enumerate}

For the pSCT, a first ``demonstrator'' version of the DIAT is interfaced to the nine backplanes that will be present in the fully upgraded camera. The DIAT demonstrator provides functionality associated with features (1)--(3) listed above. 
In the pSCT, the DIAT will be installed within the data collection trailer and connected to the camera backplane via fiber optic cable and custom designed interface boards. 

The critical role of the DIAT demonstrator for the pSCT is camera synchronization.  Synchronization of the back and front end electronics is achieved by DIAT calibration of delays and by setting local clocks to the GPS disciplined clock of the DIAT. The messages between the DIAT and the backplane tell the backplane to set the 1~ns clock to some value on the next clock edge.

The FEE module does not inherently have true 1~ns precision. Instead it synchronizes with the 8~ns clock provided by the backplane. The module then reads the 64~bit 1~ns time and uses this to address the analog memory. The module resets the counter to the next 8~ns clock edge using SYNC messages generated by the DIAT which are also passed through the backplane. The backplane in turn ensures that the 125~MHz clock is aligned to better than 80~ps.

\begin{figure}[ht]
    \centering
    \includegraphics[width=0.95\textwidth,angle=0]{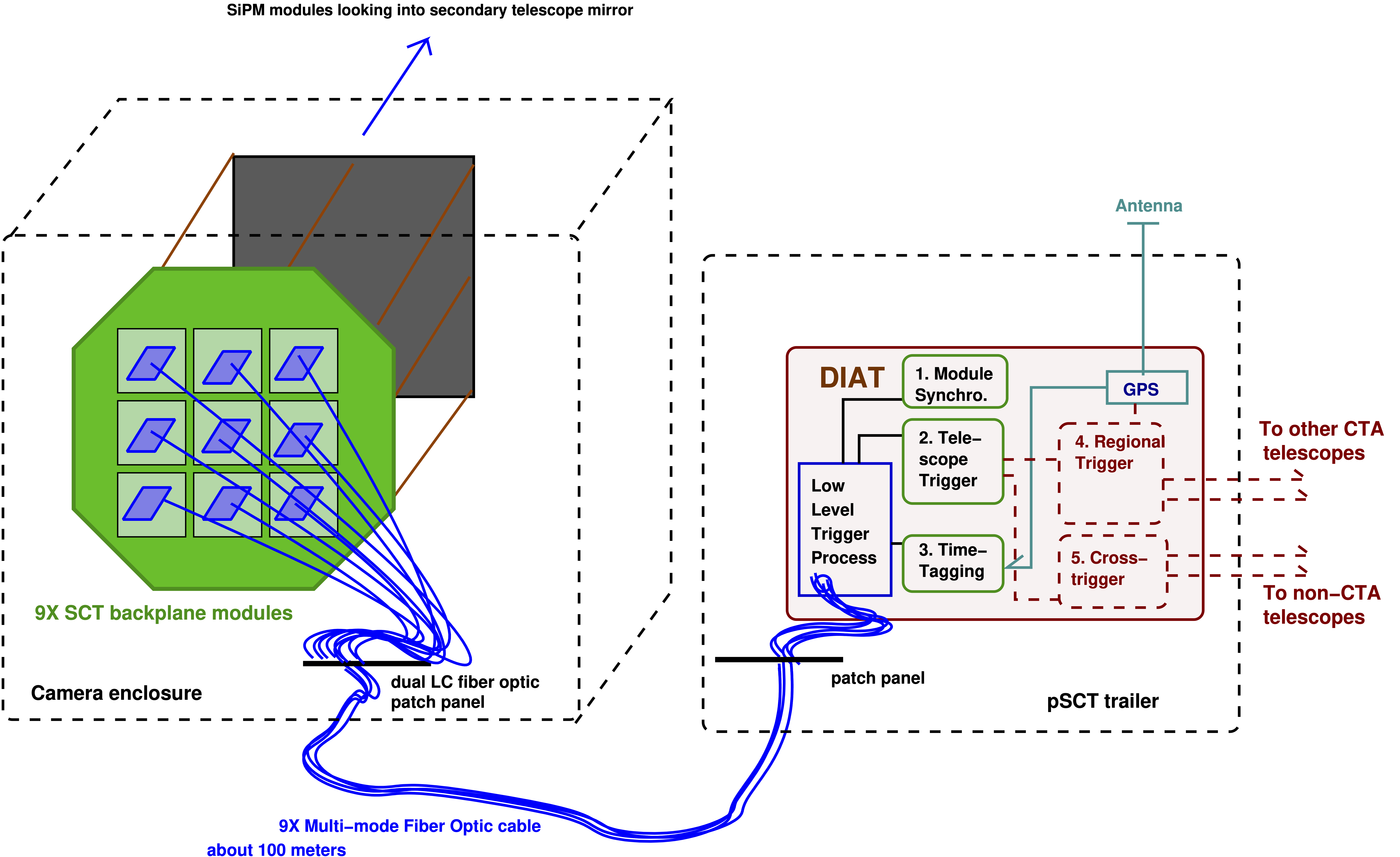}
    \caption{Conceptual schematic of the operation of the Distributed Intelligent Array Trigger (DIAT) module (right) in relation to the full SCT camera with nine backplanes (left).  Low-level trigger and timing messages from individual detector modules (FEE) are collected and relayed with timing information by each of nine backplane modules via fiber optic lines.  The DIAT module (located within the telescope trailer) has five functions: (1) synchronization of detector modules, (2) telescope Triggers, (3) GPS time-tagging, (4) regional Triggers (via shared hit-patterns) and (5) cross-triggers with other telescopes in CTA.  For the pSCT, the DIAT demonstrator module provides functions (1)--(3) as required for a single backplane.}
    \label{f:DIAT_concept}
\end{figure}

The DIAT will provide functionality for synchronization for up to nine backplanes within the SCT camera and across multiple telescopes. Figure \ref{f:DIAT_concept} shows a conceptual schematic of the DIAT in relation to the full SCT camera. The DIAT is designed to act as an intelligent trigger for both a single SCT telescope (generating telescope triggers based on information relayed from each backplane module). In principle, the DIAT can also provide functionality for cross-triggers with neighboring SCT telescopes within the larger CTA array.

The DIAT interface uses a SerDes device and SFP Gbit fiber connection to send the trigger time and hit pattern to a single DIAT module in the telescope operating trailer. If there are multiple backplanes or multiple telescopes, the DIAT derives a coincidence trigger and sends a message (a TACK command) back to each backplane in the camera. The 125~MHz backplane clock can be derived from a free-running oscillator (as it is for the prototype) or from a phase-locked clock derived by the DIAT interface board from the fiber-optic data connection. The DIAT uses time-of-flight message delays to synchronize the time on each backplane, ensuring that all backplanes and camera modules have a common absolute 1 ns time for time-tagging triggers, and localizing the region of interest around each trigger in the analog pipeline \cite{wang2015ftk}.


\section*{Acknowledgements}
We would like to acknowledge the significant contributions of Phil Moore (formerly Washington University, deceased) to the design of the backplane and data acquisition electronics.

This research is supported by grants from the U.S. National Science Foundation and the Smithsonian Institution, by the Istituto Nazionale di Fisica Nucleare (INFN) in Italy and by the Helmholtz Association in Germany. The development, construction and operation of the pSCT was supported by NSF awards 
(PHY-\-1229792, PHY-\-1229205, PHY-\-1229654, PHY-\-1913552, PHY-\-1807029, PHY-\-1510504, PHY-\-1707945, PHY-\-2013102, PHY-\-1707544, PHY-\-2011361, PHY-\-1707432, and PHY\-1806554) together with funds from
Barnard College,
California State University East Bay,
Columbia University,
Georgia Institute of Technology,
Iowa State University,
Smithsonian Institution,
Stanford University,
University of Chicago,
University of Alabama in Huntsville,
University of California,
University of Iowa,
University of Utah,
University of Wisconsin--Madison, and
Washington University in St. Louis.
ASTRI participation in this effort was supported by the Italian Ministry of University and Research (MUR) with funds specifically assigned to the Italian National Institute of Astrophysics (INAF) for the development of technologies toward the implementation of CTA. 
Support from the Japan Society for the Promotion of Science was provided by KAKENHI grant numbers JP23244051, JP25610040, JP15H02086, JP16K13801, JP17H04838 and JP18KK0384. This work was also partially supported by UNAM-PAPIIT IG101320.
We acknowledge the excellent work of the technical support staff at the Fred Lawrence Whipple Observatory and at the collaborating institutions in the construction and operation of the instrument. We also thank the VERITAS Collaboration for their cooperation in obtaining joint observations and for the use of their data. 
This paper has gone through internal review by the CTA Consortium.

\bibliographystyle{spiejour}
\bibliography{references.bib}

\begin{thebibliography}{10}

\bibitem{actis2011design}
M.~Actis, G.~Agnetta, F.~Aharonian, {\em et~al.}, ``{{Design concepts for the
  Cherenkov Telescope Array (CTA): an advanced facility for ground-based
  high-energy gamma-ray astronomy}},'' {\em Experimental Astronomy} {\bf
  32}(3), 193--316  (2011).

\bibitem{2018NIMPA}
J.~{Zorn}, R.~{White}, J.~J. {Watson}, {\em et~al.}, ``{{Characterisation and
  testing of CHEC-MA camera prototype for the small-sized telescopes of the
  Cherenkov telescope array}},'' {\em Nuclear Instruments and Methods in
  Physics Research A} {\bf 904}, 44--63  (2018).

\bibitem{consortium2017science}
G.~Tovmassian and the Consortium, {\em Science with the Cherenkov Telescope
  Array}  (2019).

\bibitem{Wachtendonk2018}
M.~Wachtendonk, ``{{Characterization and Reduction of Noise in the Prototype
  Schwarzschild-Couder Telescope Camera}},'' Master's thesis, University of
  Wisconsin - Madison  (2018).

\bibitem{vassiliev2007schwarzschild}
V.~V. Vassiliev, S.~Fegan, and P.~Brousseau, ``{Schwarzschild-Couder two-mirror
  telescope for ground-based $\gamma$-ray astronomy},'' in {\em {30th
  International Cosmic Ray Conference}},   {\bf 3}, 1445--1448  (2007).

\bibitem{vassiliev2017prototype}
V.~Vassiliev, ``{{Prototype 9.7 m Schwarzschild-Couder telescope for the
  Cherenkov Telescope Array: Project Overview}},'' {\em PoS} , 838  (2017).

\bibitem{otte2015development}
N.~Otte {\em et~al.}, ``{Development of a SiPM Camera for a
  Schwarzschild-Couder Cherenkov Telescope for the Cherenkov Telescope
  Array},'' {\em PoS} {\bf ICRC2015}, 1023  (2016).

\bibitem{Zorn:2019hgk}
J.~Zorn, ``{CHEC\textemdash{}A compact high energy camera for the Cherenkov
  Telescope Array},'' {\em Nucl. Instrum. Meth. A} {\bf 936}, 229--230  (2019).

\bibitem{adams2020verification}
C.~Adams, R.~Alfaro, G.~Ambrosi, {\em et~al.}, ``Verification of the optical
  system of the 9.7-m prototype schwarzschild-couder telescope,'' in {\em
  Optical System Alignment, Tolerancing, and Verification XIII},   {\bf 11488},
  1148805, International Society for Optics and Photonics  (2020).

\bibitem{ambrosi2017silicon}
G.~Ambrosi, E.~Bissaldi, N.~Giglietto, {\em et~al.}, ``{{Silicon
  Photomultipliers and front-end electronics performance for Cherenkov
  Telescope Array camera development}},'' {\em Nuclear Instruments and Methods
  in Physics Research Section A: Accelerators, Spectrometers, Detectors and
  Associated Equipment} {\bf 845}, 8--11  (2017).

\bibitem{nieto2017prototype}
D.~N. Casta\~no {\em et~al.}, ``{Prototype 9.7m Schwarzschild-Couder telescope
  for the Cherenkov Telescope Array: status of the optical system},'' {\em PoS}
  {\bf ICRC2017}, 815  (2018).

\bibitem{ieeenss7431131}
J.~{Biteau}, ``{{Characterization of silicon photomultipliers for the Cherenkov
  Telescope Array medium-sized telescopes}},'' in {\em 2014 IEEE Nuclear
  Science Symposium and Medical Imaging Conference (NSS/MIC)},  1--3  (2014).

\bibitem{Otte:2015}
A.~N. Otte, K.~Meagher, T.~Nguyen, {\em et~al.}, ``{Silicon photomultiplier
  integration in the camera of the mid-size Schwarzschild-Couder Cherenkov
  telescope for CTA.},'' {\em Nuclear Instruments and Methods in Physics
  Research Section A: Accelerators, Spectrometers, Detectors and Associated
  Equipment} {\bf 787}, 85--88  (2015).

\bibitem{bechtol2012target}
K.~Bechtol, S.~Funk, A.~Okumura, {\em et~al.}, ``{TARGET: A multi-channel
  digitizer chip for very-high-energy gamma-ray telescopes},'' {\em
  Astroparticle Physics} {\bf 36}(1), 156--165  (2012).

\bibitem{albert2017target}
A.~Albert, S.~Funk, H.~Katagiri, {\em et~al.}, ``{TARGET 5: A new multi-channel
  digitizer with triggering capabilities for gamma-ray atmospheric Cherenkov
  telescopes},'' {\em Astroparticle Physics} {\bf 92}, 49--61  (2017).

\bibitem{tibaldo2015target}
L.~Tibaldo {\em et~al.}, ``{TARGET: toward a solution for the readout
  electronics of the Cherenkov Telescope Array},'' {\em PoS} {\bf ICRC2015},
  932  (2016).

\bibitem{Brown2015}
A.~M. Brown, T.~Armstrong, P.~M. Chadwick, {\em et~al.}, ``{Flasher and
  muon-based calibration of the GCT telescopes proposed for the Cherenkov
  Telescope Array},'' {\em PoS} {\bf ICRC2015}, 934  (2016).

\bibitem{Hanna2010}
D.~{Hanna}, A.~{McCann}, M.~{McCutcheon}, {\em et~al.}, ``{An LED-based flasher
  system for VERITAS},'' {\em Nuclear Instruments and Methods in Physics
  Research A} {\bf 612}, 278--287  (2010).

\bibitem{Varda2008}
K.~Varda, ``Protocol buffers: Google's data interchange format,'' tech. rep.,
  Google  (2008).
\newblock Accessed: 2022-01-30.

\bibitem{2010AA}
W.~D. {Pence}, L.~{Chiappetti}, C.~G. {Page}, {\em et~al.}, ``{{Definition of
  the Flexible Image Transport System (FITS), version 3.0}},'' {\em Astronomy
  \& Astrophysics} {\bf 524}, A42  (2010).

\bibitem{adams2021detection}
C.~Adams, R.~Alfaro, G.~Ambrosi, {\em et~al.}, ``{Detection of the Crab Nebula
  with the 9.7 m prototype Schwarzschild-Couder telescope},'' {\em
  Astroparticle Physics} {\bf 128}, 102562  (2021).

\bibitem{meuresICRC2019}
C.~Adams {\em et~al.}, ``{{Upgrading the Prototype Schwarzschild-Couder
  Telescope Camera to a Wide-Field, High-Resolution Instrument}},'' {\em 36th
  International Cosmic Ray Conference} , PoS (ICRC2019) 742  (2019).

\bibitem{2018NIMPA.891....6D}
H.~{Dickinson}, F.~{Krennrich}, A.~{Weinstein}, {\em et~al.}, ``{{An
  image-based array trigger for imaging atmospheric Cherenkov telescope
  arrays}},'' {\em Nuclear Instruments and Methods in Physics Research A} {\bf
  891}, 6--17  (2018).

\bibitem{wang2015ftk}
R.~Wang, J.~Anderson, B.~Auerbach, {\em et~al.}, ``{{The FTK to Level-2
  Interface Card (FLIC) for the ATLAS experiment}},'' in {\em 2015 IEEE Nuclear
  Science Symposium and Medical Imaging Conference (NSS/MIC)},  1--4, IEEE
  (2015).

\end{thebibliography}


\pagebreak
\begin{appendices}

\section{TARGET 7 ASIC Sampling Optimization} \label{Appendix:ASIC}
The TARGET 7 ASIC Sampler is comprised of two main parts, the Sampling Array and the Storage Array. The Sampling Array is designed to correctly sample the value of an incoming signal once per nanosecond. Then these values are transferred from the Sampling Array to the Storage Array. Finally, the data are stored in the Storage Array in a known order so that it can be reconstructed at a later time. The timing of this process is determined by six ASIC signals: SSTin, SSPin, STRB1, STRB2, Incr1, and Incr2.

\subsection{ASIC Parameters and Definitions}\label{ASIC Parameters and Definitions}
\begin{description}
	\item[Sampling Array] \hfill \\ 64 capacitors which record incoming signals. It is broken up into Group  1 (capacitors 0--31) and Group 2 (capacitors 32--63).
	\item[Tracking Mode] \hfill \\ When a capacitor in the Sampling Array is allowed to take on the value of the incoming signal (The SSTin/SSpin switch is closed).
	\item[Holding Mode] \hfill \\ When a capacitor in the Sampling Array is held at some fixed value (The SSTin/SSpin switch is open).
	\item[SSTin (Sample STart input)] \hfill \\ The main clock in the ASIC. Its leading edge is defined to be 0~ns and its trailing edge occurs at 32~ns. All other initialization parameters are measured from the SSTin leading edge. The leading edge switches the capacitors from tracking mode to holding mode.
	\item[SSPin (Sample StoP input)] \hfill \\ The leading edge switches the capacitors from holding mode back to tracking mode.
	\item[Storage Array] \hfill \\ An array of 512 blocks (8 rows x 64 columns) with each block containing 32 capacitors. The data taken by each Group in the Sampling Array are transferred to the Storage Array for long term storage.
	\item[STRB (Write Strobe)] \hfill \\ The leading edge closes the switch between the sampling array and the storage array. The trailing edge opens this switch. Together the leading and trailing edges dictate both the time and the duration that the Sampling Array capacitors are connected to the Storage Array capacitors. There are two STRB parameters, one for Sampling Array Group1 (STRB1) and one for Sampling Array Group 2 (STRB2). The switch between the Sampling Array Group and Storage Array must be closed after the signal is already stored and all the capacitors in the group have been converted to holding mode.
	\item[Write Address] \hfill \\ Also known as the Block ID. This is the location within the Storage Array to which values from a Sampling Array Group will be transferred. Each of the 512 locations in the Storage Array has a unique Block ID between 0 and 511.
	\item[Incr (Increment)] \hfill \\ The leading edge increments the write address of the Storage Array by 2 (that is, it tells the STRB command the next place to connect the sampling array). This must occur before the sampling and storage arrays are connected with the STRB command (otherwise part/all of the data will be transferred to the wrong location). Incr1 increments the write address for Group 1 and Incr2 increments the write address for Group 2. Incr1 only writes to even blocks and Incr2 only writes to odd blocks.
\end{description}

\begin{figure}[p]
    \includegraphics[width=\textwidth]{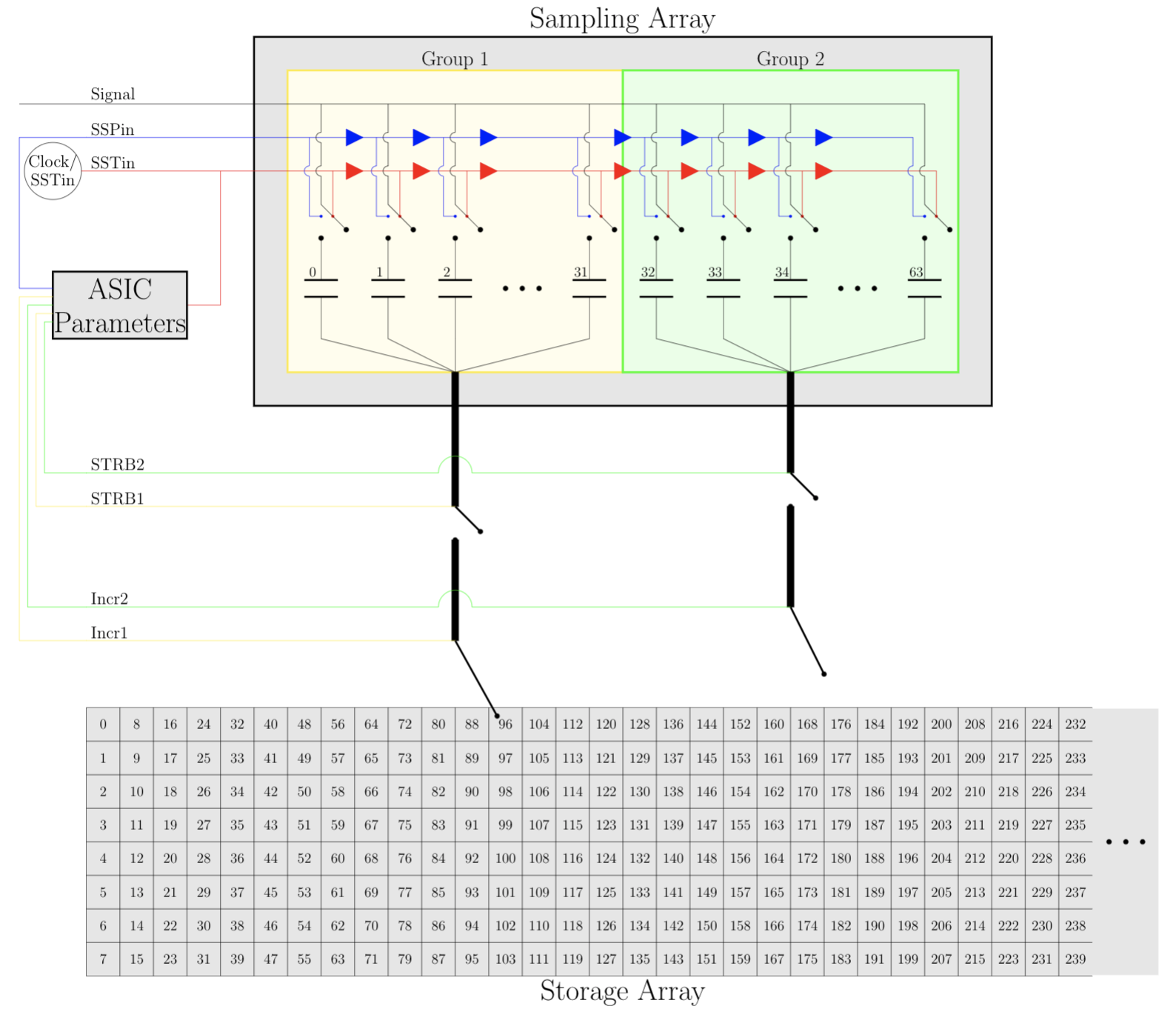}
	\caption{ ASIC Diagram including ASIC Inputs, Sampling Array, and Storage Array and how they connect. The triangles on the SSTin and SSPin lines are 1~ns delay elements. When a signal hits a delay element it is delayed by 1~ns for each element. Notice there are no delay elements on the Signal line. The incoming signal hits all the capacitors simultaneously. The Incr1 and Incr2 switches can connect to any Storage Array block, though the mechanism is more complicated than depicted here.}
	\label{fig:ASIC_Diagram}
\end{figure}

\subsection{ASIC Functions}

\begin{table}
    \centering
	\begin{tabular}{|c|c|} 
	\hline
	Parameter & Setting \\ [0.5ex] 
	\hline\hline
	SSTin\_LE & 0~ns \\
	\hline
	SSTin\_TE & 32~ns \\
	\hline
	SSPin\_LE & 50~ns \\ 
	\hline
	SSPin\_TE & 3~ns \\
	\hline
	Incr1\_LE & 3~ns \\
	\hline
	Incr1\_TE & 18~ns \\
	\hline
	STRB1\_LE & 32~ns \\
	\hline
	STRB1\_TE & 39~ns \\
	\hline
	Incr2\_LE & 35~ns \\
	\hline
	Incr2\_TE & 50~ns \\
	\hline
	STRB2\_LE & 0~ns \\
	\hline
	STRB2\_TE & 7~ns \\
	\hline
	\end{tabular}
    \caption{Optimized ASIC Initialization Parameters}
    \label{table:ASICsettings}
\end{table}

\begin{figure}[p]
\centering
	\includegraphics[width=\textwidth]{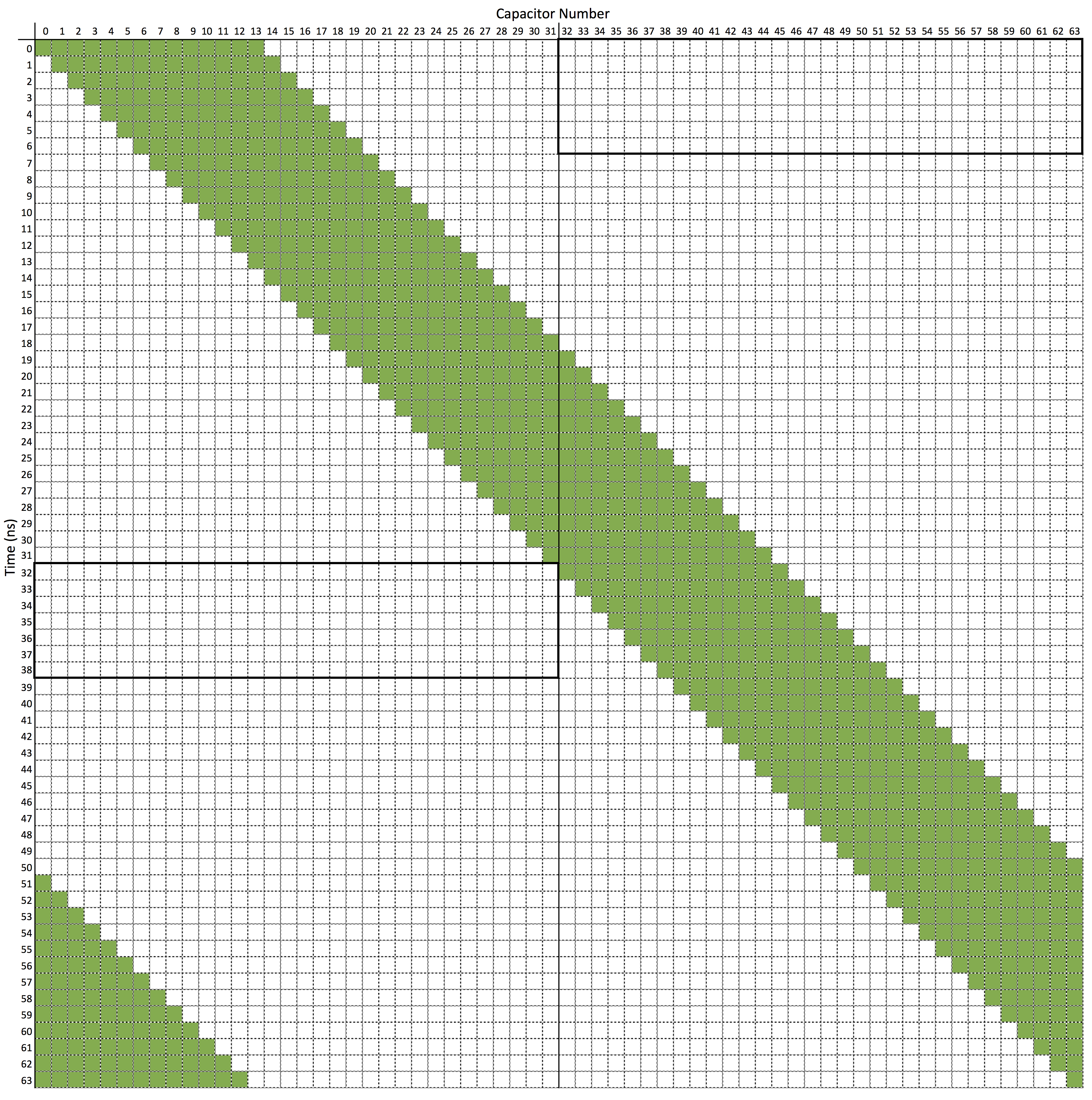}
	\caption{Holding/Tracking diagram for the optimized ASIC settings. Capacitors are labeled on the top of the diagram and as we move down the figure time progresses in steps of 1~ns. The state of each capacitor is given by the color. Green means the capacitor is in tracking mode while white means it is in holding mode. The thick black boxes indicate the duration of the STRB for each group of capacitors.}
	\label{fig:TimingRecommendedParameters}
\end{figure}

\begin{figure}[ht]
    \centering
	\includegraphics[trim=0cm 0cm 9cm 0cm, clip=true, width=0.5\textwidth]{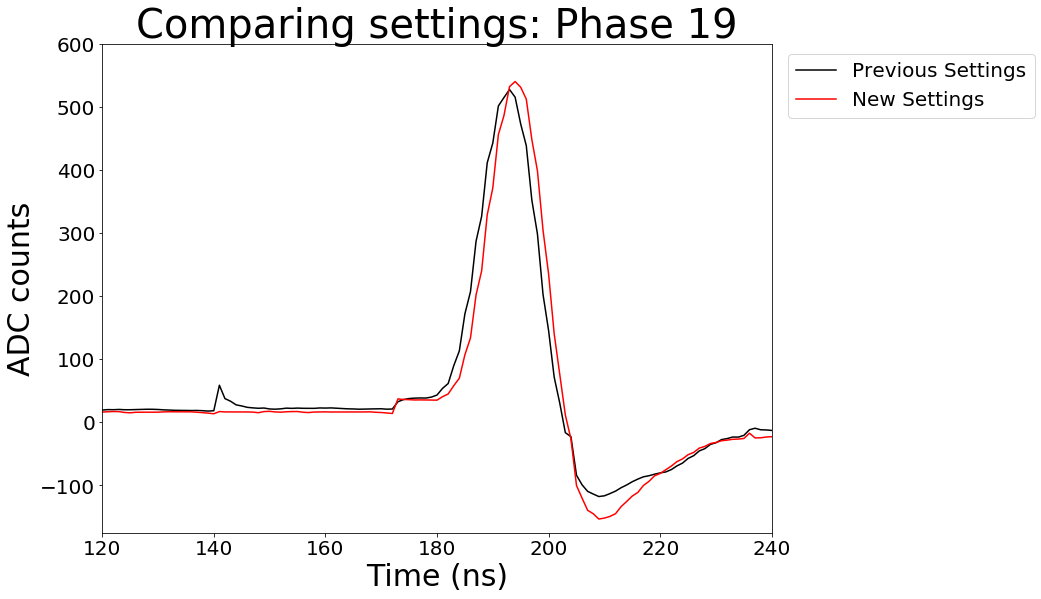}
	\caption{In order to compare the original ASIC settings (shown in black) to the optimized settings (shown in red and listed in table \ref{table:ASICsettings}), data was taken using an LED flasher with an asynchronous trigger provided by an external signal generator. The data was pedestal subtracted and averaged, once for the previous settings and once for the new settings. Because, the location of the pulse in relation to the block boundary (called the ``phase'') determines where the pre-pulse is located in the waveform the two settings were compared individually for every phase. The figure on the right shows the difference between the previous and new settings for a single phase. It is clear that the New ASIC Settings reduce the pre-pulse. Similar results are found for all phases.}
	\label{fig:Old_vs_New}
\end{figure}

The Sampling Array is designed to correctly sample the value of the incoming signal once per nanosecond and then store these values long enough that they can be transferred to long term storage. The way in which this is achieved is by switching each Sampling Array capacitor between Holding Mode and Tracking Mode at the appropriate time. This is achieved through the SSTin and SSPin leading edges. 

Both the SSTin and SSPin parameters must go through 1~ns delays after each capacitor (See Figure \ref{fig:ASIC_Diagram}). The SSTin leading edge is defined to be at 0~ns, thus at 0~ns capacitor 0 switches from Tracking to Holding Mode. However, the SSTin leading edge is delayed by 1~ns before it can reach capacitor 1. Thus capacitor 1 switches from Tracking to Holding Mode at 1~ns. This continues for subsequent capacitors, each with an additional 1~ns delay before they are affected by the SSTin leading edge.

Similarly, the SSPin leading edge hits capacitor 0 at 50~ns. However, capacitor 1 is not affected until 51~ns due to the delay. This, combined with the SSTin signal, produces a moving subsection of capacitors that are in Holding Mode as shown in the Holding/Tracking diagram (Figure \ref{fig:TimingRecommendedParameters}). Each individual capacitor is in Holding Mode for 50~ns while each Group has all capacitors simultaneously in holding mode for only 18~ns. Thus there is an 18~ns window during which one Group of the Sampling Array is storing 32~ns of signal.

Once values have been properly stored short-term in the Sampling Array the next step is to transfer those values into the Storage Array. This transfer process depends on both the Incr and STRB parameters for each group. The STRB parameter connects the Sampling and Storage Array, allowing the stored data from one Group on the Sampling Array to be transferred to one block on the Storage Array. Thus the STRB parameter must begin and end within the 18~ns window during which the Sampling Array is in holding mode. After the values have been transferred using STRB, the Write Address of the Storage Array must be incremented. This is achieved using the Incr parameter. It is important that the Write Address is incremented fully before the corresponding STRB leading edge occurs. Otherwise current values will overwrite the values from the last transfer. 

In order to ensure that the values from each Group do not interfere with one another, the following increment procedure is used. Group 1 only writes to even numbered Storage Array blocks while Group 2 only writes to odd numbered Storage Array blocks. Each Group's Incr parameter increments the Storage Array location by two. This produces a ``ping-pong'' effect in which one Group is transferring data to the Storage Array while the other is taking data and incrementing its write address. After 32~ns the two groups switch. Storing data in this way reduces the chance of error but also means that the signals are not stored in order in the Storage Array. Individual waveforms that are longer than 32~ns are comprised of several Storage Array blocks. The Write Addresses of these blocks will have a +3/-1 pattern: e.g., $0, 3, 2, 5, 4, 7, 6, 9, 8, 11,...$

Incorrect ASIC settings can produce waveforms where part of the signal is located in the incorrect portion of the readout. This often looks like a ``pre-pulse'' occuring before the bulk of the signal. To reduce this effect the ASIC settings were optimized. A comparison of the settings pre and post optimization are shown in Figure \ref{fig:Old_vs_New}. Optimized ASIC settings are shown in Figure \ref{fig:TimingRecommendedParameters} and in Table \ref{table:ASICsettings}.

\newpage
\section{Component Photos}
\label{Appendix:Photos}
\begin{figure}[hb]
    \centering
	\includegraphics[width=\textwidth]{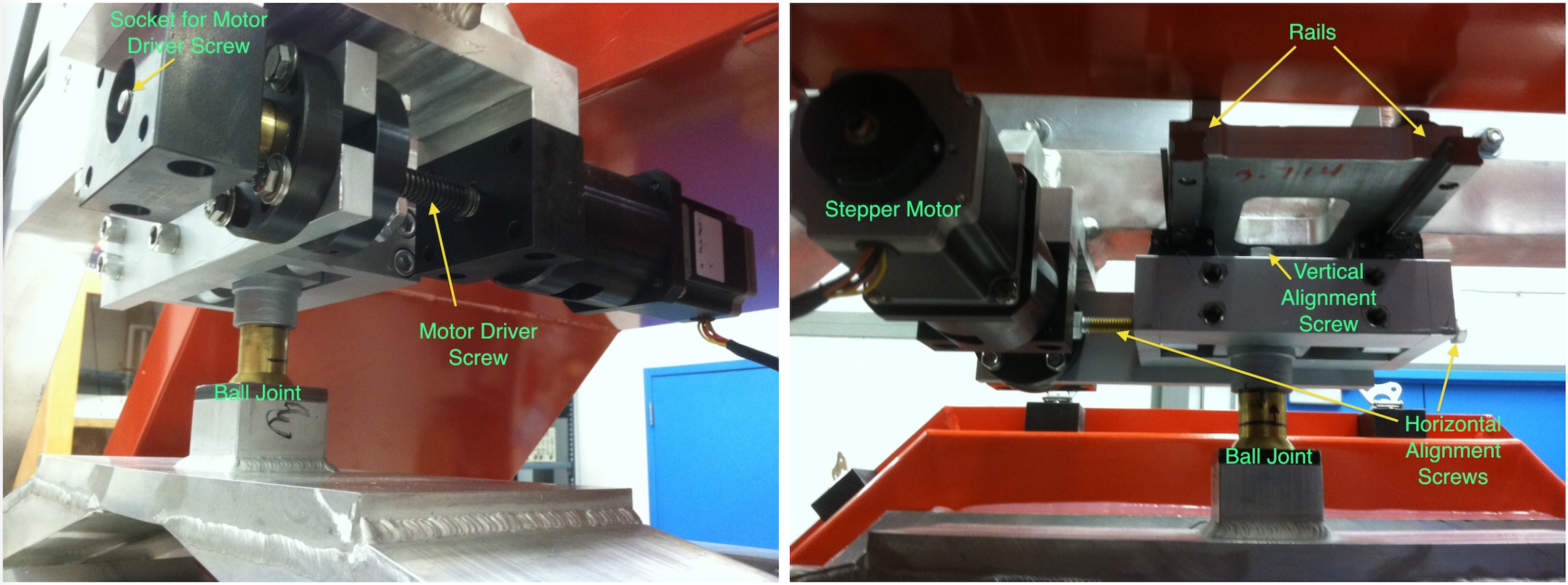}
	\caption{Photo of Motor assembly C, located at the top of the camera. Labeled are the ball joint, motor driver screw, rails, vertical and horizontal alignment screws, and the stepper motor.}
	\label{fig:MotorAssembly}
\end{figure}

\begin{figure}[hb]
    \centering
	\includegraphics[width=\textwidth]{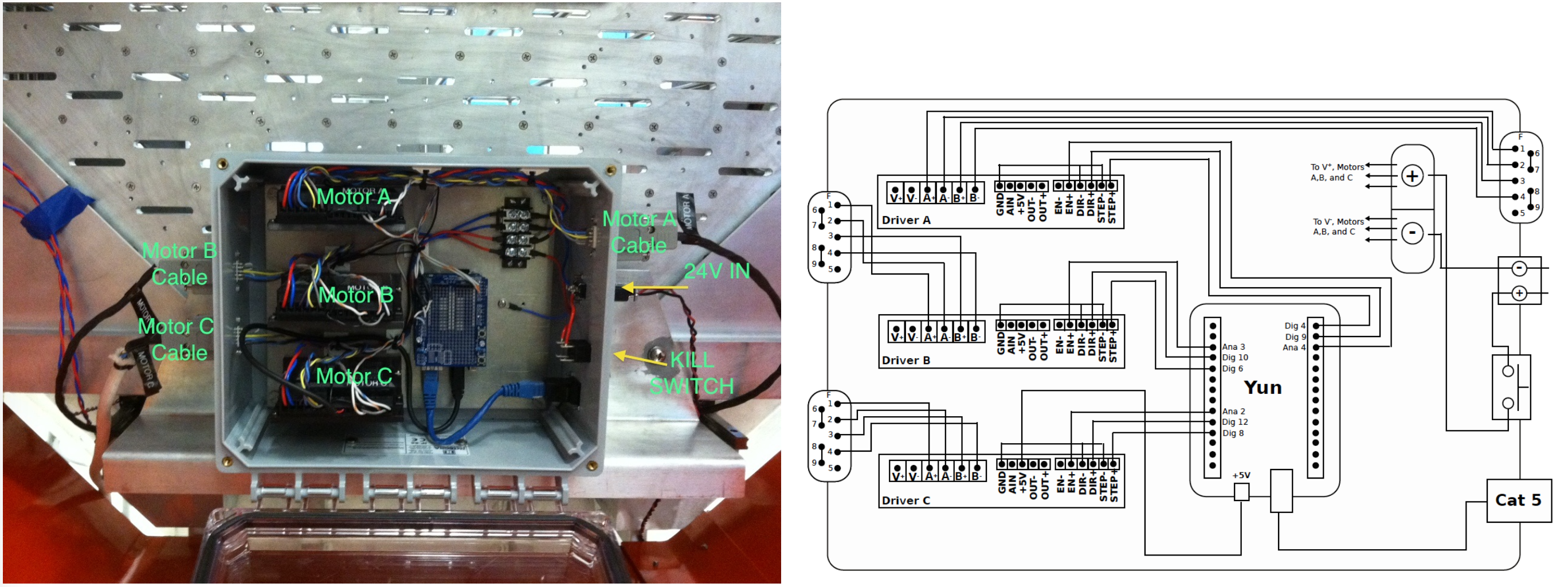}
	\caption{Photo (left) and circuit diagram (right) of the motor electronics box located in the back of the camera. This box includes an Arduino and three motors which control the movement of each stepper motor.}
	\label{fig:MotorBox}
\end{figure}

\begin{figure}[hb]
    \centering
    \includegraphics[width=0.5\textwidth,angle=0]{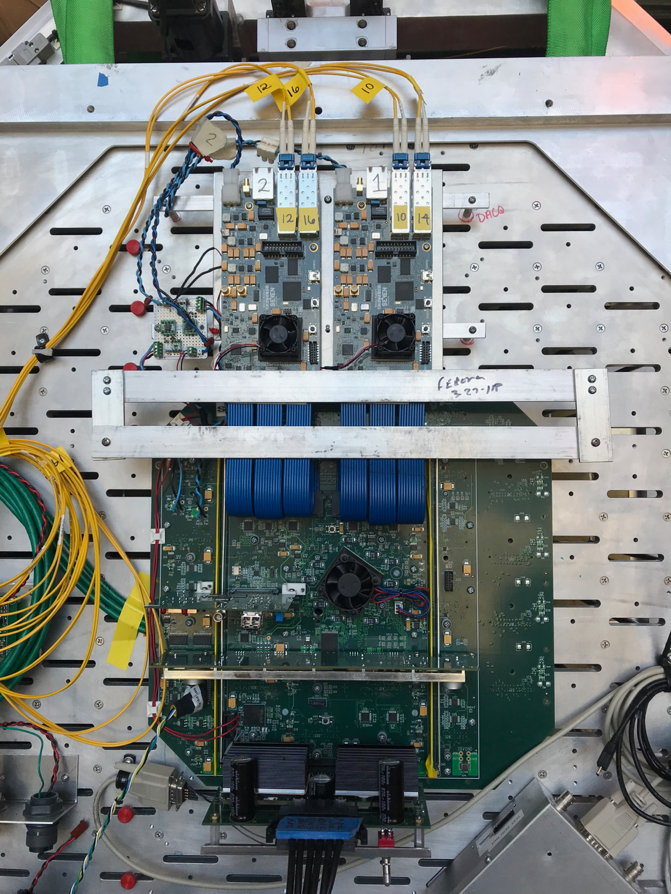}
    \caption{Photo of the backplane. The back bulkhead has two holes per module, one long slit for the module backplane connector and one for the module securing screw. The backplane is mounted onto this back bulkhead. Each module connects to the backplane through a backplane connector and is secured with a screw running through the backplane, bulkhead and module. On top of the backplane are mounted two DACQ boards (top), one power board (middle) and two power connectors (bottom). The DACQ boards have two fiber cables each (yellow) leading to the network switch. The green cables allow a central alignment device, which can be installed instead of the central module, to connect to a positioner. This device is used to align the camera and optics. The Raspberry Pi is in an aluminum box in the lower right hand corner and is connected to the backplane via an SPI link through a D-shell connector. The fans are connected to the backplane via a small board in the upper left hand corner.}
    \label{fig:BackplanePhoto}
\end{figure}

\begin{figure}[hb]
    \centering
    \includegraphics[width=0.7\textwidth,angle=0]{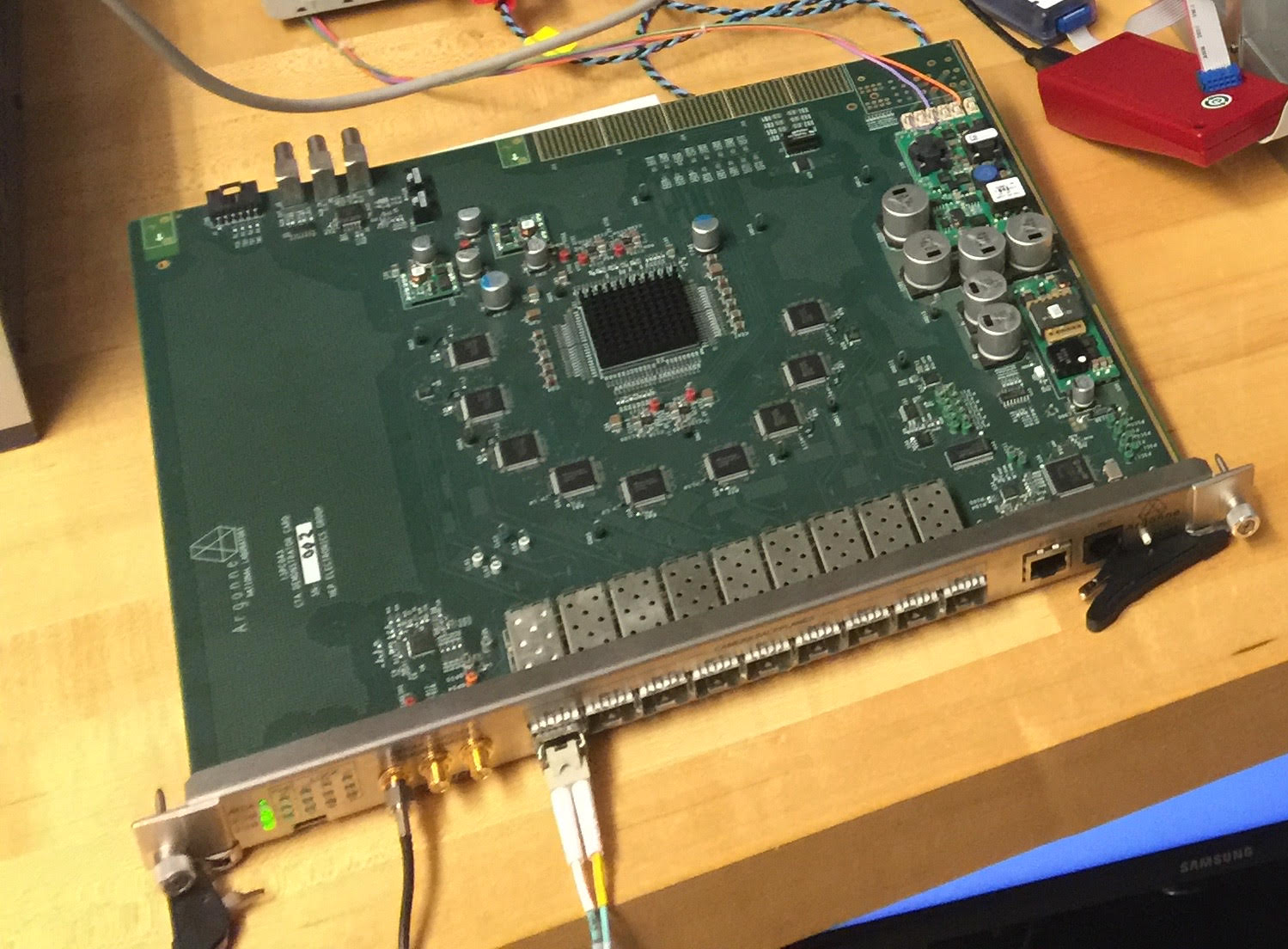}
    \caption{Photo of the demonstrator Distributed Intelligent Array Trigger (DIAT) on a laboratory test stand, connected to the pSCT camera backplane.  The DIAT is a single-board custom-designed module for time-tagging, timing synchronization, hit pattern processing and cross-triggering.  Fiber optic ports on the front panel connect to up to nine SCT backplanes for the full-sized camera.}
    \label{f:DIAT_phot}
\end{figure}

\begin{figure}[hb]
    \centering
    \includegraphics[width=\textwidth]{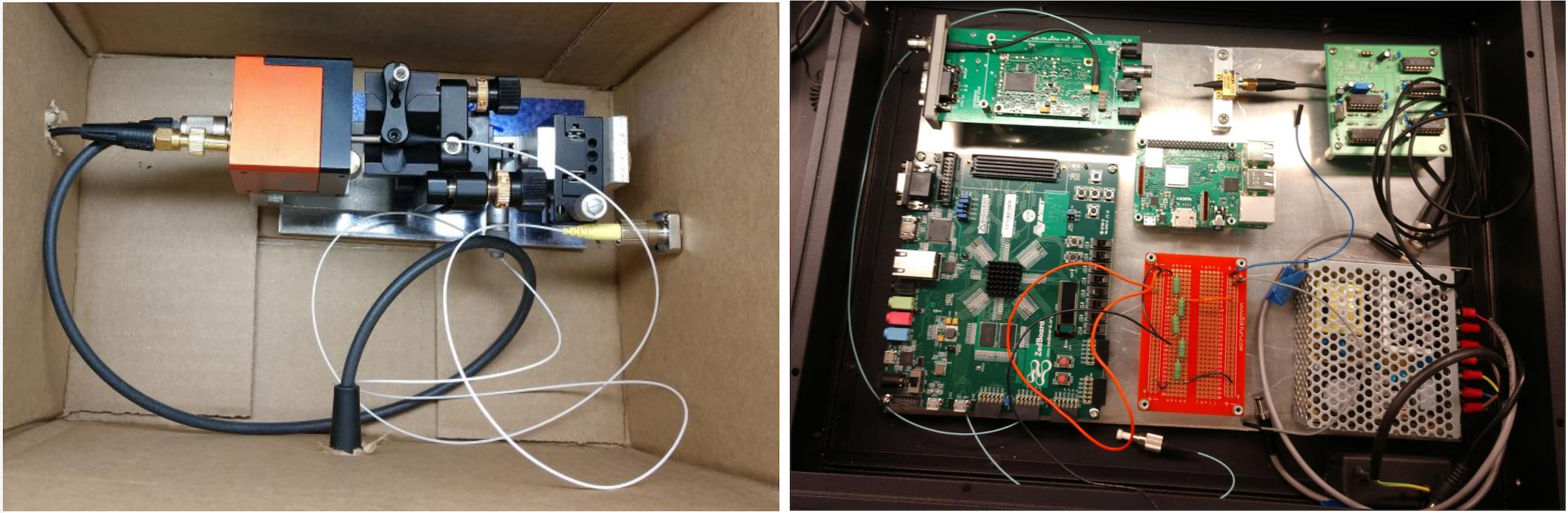}
    \caption{Photos showing main components of the stand-alone timing system:  the laser-diode module which is installed near the telescope camera backplane (left) and the photo-diode and GPS timing module which is installed in the pSCT trailer (right).}
    \label{fig:ttag_photo}
\end{figure}

\begin{figure}[hb]
    \centering
    \includegraphics[width=0.5\textwidth,angle=0]{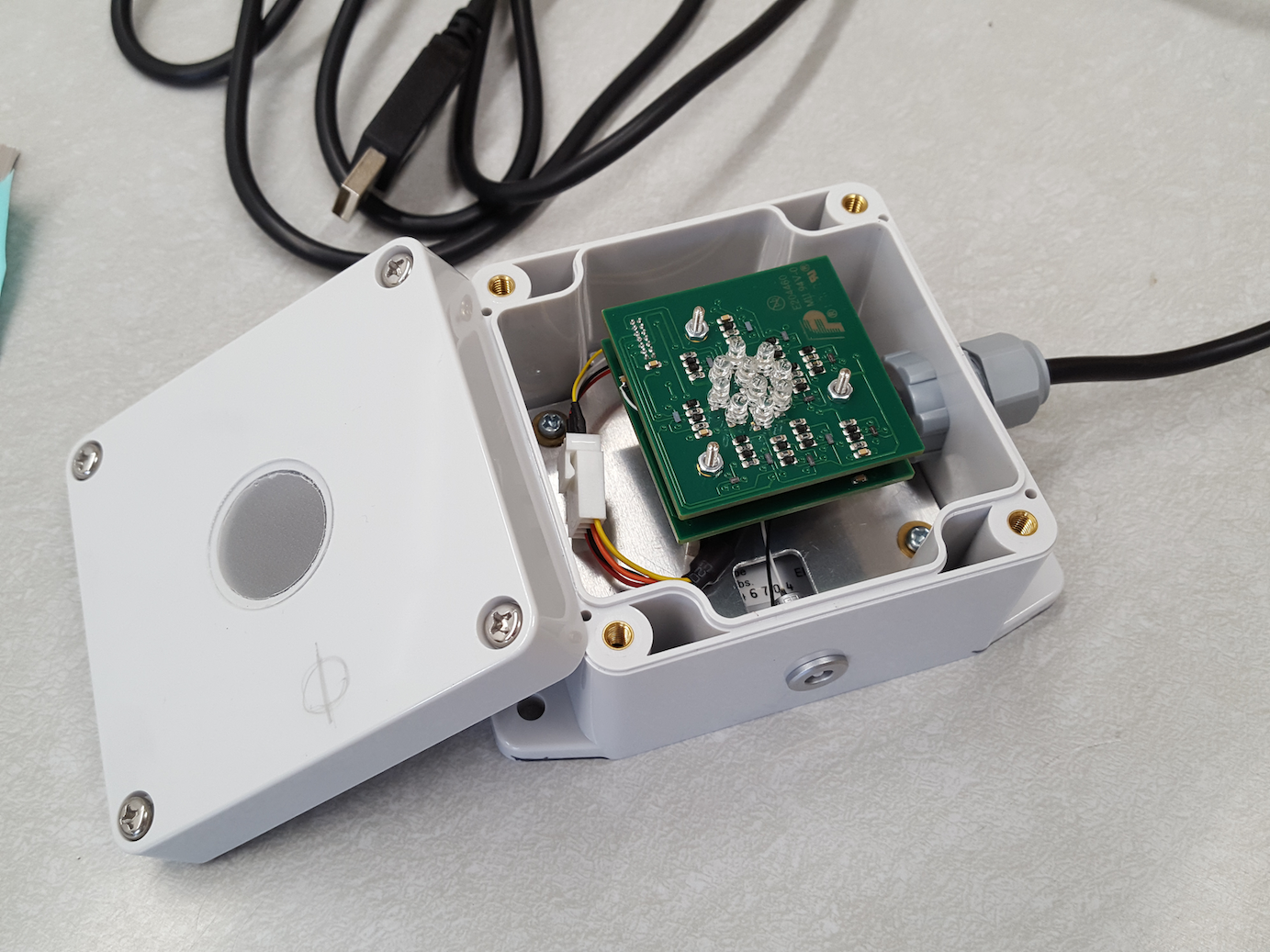}
    \caption{Photograph of one of three LED flashers currently installed in the pSCT.  The cover containing the diffuser is open to expose the LEDs and circuit inside the module. The flasher enclosure is 8~cm x 8~cm x 5.5~cm and the diffuser lens has a diameter of 2.5~cm}
    \label{fig:LEDphoto}
\end{figure}

Figures \ref{fig:MotorAssembly}-\ref{fig:LEDphoto} show photos of several pSCT components.

\end{appendices}

\FloatBarrier
\listoffigures

\end{spacing}
\end{document}